%% file: v1.tex
\documentclass[12pt,bibliography=totoc,DIV=19]{scrartcl}
\pdfoutput=1

\input{paperdef}

\begin{document}


\thispagestyle{empty}

\def\thefootnote{\fnsymbol{footnote}}

\begin{flushright}
DESY--18--084\\
IFT--UAM/CSIC--17--125
\end{flushright}

\vfill

\begin{center}

\mytitle{
Decays of the neutral Higgs bosons\\
into SM fermions and gauge bosons\\
in the \cp-violating NMSSM
}

\vspace{1cm}

Florian Domingo$^{1,2}$\footnote{email: florian.domingo@csic.es},
Sven Heinemeyer$^{1,2,3}$\footnote{email: sven.heinemeyer@cern.ch},
Sebastian Pa{\ss}ehr$^{4,5}$\footnote{email: passehr@lpthe.jussieu.fr}
and
Georg Weiglein$^4$\footnote{email: georg.weiglein@desy.de}

\vspace*{1cm}

\textsl{
$^1$Instituto de Física Teórica (UAM/CSIC), Universidad Autónoma de Madrid,\\
Cantoblanco, E--28049 Madrid, Spain
}

\medskip
\textsl{
$^2$Instituto de Física de Cantabria (CSIC-UC),
E--39005 Santander, Spain
}

\medskip
\textsl{
$^3$Campus of International Excellence UAM+CSIC,
Cantoblanco, E--28049, Madrid, Spain
}

\medskip
\textsl{
$^4$Deutsches Elektronensynchrotron DESY,
Notkestraße 85, D--22607 Hamburg, Germany
}

\medskip
\textsl{
$^5$Sorbonne Université, CNRS,
Laboratoire de Physique Théorique et Hautes Énergies (LPTHE),\\
4 Place Jussieu, F--75252 Paris CEDEX~05, France
}

\end{center}

\vfill

\begin{abstract}{}
\input{00_Abstract}
\end{abstract}

\vfill

\def\thefootnote{\arabic{footnote}}
\setcounter{page}{0}
\setcounter{footnote}{0}
\newpage
\hypersetup{linkcolor=black}
\tableofcontents\label{TOC}
\hypersetup{linkcolor=blue}

\input{01_Introduction}
\input{02_NMSSM}
\input{03_NumericalAnalysis}
\input{04_Conclusions}

\section*{\tocref{Acknowledgments}}

We thank T.~Hahn, M.~Mühlleitner, P.~Slavich, M.~Spira and
D.~Stöckinger for fruitful discussions. The work of F.~D. and
S.~H. was supported in part by the MEINCOP~(Spain) under contract
\mbox{FPA2016-78022-P}, in part by the ``Spanish Agencia Estatal de
Investigaci\'on''~(AEI) and the EU~``Fondo Europeo de Desarrollo
Regional''~(FEDER) through the project \mbox{FPA2016-78022-P}, and in
part by the~AEI through the grant IFT~Centro de Excelencia Severo
Ochoa \mbox{SEV-2016-0597}. In addition, the work of S.~H. is
supported in part by the ``Spanish Red Consolider MultiDark''
\mbox{FPA2017-90566-REDC}. During different stages of the project,
S.~P. acknowledges support by the \mbox{Collaborative Research
  Center~SFB676} of the~DFG, ``Particles, Strings and the early
Universe'' and by the~ANR grant \mbox{``HiggsAutomator''}
\mbox{(ANR-15-CE31-0002)}. G.~W. acknowledges support by the
Collaborative Research Center~SFB676 of the~DFG, ``Particles, Strings
and the early Universe''.

\begingroup
\let\secfnt\undefined
\newfont{\secfnt}{ptmb8t at 10pt}
\setstretch{.5}
\bibliographystyle{h-physrev}     
\bibliography{literature}
\endgroup


\end{document}

%% file: paperdef.tex
\usepackage[T1]{fontenc}
\usepackage[utf8]{inputenc}
\newcommand{\unicodelesssim}{^^e2^^89^^b2}
\newcommand{\unicodemathcalC}{^^f0^^9d^^93^^92}
\newcommand{\unicodemathcalP}{^^f0^^9d^^93^^9f}

\usepackage{amsmath,amsfonts}
\usepackage[mathscr]{euscript}
\usepackage{booktabs}
\usepackage{cite}
\usepackage{graphicx, subfigure, ulem, hhline}
\usepackage[table]{xcolor}
\usepackage{siunitx}
\usepackage{soul}
\usepackage{wasysym}              
\usepackage{hyphenat}             

\usepackage{etoolbox,xstring,xspace,calc,xifthen}

\let\theparentequation\theequation
\patchcmd{\theparentequation}{equation}{parentequation}{}{}

\renewenvironment{subequations}[1][]{
  \refstepcounter{equation}%
  \setcounter{parentequation}{\value{equation}}
  \setcounter{equation}{0}
  \def\theequation{\theparentequation\alph{equation}}%
  \let\parentlabel\label
  \ifx\\#1\\\relax\else\label{#1}\fi
  \ignorespaces
}{%
  \setcounter{equation}{\value{parentequation}}
  \ignorespacesafterend
}

\newcommand*{\nextParentEquation}[1][]{
  \refstepcounter{parentequation}
  \setcounter{equation}{0}
  \ifx\\#1\\\relax\else\parentlabel{#1}\fi
}

\makeatletter
\renewcommand\@makefntext[1]{\leftskip=.8em\hskip-.62em\@makefnmark#1}
\makeatother

\usepackage{setspace}

\usepackage{jheppub}
\usepackage[all]{hypcap}
\usepackage{footnotebackref}

\usepackage[format=plain, margin=0.5cm, font=footnotesize, labelfont=bf]{caption}
\DeclareCaptionSubType*[arabic]{figure}
\DeclareCaptionSubType*[arabic]{table}
\let\theparentequation\theequation
\patchcmd{\theparentequation}{equation}{parentequation}{}{}

\apptocmd{\thebibliography}{\scriptsize}{}{}
\let\OLDthebibliography\thebibliography
\renewcommand\thebibliography[1]{
  \OLDthebibliography{#1}
  \setlength{\parskip}{1pt}
  \setlength{\itemsep}{1pt plus 0.3ex}
}

\newcommand{\citere}[1]{Ref.\,\cite{#1}}
\newcommand{\citeres}[1]{Refs.\,\cite{#1}}
\newcommand{\simord}{\mathord{\sim}\,}
\newcommand{\simeqord}{\mathord{\simeq}\,}
\newcommand{\lsimord}{\mathord{\lesssim}\,}
\newcommand{\gsimord}{\mathord{\gtrsim}\,}
\newcommand{\leqord}{\mathord{\leq}\,}

\def\reffi#1{\mbox{Fig.\,\ref{#1}}}

\def\refta#1{\mbox{Tab.\,\ref{#1}}}

\def\ga{\gamma}

\newcommand{\Amp}[4][\mathcal{A}]{#1^{\mbox{\tiny #2}}_{\mbox{\tiny #3}}\ifthenelse{\isempty{#4}}{}{{\left[#4\right]}}}
\newcommand{\Un}[1]{{\left(U_n\right)}_{#1}}
\newcommand{\IE}{\textit{i.\,e.}}
\newcommand{\EG}{\textit{e.\,g.}}
\newcommand{\I}{\imath}

\newcommand{\Bnull}[1]{B_{0}{\left(#1\right)}}

\newcommand{\mueff}{\ensuremath{\mu_{\text{eff}}}}

\newcommand{\cp}{\ensuremath{{\cal CP}}}

\newcommand{\FH}{\texttt{FeynHiggs}}

\newcommand{\DRbar}{\ensuremath{\overline{\mathrm{DR}}}}
\newcommand{\MSbar}{\ensuremath{\overline{\mathrm{MS}}}}

\def\refeq#1{\mbox{Eq.\,\eqref{#1}}}

\makeatletter
\DeclareRobustCommand\em{%
  \@nomath\em \ifdim \fontdimen\@ne\font >\z@\scshape
  \else \slshape \fi}
\makeatother

\renewcommand{\emph}[1]{{\em #1}}

\addtokomafont{disposition}{\rmfamily}
\BeforeTOCHead[toc]{{\pdfbookmark[1]{\contentsname}{toc}}}

\usepackage{diagbox}

\usepackage{array}

\hypersetup{
  pdfauthor={Florian Domingo, Sven Heinemeyer, Sebastian Paßehr and Georg Weiglein}
  ,pdftitle={Decays of the neutral Higgs bosons into SM fermions and
    gauge bosons in the \unicodemathcalC\unicodemathcalP-violating NMSSM}
  ,pdfsubject={}
  ,pdfkeywords={}
}

\newcommand*{\mytitle}[1]{%
  \parbox{\linewidth}{\setstretch{1.5}\centering\Large\textsc{\textbf{\boldmath #1}}}
}



%% file: 00_Abstract.tex
The Next-to-Minimal Supersymmetric Standard Model~(NMSSM) offers a
rich framework embedding physics beyond the Standard Model as well as
consistent interpretations of the results about the Higgs signal
detected at the~LHC. We investigate the decays of neutral Higgs states
into Standard Model~(SM) fermions and gauge bosons. We perform full
one-loop calculations of the decay widths and include leading
higher-order QCD~corrections. We first discuss the technical aspects
of our approach, before confronting our predictions to those of
existing public tools, performing a numerical analysis and discussing
the remaining theoretical uncertainties. In particular, we find that
the decay widths of doublet-dominated heavy Higgs bosons into
electroweak gauge bosons are dominated by the radiative corrections,
so that the tree-level approximations that are often employed in
phenomenological analyses fail. Finally, we focus on the
phenomenological properties of a mostly singlet-like state with a mass
below the one at~125$\,$GeV, a scenario that appears commonly within
the~NMSSM. In fact, the possible existence of a singlet-dominated
state in the mass range around or just below~100$\,$GeV would have
interesting phenomenological implications. Such a scenario could
provide an interpretation for both the~$2.3\,\sigma$ local excess
observed at~LEP in the~\mbox{$e^+e^-\to Z(H\to b\bar{b})$}~searches
at~$\simord 98$\,GeV and for the local excess in the diphoton searches
recently reported by~CMS in this mass range, while at the same time it
would reduce the ``Little Hierarchy'' problem.

%% file: 01_Introduction.tex
\section[Introduction]{\tocref{\label{sec:intro}Introduction}}

The signal that was discovered in the Higgs searches at~ATLAS and~CMS
at a mass of~$\simord
125$\,GeV\,\cite{Aad:2012tfa,Chatrchyan:2012ufa,Khachatryan:2016vau}
is, within the current theoretical and experimental uncertainties,
compatible with the properties of the Higgs boson predicted within
Standard-Model~(SM) of particle physics. No conclusive signs of
physics beyond the~SM have been reported so far. However, the
measurements of Higgs signal strenghts for the various channels leave
considerable room for Beyond Standard Model~(BSM)
interpretations. Consequently, the investigation of the precise
properties of the discovered Higgs boson will be one of the prime
goals at the~LHC and beyond.  While the mass of the observed particle
is already known with excellent
accuracy\,\cite{Aad:2015zhl,Sirunyan:2017exp}, significant
improvements of the information about the couplings of the observed
state are expected from the upcoming runs of
the~LHC\,\cite{CMS-HL,ATLAS-HL,Khachatryan:2016vau,Testa:2017,Cepeda:2017}
and even more so from the high-precision measurements at a future
$e^+e^-$~collider\,\cite{Baer:2013cma,Fujii:2015jha,Fujii:2017ekh,Moortgat-Picka:2015yla}.

Motivated by the ``Hierarchy Problem'', Supersymmetry~(SUSY)-inspired
extensions of the~SM play a prominent role in the investigations of
possible new physics. As such, the Minimal Supersymmetric Standard
Model~(MSSM)\,\cite{Nilles:1983ge,Haber:1984rc} or its singlet
extension,
the~Next-to-MSSM~(NMSSM)\,\cite{Ellwanger:2009dp,Maniatis:2009re},
have been the object of many studies in the last decades. Despite this
attention, these models are not yet prepared for an era of precision
tests as the uncertainties at the level of the Higgs-mass
calculation\,\cite{Degrassi:2002fi,Staub:2015aea,Drechsel:2016htw} are
about one order of magnitude larger than the experimental
uncertainty. At the level of the decays, the theoretical uncertainty
arising from unknown higher-order corrections has been estimated for
the case of the Higgs boson of the~SM (where the Higgs mass is treated
as a free input parameter) in
Refs.\,\cite{Denner:2011mq,Heinemeyer:2013tqa} and updated in
Ref.\,\cite{deFlorian:2016spz}: depending on the channel and the Higgs
mass, it typically falls in the range of~$\simord 0.5$--$5\%$. To our
knowledge, no similar analysis has been performed in SUSY-inspired
models, but one can expect the uncertainties from missing higher-order
corrections to be larger in general---with many nuances depending on
the characteristics of the Higgs state and the considered point in
parameter space: we provide some discussion of this issue at the end
of this paper. In addition, parametric uncertainties that are induced
by the experimental errors of the input parameters should be taken
into account as well. For the case of the~SM decays those parametric
uncertainties have been discussed in the references above. In
the~SUSY~case the parametric uncertainties induced by the (known)~SM
input parameters can be determined in the same way as for the~SM,
while the dependence on unknown~SUSY parameters can be utilised for
setting constraints on those parameters. While still competitive
today, the level of accuracy of the theoretical predictions of
Higgs-boson decays in~SUSY~models should soon become outclassed by the
achieved experimental precision on the decays of the observed Higgs
signal. Without comparable accuracy of the theoretical predictions,
the impact of the exploitation of the precision data will be
diminished---either in terms of further constraining the parameter
space or of interpreting deviations from the~SM results. Further
efforts towards improving the theoretical accuracy are therefore
necessary in order to enable a thorough investigation of the
phenomenology of these models. Besides the decays of the~SM-like state
at~125$\,$GeV of a~SUSY~model---where the goal is clearly to reach an
accuracy that is comparable to the case of the~SM---it is also of
interest to obtain reliable and accurate predictions for the decays of
the other Higgs bosons in the spectrum. The decays of the non-SM-like
Higgs bosons can be affected by large higher-order corrections as a
consequence of either large enhancement factors or a suppression of
the lowest-order contribution. Confronting accurate predictions with
the available search limits yields important constraints on the
parameter space.

In this paper we present an evaluation of the decays of the neutral
Higgs bosons of the $\mathbb{Z}_3$-conserving~NMSSM into
SM~particles. The extension of the~MSSM by a gauge-singlet superfield
was originally motivated by the~`$\mu$~problem'\,\cite{Kim:1983dt},
but also leads to a richer phenomenology in the Higgs sector (see
\EG\ the introduction of \citere{Domingo:2017rhb} for a recent summary
of related activities). Several public tools provide an implementation
of Higgs decays in the NMSSM:
\texttt{HDECAY}\,\cite{Djouadi:1997yw,Butterworth:2010ym,Djouadi:2018xqq},
focussing on the~SM and~MSSM, was the object of various extensions to
the~\cp-conserving or~-violating~NMSSM for
\texttt{NMSSMTools}\,\cite{Ellwanger:2004xm,
  Ellwanger:2005dv,Domingo:2015qaa,NMSSMTOOLS-www} and
\texttt{NMSSMCALC}\,\cite{Baglio:2013iia, NMSSMCALC-www};
\texttt{SOFTSUSY}\,\cite{Allanach:2001kg,Allanach:2013kza} recently
released its own set of routines\,\cite{Allanach:2017hcf}, which
generally are confined to the leading order or leading
QCD~corrections; a one-loop evaluation of two-body decays in
the~\DRbar~$\left(\MSbar\right)$~scheme for generic
models\,\cite{Goodsell:2017pdq} has been recently presented
for~\texttt{SPHENO}\,\cite{Porod:2003um,Porod:2011nf,Goodsell:2014bna,spheno-www},
which employs
\texttt{SARAH}\,\cite{Staub:2009bi,Staub:2010jh,Staub:2012pb,Staub:2013tta}.
Moreover, the non-public code~\texttt{SloopS}\,\cite{sloops-www} has
been extended to the~NMSSM\,\cite{Belanger:2016tqb,Belanger:2017rgu}
and applied to the calculation of Higgs
decays\,\cite{Belanger:2014roa,Belanger:2017rgu}.

The current work focussing on NMSSM Higgs decays is part of the effort
for developing a version
of~\texttt{FeynHiggs}\,\cite{Heinemeyer:1998np,Heinemeyer:1998yj,
  Degrassi:2002fi, Frank:2006yh, Hahn:2013ria, Bahl:2016brp,
  Bahl:2017aev, FH-www} dedicated to
the~NMSSM\,\cite{Drechsel:2016jdg,Domingo:2017rhb}. The general
methodology relies on a Feynman-diagrammatic calculation of radiative
corrections, which
employs~\texttt{FeynArts}\,\cite{Kublbeck:1990xc,Hahn:2000kx},
\texttt{FormCalc}\,\cite{Hahn:1998yk}
and~\texttt{LoopTools}\,\cite{Hahn:1998yk}. The implementation of the
renormalization scheme within the~NMSSM\,\cite{Domingo:2017rhb} has
been done in such a way that the result in the~MSSM~limit of the~NMSSM
exactly coincides with the~MSSM result obtained from
\texttt{FeynHiggs} without any further adjustments of parameters
(cases where the~NMSSM~result is more complete than the current
implementation of the~MSSM~result will be discussed below).
Concerning the Higgs decays, our routines, in their current status,
contain an evaluation of all the two-body decays of neutral and
charged Higgs bosons. In the present paper, we wish to focus on decays
of the neutral Higgs bosons into SM~final states, where we have
obtained results including higher-order contributions as detailed
below as well as further refinements. The channels of the type
Higgs-to-Higgs and Higgs-to-SUSY are currently only implemented at
leading order.\footnote{Within the~MSSM (including complex parameters)
  the Higgs-to-Higgs decays are implemented into \FH\ at the full
  one-loop level\,\cite{Williams:2007dc,Williams:2011bu}, and
  Higgs-to-SUSY decays have been calculated at the full one-loop level
  in \citeres{Heinemeyer:2014yya,Heinemeyer:2015pfa} (see also
  \citere{Fowler:2009ay}).  Moreover, the~$Z$~factors (as implemented
  in \FH) include corrections beyond one loop, corresponding to the
  Higgs-boson mass calculation.}
  
For the evaluation of the decays of the neutral Higgs bosons of
the~NMSSM into SM~final states we have followed the same general
approach as for the implementation of the~MSSM Higgs decays
in~\texttt{FeynHiggs}, which have been described in
\citeres{Heinemeyer:2000fa,Williams:2011bu}. At the level of the
external Higgs fields, mixing effects are consistently taken into
account as explained in \citeres{Fuchs:2016swt,Domingo:2017rhb}. Full
one-loop contributions are considered in the fermionic decay channels,
supplemented by~QCD~higher-order corrections. For the bosonic decay
modes generated at the radiative order, leading QCD~corrections are
taken into account. For decays into massive electroweak gauge bosons,
the implementation in the~MSSM is such that \texttt{FeynHiggs} first
extracts the loop-corrected width
that~\texttt{Prophecy4f}\,\cite{Bredenstein:2006rh,Bredenstein:2006nk,Bredenstein:2006ha}
calculates in the~SM for a Higgs boson at a given mass, and then
rescales this result by the squared coupling of the~MSSM Higgs boson
to~$WW$ and~$ZZ$, normalized to the~SM~value. We go beyond this
approach and include full one-loop on-shell results for these decay
widths---however, the on-shell kinematical factor of the tree-level
contribution is replaced by its off-shell counterpart, leading to a
tree-level estimate below threshold.\footnote{As the on-shell
  kinematical factor multiplying the contributions of one-loop order
  vanishes at threshold, the predicted width remains a continuous
  function of the Higgs mass.} More generally, the refinements that we
implement, \EG~the inclusion of higher-order corrections, often
surpass the assumptions made in public codes dedicated to
the~NMSSM. However, we stress that we strictly confine ourselves to
the free-particle approximation for the final states. Accordingly,
dedicated analyses would be needed for a proper treatment of decays
close to threshold, since finite-width effects need to be taken into
account in this region, and effects of the interactions among the
final states can be very
large\,\cite{Bigi:1991mi,Melnikov:1994jb,Drees:1989du,Domingo:2011rn,Domingo:2016yih}.

The case where the Higgs spectrum contains a singlet-dominated Higgs
state with a mass below the one of the signal detected at~125$\,$GeV
is a particularly interesting scenario that commonly emerges in
the~NMSSM. Among the appealing features of this class of scenarios it
should be mentioned that the somewhat high mass (in an MSSM~context)
of the SM-like Higgs state observed at the~LHC could be understood
more naturally as the result of the mixing with a lighter singlet
state---see \EG~\citere{Domingo:2015eea} for a list of references and
an estimate of the possible uplift in mass. Furthermore, the existence
of a mostly singlet-like state in the range of~$\lsimord 100$\,GeV has
been suggested\,\cite{Dermisek:2007ah,Belanger:2012tt} as a possible
explanation of the~$2.3\,\sigma$ local excess observed at~LEP in
the~\mbox{$e^+e^-\to Z(H\to
  b\bar{b})$}~searches\,\cite{Barate:2003sz}. More recently,
the~CMS~collaboration has reported a local excess in its diphoton
Higgs searches for a mass in the vicinity of~$\simord
96$\,GeV\,\cite{CMS:2017yta}. This excess reaches
the~(local)~$2.9\,\sigma$~level in the~Run\,II~data and has already
received attention from the particle-physics
community\,\cite{Mariotti:2017vtv,Crivellin:2017upt,Fox:2017uwr,Cacciapaglia:2017iws}.
A similar excess~($2.0\,\sigma$) was already present in
the~Run\,I~data in the same mass range. In an~NMSSM~context, the
possibility of large diphoton signals is
well-known\,\cite{Ellwanger:2010nf,Benbrik:2012rm,Ellwanger:2015uaz}. Here
we show that it is in fact possible to describe simultaneously the~LEP
and the~CMS~excesses within the~NMSSM. However, it should be kept in
mind that the excesses that were observed at~LEP and~CMS at~$\lsimord
100$\,GeV could of course just be statistical fluctuations of the
background and that their possible explanation in terms of an~NMSSM
Higgs state remains somewhat speculative. In particular, the diphoton
excess observed at~CMS would of course require confirmation from
the~\mbox{ATLAS}~data as well\,\cite{ATLAS:2018xad}.

In the following section, we discuss the technical aspects of our
calculation, describing the conventions, assumptions and higher-order
corrections that we address. Sect.\,\ref{sec:numerics} illustrates the
workings of our decay routines in several scenarios of the~NMSSM, and
we perform comparisons with existing public tools. We also investigate
the~NMSSM~scenario with a mostly singlet-like state with mass close
to~$100$\,GeV. Furthermore we discuss the possible size of the remaining 
theoretical uncertainties from unknown higher-order corrections. 
The conclusions are summarized in Sect.\,\ref{sec:conclusion}.

%% file: 02_NMSSM.tex
\section[\boldmath Higgs decays to SM particles in the \cp-violating NMSSM]{\tocref{\label{sec:theory}\boldmath Higgs decays to SM particles in the \cp-violating NMSSM}}

In this section, we describe the technical aspects of our calculation
of the Higgs decays. Our notation and the renormalization scheme that
we employ for the~$\mathbb{Z}_3$-conserving~NMSSM in the general case
of complex parameters were presented in
Sect.\,\href{https://arxiv.org/pdf/1706.00437.pdf#section.2}{$2$} of
\citere{Domingo:2017rhb}, and we refer the reader to this article for
further details.

\subsection[Decay amplitudes for a physical (on-shell) Higgs state -- Generalities]{\tocref{Decay amplitudes for a physical (on-shell) Higgs state -- Generalities\label{sec:general}}}

\paragraph{On-shell external Higgs leg}

In this article, we consider the decays of a physical Higgs state,
\IE~an eigenstate of the inverse propagator matrix for the Higgs
fields evaluated at the corresponding pole eigenvalue. The connection
between such a physical state and the tree-level Higgs~fields entering
the Feynman diagrams is non-trivial in general since the higher-order
contributions induce mixing among the Higgs states and between the
Higgs states and the gauge bosons (as well as the associated Goldstone
bosons). The~LSZ~reduction fully determines the (non-unitary)
transition matrix~$\mathbf{Z}^{\mbox{\tiny mix}}$ between the
loop-corrected mass eigenstates and the lowest-order states. Then, the
amplitude describing the decay of the physical state~$h_i^{\mbox{\tiny
    phys}}$ (we shall omit the superscript~`phys' later on), into
\EG~a fermion pair~$f\bar{f}$, relates to the amplitudes in terms of
the tree-level states~$h_j^0$ according to (see below for the mixing
with gauge bosons and Goldstone bosons):
\begin{align}
  \label{eq:RelationPhysAmplitude}
  \Amp{}{}{h_i^{\mbox{\tiny  phys}}\to f\bar{f}\,}&=
  Z^{\mbox{\tiny mix}}_{ij}\,\Amp{}{}{h_j^0\to f\bar{f}\,}\,.
\end{align}
Here, we characterize the physical Higgs states according to the
procedure outlined in \citere{Domingo:2017rhb} (see also
\citeres{Frank:2006yh,Williams:2011bu,Fuchs:2016swt}):
\begin{itemize}
\item the Higgs self-energies include full one-loop and
  leading~$\mathcal{O}{(\alpha_t\alpha_s,\alpha_t^2)}$ two-loop
  corrections (with two-loop effects obtained in
  the~MSSM~approximation via the public
  code~\texttt{FeynHiggs}\footnote{The Higgs masses
    in~\texttt{FeynHiggs} could be computed with additional
    improvements such as additional fixed-order
    results\,\cite{Passehr:2017ufr,Borowka:2018anu} or the resummation
    of large logarithms for very heavy SUSY
    particles\,\cite{Hahn:2013ria,Bahl:2016brp,Bahl:2017aev}. For
    simplicity we do not take such refinements into account in the
    present article.});
\item the pole masses correspond to the zeroes of the determinant of
  the inverse-propagator matrix;
\item the~$(5 \times 5)$~matrix $\mathbf{Z}^{\mbox{\tiny mix}}$ is
  obtained in terms of the solutions of the eigenvector equation for
  the effective mass matrix evaluated at the poles, and satisfying the
  appropriate normalization conditions (see
  Sect.\,\href{https://arxiv.org/pdf/1706.00437.pdf#subsection.2.6}{$2.6$}
  of \citere{Domingo:2017rhb}).
\end{itemize}
In correcting the external Higgs legs by the full
matrix~$\mathbf{Z}^{\mbox{\tiny mix}}$---instead of employing a simple
diagrammatic expansion---we resum contributions to the transition
amplitudes that are formally of higher loop order. This resummation is
convenient for taking into account numerically relevant leading
higher-order contributions. It can in fact be crucial for the frequent
case where radiative corrections mix states that are almost
mass-degenerate in order to properly describe the resonance-type
effects that are induced by the mixing. On the other hand, care needs
to be taken to avoid the occurrence of non-decoupling terms when Higgs
states are well-separated in mass, since higher-order effects can
spoil the order-by-order cancellations with vertex corrections.

We stress that all public tools, with the exception of \FH, neglect
the full effect of the transition to the physical Higgs states encoded
within~$\mathbf{Z}^{\mbox{\tiny mix}}$, and instead employ the unitary
approximation~$\mathbf{U}^0$ neglecting external momenta (which is in
accordance with leading-order or QCD-improved leading-order
predictions). We refer the reader to
\citeres{Frank:2006yh,Fuchs:2016swt,Domingo:2017rhb} for the details
of the definition of~$\mathbf{U}^0$ or~$\mathbf{U}^m$ (another unitary
approximation) as well as a discussion of their impact at the level of
Higgs decay widths.

\paragraph{Higgs--electroweak mixing}

For the mass~determination, we do not take into account contributions
arising from the mixing of the Higgs fields with the neutral Goldstone
or $Z$~bosons since these corrections enter at the sub-dominant
two-loop level (contributions of this kind can also be compensated by
appropriate field-renormalization
conditions\,\cite{Hollik:2002mv}). We note that, in the~\cp-conserving
case, only external~\cp-odd Higgs components are affected by such a
mixing. Yet, at the level of the decay amplitudes, the Higgs mixing
with the Goldstone and $Z$~bosons enters already at the one-loop order
(even if the corresponding self-energies are cancelled by an
appropriate field-renormalization condition, this procedure would
still provide a contribution to
the~$h_if\bar{f}$~counterterm). Therefore, for a complete one-loop
result of the decay amplitudes it is in general necessary to
incorporate Higgs--Goldstone and Higgs--$Z$ self-energy transition
diagrams\,\cite{Williams:2007dc,Fowler:2009ay,Williams:2011bu}. In the
following, we evaluate such contributions to the decay amplitudes in
the usual diagrammatic fashion (as prescribed by the~LSZ~reduction)
with the help of the \texttt{FeynArts} modelfile for
the~\cp-violating~NMSSM\,\cite{Domingo:2017rhb}. The corresponding
one-loop amplitudes (including the associated counterterms) will be
symbolically denoted as~$\Amp{1L}{$G/Z$}{}$. These amplitudes can be
written in terms of the self-energies~$\Sigma_{h_iG/Z}$ with Higgs and
Goldstone$/Z$ bosons in the external legs. In turn, these
self-energies are connected by a Slavnov--Taylor identity (see
\EG~App.\,\href{https://arxiv.org/pdf/0807.4668.pdf#page=29}{A} of
\citere{Baro:2008bg}):\footnote{We denote the imaginary unit by~$\I$.}
\begin{subequations}
\begin{align}\label{eq:hG-hZ}
  \begin{split}
    0 &=
    M_Z\,\Sigma_{h_iG}{\left(p^2\right)} + \I\,p^2\,\Sigma_{h_iZ}{\left(p^2\right)}
    + M_Z\left(p^2-m_{h_i}^2\right)f{\left(p^2\right)}\\
    &\quad -\frac{e}{2\,s_{\text{w}}\,c_{\text{w}}}\sum_j\left[
      \Un{i1}\Un{j4}-\Un{i2}\Un{j5}-\Un{j1}\Un{i4}+\Un{j2}\Un{i5}
      \right]T_{h_j}\,,
  \end{split}\\
  \begin{split}
    f{\left(p^2\right)} &\equiv
    \begin{aligned}[t]
      -\frac{\alpha}{16\,\pi\,s_{\text{w}}\,c_{\text{w}}}\sum_j &
      \left[\Un{i1}\Un{j4}-\Un{i2}\Un{j5}-\Un{j1}\Un{i4}+\Un{j2}\Un{i5}\right]\\
      &\times\left[c_\beta\,\Un{j1} + s_\beta\,\Un{j2}\right]\Bnull{p^2,m_{h_j}^2,M_Z^2},
    \end{aligned}
  \end{split}
\end{align}
\end{subequations}
where the~$T_{h_i}$ correspond to the tadpole terms of the Higgs
potential, and~$\Un{ij}$ are the elements of the transition matrix
between the gauge- and tree-level mass-eigenstate bases of the Higgs
bosons---the notation is introduced in
Sect.\,\href{https://arxiv.org/pdf/1706.00437.pdf#subsection.2.1}{2.1}
of \citere{Domingo:2017rhb}. Similar relations in the~MSSM are also
provided in
Eqs.\,(\href{https://arxiv.org/pdf/1103.1335.pdf#page=69}{127}) of
\citere{Williams:2011bu}. We checked this identity at the numerical
level.

\paragraph{Inclusion of one-loop contributions}

The wave function normalization factors contained
in~$\mathbf{Z}^{\mbox{\tiny mix}}$ together with the described
treatment of the mixing with the Goldstone and $Z$~bosons ensure the
correct on-shell properties of the external Higgs leg in the decay
amplitude, so that no further diagrams correcting this external leg
are needed. Moreover, the SM~fermions and gauge bosons are also
treated as on-shell particles in our renormalization scheme. Beyond
the transition to the loop-corrected states incorporated
by~$\mathbf{Z}^{\mbox{\tiny mix}}$, we thus compute the decay
amplitudes at the one-loop order as the sum of the tree-level
contribution~$\Amp{tree}{}{}$ (possibly equal to zero), the
Higgs--electroweak one-loop mixing~$\Amp{1L}{$G/Z$}{}$ and the
(renormalized) one-loop vertex corrections~$\Amp{1L}{vert}{}$
(including counterterm contributions)---we note that each of these
pieces of the full amplitude is separately ultraviolet-finite. In the
example of the~$f\bar{f}$~decay, the amplitudes with a tree-level
external Higgs field~$h_j^0$---on the right-hand side of
\refeq{eq:RelationPhysAmplitude}---thus symbolically read:
\begin{align}
  \label{eq:1LAmplitude}
  \Amp{}{}{h_j^0\to f\bar{f}\,} &=
  \Amp{tree}{}{h_j^0\to f\bar{f}\,}+
  \Amp{1L}{$G/Z$}{h_j^0\to f\bar{f}\,}+
  \Amp{1L}{vert}{h_j^0\to f\bar{f}\,}\,.
\end{align}
All the pieces on the right-hand side of this equation are computed
with the help of
\texttt{FeynArts}\,\cite{Kublbeck:1990xc,Hahn:2000kx},
\texttt{FormCalc}\,\cite{Hahn:1998yk} and
\texttt{LoopTools}\,\cite{Hahn:1998yk}, according to the prescriptions
that are encoded in the modelfile for
the~\cp-violating~NMSSM. However, we use a specific treatment for some
of the contributions, such as~QED and~QCD~one-loop corrections to
Higgs decays into final state particles that are electrically and/or
color~charged, or include certain higher-order corrections. We
describe these channel-specific modifications in the following
subsections.

\paragraph{Goldstone-boson couplings}

The cubic Higgs--Goldstone-boson vertices can be expressed as
\begin{multline}\label{eq:treelevelG}
  \mathcal{L}\ni-\frac{1}{\sqrt{2}\,v}\left\{
  \sum_j m^2_{h_j}\left[\cos\beta\,\Un{j1}+\sin\beta\,\Un{j2}\right]
  h_j^0\left[G^+G^-+\tfrac{1}{2}\left(G^{0}\right)^2\right]\right.\\
  +\Big[\sum_j\left(m^2_{H^{\pm}}-m^2_{h_j}\right)
    \left(\sin\beta\left[\Un{j1}+\I\,\Un{j4}\right]
    -\cos\beta\left[\Un{j2}-\I\,\Un{j5}\right]\right)
    h_j^0H^+G^-+\text{h.\,c.}\Big]\\
  \left.+\frac{1}{2}\sum_{j,\,k}\left(m^2_{h_k}-m^2_{h_j}\right)
  \left[\Un{j1}\Un{k4}-\Un{j2}\Un{k5}-\left(j\leftrightarrow k\right)\right]
  h_j^0h_k^0G^0\right\}.
\end{multline}
The doublet vacuum expectation
value~(vev),~\mbox{$v=\,M_W\,s_{\text{w}}/\sqrt{2\,\pi\,\alpha}$}, is
expressed in terms of the gauge-boson masses~$M_W$ and~$M_Z$
\mbox{$\big(s_{\text{w}}=\sqrt{1-M_W^2/M_Z^2}\big)$}, as well as the
electromagnetic coupling~$\alpha$. The symbol~$m_{h_j}^2$,
($j=1,\ldots,5$), represents the tree-level mass squared of the
neutral Higgs state~$h_j^0$, and~$m^2_{H^{\pm}}$ the mass squared of
the charged Higgs state.

The use of the tree-level couplings of \refeq{eq:treelevelG} together
with a physical (loop-corrected) external Higgs
leg~\mbox{$h_i=\sum_jZ^{\mbox{\tiny mix}}_{ij}\,h^0_j$} is potentially
problematic regarding the gauge properties of the matrix elements. The
structure of the gauge theory and its renormalization indeed guarantee
that the gauge identities are observed at the order of the calculation
(one loop). However, the evaluation of Feynman amplitudes is not
protected against a violation of the gauge identities at the
(incomplete) two-loop order. We detected such gauge-violating effects
of two-loop order at several points in our calculation of the
neutral-Higgs decays, \EG:
\begin{itemize}
\item the Ward identity in~\mbox{$h_i\to\gamma\gamma$} is not
  satisfied (see also Ref.\,\cite{Benbrik:2012rm});
\item infrared~(IR) divergences of the virtual corrections
  in~\mbox{$h_i\to W^+W^-$} do not cancel their counterparts in the
  bremsstrahlung process~\mbox{$h_i\to W^+W^-\gamma$} (see also
  Ref.\,\cite{Gonzalez:2012mq});
\item computing~\mbox{$h_i\to f\bar{f}$} in an~$R_{\xi}$~gauge entails
  non-vanishing dependence of the amplitudes on the electroweak
  gauge-fixing parameters~$\xi_Z$ and~$\xi_W$.
\end{itemize}
As these gauge-breaking effects could intervene with sizable and
uncontrolled numerical impact, it is desirable to add two-loop order
terms restoring the gauge identities at the level of the matrix
elements. Technically, there are different possible procedures to
achieve this: one would amount to replace the kinematic Higgs masses
that appear in Higgs--gauge-boson couplings by tree-level Higgs
masses; we prefer the alternative procedure consisting in changing the
Higgs--Goldstone-boson couplings of \refeq{eq:treelevelG}: for the
Higgs mass associated to the external Higgs leg the loop-corrected
Higgs mass~$M_{h_i}$ is used instead of the tree-level one. This is
actually the form of the Higgs--Goldstone-boson couplings that would
be expected in an effective field theory of the physical Higgs
boson~$h_i$. Using the definition of~$Z^{\mbox{\tiny mix}}_{ij}$ as an
eigenvector of the loop-corrected mass matrix for the
eigenvalue~$M^2_{h_i}$---see
Sect.\,\href{https://arxiv.org/pdf/1706.00437.pdf#subsection.2.6}{$2.6$}
of \citere{Domingo:2017rhb}---one can verify that the effective
Higgs--Goldstone-boson vertices employing the physical Higgs mass
differ from their tree-level counterparts by a term of one-loop order
(proportional to the Higgs self-energies) so that the alteration of
the one-loop amplitudes is indeed of two-loop order. Employing this
shift of the Higgs--Goldstone couplings cures the gauge-related issues
that we mentioned earlier.

Another issue with gauge invariance appears in connection with the
amplitudes~$\Amp{1L}{$G/Z$}{}$. The Goldstone and $Z$-boson
propagators generate denominators with pole~$M_Z^2$ (or~$\xi_Z\,M_Z^2$
in an~$R_{\xi}$~gauge): in virtue of the Slavnov--Taylor identity of
Eq.\,\eqref{eq:hG-hZ} these terms should cancel one another in the
total amplitude at the one-loop order---we refer the reader to
Sect.\,\href{https://arxiv.org/pdf/1103.1335.pdf#page=11}{4.3} of
\citere{Williams:2011bu} for a detailed discussion.  However, the
term~\mbox{$\left(p^2-M_Z^2\right)^{-1}$}
multiplying~$f{\left(p^2\right)}$ of Eq.\,\eqref{eq:hG-hZ} only
vanishes if~\mbox{$p^2=m^2_{h_i}$}: if we
employ~\mbox{$p^2=M^2_{h_i}$} (the loop-corrected Higgs mass), the
cancellation is spoilt by a term of two-loop order. In order to
address this problem, we re-define~$\Amp{1L}{$G/Z$}{}$ by adding a
two-loop term:
\begin{align}\label{EWHmix}
  \Amp[\tilde{\mathcal{A}}]{1L}{$G/Z$}{h_i\to f\bar{f}\,} &\equiv
  Z^{\text{mix}}_{ij}\cdot\Amp{1L}{$G/Z$}{h_j^0\to f\bar{f}\,}
  +\frac{\Gamma^{\text{tree}}_{Gf\bar{f}}}{M_{h_i}^2}\sum_{j,\,k}
  \hat{\Sigma}_{h_jh_k}{\left(M_{h_i}^2\right)}\cdot Z^{\text{mix}}_{ik}\,
  \frac{f{\left(M_{h_i}^2\right)}\,\xi_Z\,M_Z^2}{M_{h_i}^2-\xi_Z\,M_Z^2} ,
\end{align}
where~$\Gamma^{\text{tree}}_{Gf\bar{f}}$ represents the tree-level
vertex of the neutral Goldstone boson with the fermion~$f$ (in the
particular example of a Higgs decay into~$f\bar{f}$). Then, it is
straightforward to check that~$\Amp[\tilde{\mathcal{A}}]{1L}{$G/Z$}{}$
is gauge-invariant. The transformation of Eq.\,\eqref{EWHmix} can also
be interpreted as a two-loop shift re-defining~$\Sigma_{h_iZ}$, so
that it satisfies a generalized Slavnov--Taylor identity of the form
of Eq.\,\eqref{eq:hG-hZ}, but applying to a physical (loop-corrected)
Higgs field, with the
term~$\left(p^2-m^2_{h_i}\right)f{\left(p^2\right)}$ of
Eq.\,\eqref{eq:hG-hZ} replaced
by~$\left(p^2-M^2_{h_i}\right)f{\left(p^2\right)}$.

\paragraph{Numerical input in the one-loop corrections}

As usual, the numerical values of the input parameters need to reflect
the adopted renormalization scheme, and the input parameters
corresponding to different schemes differ from each other by shifts of
the appropriate loop order (at the loop level there exists some
freedom to use a numerical value of an input parameter that differs
from the tree-level value by a one-loop shift, since the difference
induced in this way is of higher order). Concerning the input values
of the relevant light quark masses, we follow in our evaluation the
choice of~\texttt{FeynHiggs} and employ $\overline{\mbox{MS}}$~quark
masses with three-loop~QCD corrections evaluated at the scale of the
mass of the decaying Higgs,~$m_q^{\overline{\mbox{\tiny
      MS}}}(M_{h_i})$, in the loop functions and the definition of the
Yukawa couplings. In addition, the input value for the pole top mass
is converted to~$m_t^{\overline{\mbox{\tiny MS}}}(m_t)$ using up to
two-loop~QCD and one-loop top Yukawa/electroweak corrections
(corresponding to the higher-order corrections included in the
Higgs-boson mass calculation). Furthermore, the $\tan\beta$-enhanced
contributions are always included in the defining relation between the
bottom Yukawa coupling and the bottom mass (and similarly for all
other down-type quarks). Concerning the Higgs~vev appearing in the
relation between the Yukawa couplings and the fermion masses, we
parametrize it in terms of~$\alpha(M_Z)$. Finally, the strong coupling
constant employed in~SUSY-QCD~diagrams is set to the scale of the
supersymmetric particles entering the loop. We will comment on
deviations from these settings if needed.

\subsection[Higgs decays into SM fermions]{\tocref{Higgs decays into SM fermions}}

Our calculation of the Higgs decay amplitudes into SM~fermions closely
follows the procedure outlined in the previous subsection. However, we
include the~QCD and~QED~corrections separately, making use of
analytical formulae that are well-documented in the
literature\,\cite{Braaten:1980yq,Drees:1990dq}. We also employ an
effective description of the~Higgs--$b\bar{b}$~interactions in order
to resum potentially large effects for large values
of~$\tan\beta$. Below, we comment on these two issues and discuss
further the derivation of the decay widths for this class of channels.

\paragraph{Tree-level amplitude}

At the tree level, the decay~$h^0_j\to f\bar{f}$ is determined by the
Yukawa coupling~$Y_f$ and the decomposition of the tree-level
state~$h^0_j$ in terms of the Higgs-doublet components:
\begin{align}
  \label{eq:hfftree}
  \begin{split}
  \Amp{tree}{}{h^0_j\to f\bar{f}\,} &=
  \begin{aligned}[t]
  -\I\,\frac{Y_f}{\sqrt{2}}\,\bar{u}_f{\left(p_f\right)}\,\Big\{
  &\delta_{f,\,d_k/e_k}\Un{j1} + \delta_{f,\,u_k}\Un{j2}\\
  &-\I\,\gamma_5\left[\delta_{f,\,d_k/e_k}\Un{j4}+\delta_{f,\,u_k}\Un{j5}\right]
  \!\Big\}\,v_f{\left(p_{\bar{f}}\right)}
  \end{aligned}\\
  &\equiv -\I\,\bar{u}_f{\left(p_f\right)}
  \left\{g_{h_jff}^S-\gamma_5\,g_{h_jff}^P\right\}v_f{\left(p_{\bar{f}}\right)}\,.
\end{split}
\end{align}
The~$\delta$-s are Kronecker symbols selecting the appropriate Higgs
matrix element for the fermionic final state,~$u_k=u,c,t$,
\mbox{$d_k=d,s,b$} or~$e_k=e,\mu,\tau$. We have written the amplitude
in the Dirac-fermion convention, separating the scalar
piece~$g_{h_jff}^S$ (first two terms between curly brackets in the
first line) from the pseudoscalar one~$g_{h_jff}^P$ (last two
terms). The fermion and antifermion spinors are denoted
as~$\bar{u}_f(p_f)$ and~$v_f(p_{\bar{f}})$, respectively.

\paragraph{\boldmath Case of the $b\bar{b}$ final state: $\tan\beta$-enhanced corrections}

In the case of a decay to~$b\bar{b}$~(and analogously for down-type
quarks of first and second generation, but with smaller numerical
impact), the loop contributions that receive a~$\tan\beta$~enhancement
may have a sizable impact, thus justifying an effective description of
the Higgs--$b\bar{b}$~vertex that provides a resummation of large
contributions\,\cite{Banks:1987iu, Hall:1993gn, Hempfling:1993kv,
  Carena:1994bv, Carena:1999py, Eberl:1999he, Williams:2011bu,
  Baglio:2013iia}. We denote the neutral components of~$\mathcal{H}_1$
and~$\mathcal{H}_2$ from
Eq.\,(\href{https://arxiv.org/pdf/1706.00437.pdf#equation.2.2}{2.2})
of \citere{Domingo:2017rhb} by~$H_d^0$ and~$H_u^0$, respectively. The
large~$\tan\beta$-enhanced effects arise from contributions to
the~$\left(H_u^{0}\right)^*\,\bar{b}\,P_L\,{b}$ operator---$P_{L,R}$
are the left- and right-handed projectors in the Dirac description of
the~$b$~spinors---and can be parametrized in the following fashion:
\begin{align}
\label{eq:Leffbb}
  \mathcal{L}^{\text{eff}} &= -Y_b\,\bar{b}\left[H_d^0 +
    \frac{\Delta_b}{\tan\beta}\left(\frac{\lambda}{\mueff}\,S\,H_u^0\right)^*
    \right] P_L\,b + \text{h.c.}
  \equiv -\sum_j g_{h_jbb}^{L\,\text{eff}}\,h^0_j\,\bar{b}\,P_L\,b + \text{h.\,c.}
\end{align}
Here,~$\Delta_b$ is a coefficient that is determined via the
calculation of the relevant~($\tan\beta$-enhanced) one-loop diagrams
to the Higgs--$b\bar{b}$~vertex, involving gluino--sbottom,
chargino--stop and neutralino--sbottom loops.\footnote{Two-loop
  corrections to~$\Delta_b$ have also been studied in
  \citeres{Noth:2008tw,Noth:2010jy}.} The symbol~$\mu_{\text{eff}}$
represents the effective~$\mu$~term that is generated when the singlet
field acquires a vev. The specific form of the
operator,~$\left(S\,H_u^0\right)^*\,\bar{b}\,P_L\,b$, is designed so
as to preserve the~$\mathbb{Z}_3$~symmetry, and it can be shown that
this operator is the one that gives rise to leading contributions to
the~$\tan\beta$-enhanced effects. We evaluate~$\Delta_b$ at a scale
corresponding to the arithmetic mean of the masses of the
contributing~SUSY~particles: this choice is consistent with the
definition of~$\Delta_b$ employed for the Higgs-mass calculation.

From the parametrization of \refeq{eq:Leffbb}, one can derive the
non-trivial relation between the `genuine' Yukawa coupling~$Y_b$ and
the effective bottom~mass~$m_b$: \mbox{$Y_b=\frac{m_b}{v_1\left(1 +
    \Delta_b\right)}$}\,. Then, the effective couplings of the neutral
Higgs fields to~$b\bar{b}$ read:
\begin{align}
  \label{eq:bbeffcoup}
  g_{h_jbb}^{L\,\text{eff}} &= 
  \frac{m_b}{\sqrt{2}\,v_1\left(1 + \Delta_b\right)}\left\{
  \Un{j1} + \I\,\Un{j4}
  +\frac{\Delta_b}{\tan\beta}\left(
  \Un{j2} - \I\,\Un{j5} +
  \tfrac{\lambda^*\,v_2}{\mueff^*}\left[\Un{j3} - \I\,\Un{j6}\right]
  \right)\right\}.
\end{align}
This can be used to substitute~$\Amp{tree}{}{h^0_j\to
  b\bar{b}\,}$ in \refeq{eq:1LAmplitude} by:
\begin{align}\label{eq:effHbb}
  \Amp{eff}{}{h^0_j\to b\bar{b}\,}&=-\I\,\bar{u}_b{\left(p_b\right)}\left[
    g_{h_jbb}^{L\,\text{eff}}\,P_L + g_{h_jbb}^{L\,\text{eff}\,*}\,P_R
    \right] v_b{\left(p_{\bar{b}}\right)}\,,
\end{align}
where this expression resums the effect of~$\tan\beta$-enhanced
corrections to the~$h^0_jb\bar{b}$ vertex. However, if one now adds
the one-loop amplitude~$\Amp{1L}{vert}{}$, the one-loop effects
associated with the~$\tan\beta$-enhanced contributions would be
included twice.  To avoid this double counting, the terms that are
linear in~$\Delta_b$ in \refeq{eq:bbeffcoup} need to be subtracted.
Employing the `subtraction' couplings
\begin{align}
  g_{h_jbb}^{L\,\text{sub}} &= \frac{m_b\,\Delta_b}{\sqrt{2}\,v_1}\left\{
    \Un{j1} + \I\,\Un{j4} -
    \frac{1}{\tan\beta}\left(\Un{j2} - \I\,\Un{j5} +
    \frac{\lambda^*\,v_u}{\mueff^*}\left[\Un{j3} - \I\,\Un{j6}\right]
    \right)\right\}
    \hspace{-.2em}
\end{align}
we define the following `tree-level' amplitude for the Higgs decays
into bottom quarks:
\begin{subequations}
\begin{align}
  \Amp{tree}{}{h^0_j\to b\bar{b}\,} &=
  \Amp{eff}{}{h^0_j\to b\bar{b}\,}+\Amp{sub}{}{h^0_j\to b\bar{b}\,}\,,\\
  \Amp{sub}{}{h^0_j\to b\bar{b}\,} &\equiv
  -\I\,\bar{u}_b{\left(p_b\right)}\left[
    g_{h_jbb}^{L\,\text{sub}}\,P_L + g_{h_jbb}^{L\,\text{sub}\,*}\,P_R
    \right]v_b{\left(p_{\bar{b}}\right)}\,.
\end{align}
\end{subequations}

\paragraph{QCD and QED corrections}

The inclusion of~QCD and~QED~corrections requires a proper treatment
of~IR~effects in the decay amplitudes. The~IR-divergent parts of the
virtual contributions by gluons or photons in~$\Amp{1L}{vert}{}$ are
cancelled by their counterparts in processes with radiated photons or
gluons. We employ directly the~QCD and~QED~correction factors that are
well-known analytically~(see below) and therefore omit the Feynman
diagrams involving a photon or gluon propagator when computing with
\texttt{FeynArts} and \texttt{FormCalc} the one-loop corrections to
the~$h^0_jf\bar{f}$~vertex and to the fermion-mass and wave-function
counterterms.  The~QCD- and~QED-correction factors applying to the
fermionic decays of a~\cp-even Higgs state were derived in
\citere{Braaten:1980yq}. The~\cp-odd case was addressed later in
\citere{Drees:1990dq}. In the~\cp-violating case, it is useful to
observe that the~$h_jf\bar{f}$~scalar and pseudoscalar operators do
not interfere, so that the~\cp-even and~\cp-odd correction factors can
be applied directly at the level of the amplitudes---although they
were obtained at the level of the squared amplitudes:
\begin{subequations}
\label{eq:hffQCD}
\begin{align}
  \Amp{tree+QCD/QED}{}{h^0_j\to f\bar{f}\,} &=
  -\I\,\frac{m_f^{\overline{\mbox{\tiny MS}}}{\left(M_{h_i}\right)}}{m_f}\,
  \bar{u}_f{\left(p_f\right)}\left\{
  g_{h_jff}^S\,c_S-\gamma_5\,g_{h_jff}^P\,c_P
  \right\}v_f{\left(p_{\bar{f}}\right)}\,,\\[-.2ex]
  c_{S,P} &= \sqrt{1+c_{S,P}^{\mbox{\tiny QED}}+c_{S,P}^{\mbox{\tiny QCD}}}\,,\\[-.3ex]
  c_{S,P}^{\mbox{\tiny QED}} &\equiv
  \frac{\alpha}{\pi}\,Q_f^2\,\Delta_{S,P}{\left(\sqrt{1-\tfrac{4\,m^2_f}{M^2_{h_i}}}\right)},\\[-.2ex]
  c_{S,P}^{\mbox{\tiny QCD}} &\equiv
  \frac{\alpha_s{\left(M_{h_i}\right)}}{\pi}\,C_2{(f)}\left[
    \Delta_{S,P}{\left(\sqrt{1-\tfrac{4\,m^2_f}{M^2_{h_i}}}\right)}+
    2+3\log{\left(\frac{M_{h_i}}{m_f}\right)}
  \right].
\end{align}
\end{subequations}
Here, $Q_f$ is the electric charge of the fermion~$f$, $C_2(f)$ is
equal to~$4/3$ for quarks and equal to~$0$ for leptons, $M_{h_i}$
corresponds to the kinematic (pole) mass in the Higgs decay under
consideration and the functions~$\Delta_{S,P}$ are explicated in
\EG~Sect.\,\href{https://arxiv.org/pdf/hep-ph/9503443.pdf#page=9}{$4$}
of \citere{Dabelstein:1995js}. In the limit of~$M_{h_i}\gg m_f$,
both~$\Delta_{S,P}$ reduce
to~$\left[-3\log{\left(M_{h_i}/m_f\right)}+\frac{9}{4}\right]$. As
noticed already in \citere{Braaten:1980yq}, the leading logarithm in
the~QCD-correction factor can be absorbed by the introduction of a
running~\MSbar~fermion mass in the definition of the Yukawa
coupling~$Y_f$. Therefore, it is motivated to
factorize~$m_f^{\overline{\mbox{\tiny MS}}}(M_{h_i})$, with higher
orders included in the definition of the~QCD~beta~function.

The~QCD (and~QED) correction factors generally induce a sizable shift
of the tree-level width of as much as~$\simord 50\%$. While these
effects were formally derived at the one-loop order, we apply them
over the full amplitudes (without the~QCD and~QED corrections), \IE~we
include the one-loop vertex amplitude
without~QCD$/$QED~corrections~$\Amp{1L wo.\ QCD/QED}{vert}{}$
and~$\Amp{1L}{$G/Z$}{}$ in the definitions of the
couplings~$g_{h_jff}^{S,P}$ that are employed in
\refeq{eq:hffQCD}---we will use the
notation~$g_{h_jff}^{S,P\,\text{1L}}$ below. The adopted factorization
corresponds to a particular choice of the higher-order contributions
beyond the ones that have been explicitly calculated.

\paragraph{Decay width}

Putting together the various pieces discussed before, we can express
the decay amplitude at the one-loop order as:
\begin{subequations}
\begin{align}
  \Amp{}{}{h_i\to f\bar{f}\,} &=
  -\I\,\frac{m_f^{\overline{\mbox{\tiny MS}}}{\left(M_{h_i}\right)}}{m_f}\,
  Z^{\mbox{\tiny mix}}_{ij}\,\bar{u}_f{\left(p_f\right)}\left\{
  g_{h_jff}^{S\,\text{1L}}\,c_S-\gamma_5\,g_{h_jff}^{P\,\text{1L}}\,c_P
  \right\}v_f{\left(p_{\bar{f}}\right)}\,,\\
  -\I\,\bar{u}_f{\left(p_f\right)}\left\{g_{h_jff}^{S\,\text{1L}}-\gamma_5\,g_{h_jff}^{P\,\text{1L}}\right\}v_f{\left(p_{\bar{f}}\right)} \equiv
  \left(\Amp{tree}{}{}+\Amp{1L wo.\ QCD/QED}{vert}{}+\Amp{1L}{$G/Z$}{}\right)
  \!\!{\left[h_j\to f\bar{f}\,\right]}\,.\span\omit
\end{align}
\end{subequations}
Summing over spinor and color degrees of freedom, the decay width is
then obtained as:
\begin{align}
  \Gamma{\left[h_i\to f\bar{f}\,\right]} &=
  \frac{1}{16\,\pi\,M_{h_i}}\sqrt{1-\frac{4m_f^2}{M_{h_i}^2}}
  \sum_{\parbox{\widthof{\tiny polarization,}}{\centering\tiny polarization,\\\tiny color}}
  \left|\Amp{}{}{h_i^{\mbox{\tiny phys.}}\to f\bar{f}\,}\right|^2\,.
\end{align}
At the considered order, we could dismiss the one-loop squared terms
in~$\left|\Amp{}{}{h_i\to f\bar{f}\,}\right|^2$. However, in order to
tackle the case where the contributions from irreducible one-loop
diagrams are numerically larger than the tree-level amplitude, we keep
the corresponding squared terms in the expression above (it should be
noted that the~QCD and~QED~corrections have been stripped off from the
one-loop amplitude that gets squared).  The approach of incorporating
the squared terms should give a reliable result in a situation where
the tree-level result is significantly suppressed, since the other
missing contribution at this order consisting of the tree-level
amplitude times the two-loop amplitude would be suppressed due to the
small tree-level result. In such a case, however, the higher-order
uncertainties are expected to be comparatively larger than in the case
where one-loop effects are subdominant to the tree level.

The kinematic masses of the fermions are easily identified in the
leptonic case. For decays into top quarks the `pole' mass~$m_t$ is
used, while for all other decays into quarks we employ
the~$\overline{\mbox{MS}}$~masses evaluated at the scale of the Higgs
mass~$m_q^{\overline{\mbox{\tiny MS}}}(M_{h_i})$. We note that these
kinematic masses have little impact on the decay widths, as long as
the Higgs state is much heavier. In the~NMSSM, however, singlet-like
Higgs states can be very light, in which case the choice of
an~$\overline{\mbox{MS}}$~mass is problematic. Yet, in this case the
Higgs state is typically near threshold so that the free-parton
approximation in the final state is not expected to be reliable. Our
current code is not properly equipped to address decays directly at
threshold independently of the issue of running kinematic
masses. Improved descriptions of the hadronic decays of Higgs states
close to the~$b\bar{b}$~threshold or in the chiral limit have been
presented in
\EG~\citeres{Drees:1989du,Fullana:2007uq,McKeen:2008gd,Domingo:2011rn,Dolan:2014ska,Domingo:2016yih}.

\subsection[Decays into SM gauge bosons]{\tocref{Decays into SM gauge bosons}}
 
Now we consider Higgs decays into the gauge bosons of the~SM. Almost
each of these channels requires a specific processing in order to
include higher-order corrections consistently or to deal with
off-shell effects.

\paragraph{Decays into electroweak gauge bosons}

Higgs decays into on-shell~$W$-s and~$Z$-s can be easily included at
the one-loop order in comparable fashion to the fermionic
decays. However, the notion of~$WW$ or~$ZZ$~final states usually
includes contributions from off-shell gauge bosons as well,
encompassing a wide range of four-fermion final states. Such off-shell
effects mostly impact the decays of Higgs bosons with a mass below
the~$WW$ or~$ZZ$~thresholds. Instead of a full processing of the
off-shell decays at one-loop order, we pursue two distinct evaluations
of the decay widths in these channels.

Our first approach is that already employed in~\texttt{FeynHiggs} for
the corresponding decays in the~MSSM. It consists in exploiting the
precise one-loop results of~\texttt{Prophecy4f} for the~SM-Higgs
decays into four
fermions\,\cite{Bredenstein:2006rh,Bredenstein:2006nk,Bredenstein:2006ha}. For
an~(N)MSSM~Higgs boson~$h_i$, the~SM~decay width is thus evaluated at
the mass~$M_{h_i}$ and then rescaled by the squared ratio of the
tree-level couplings to gauge bosons for~$h_i$ and an~SM~Higgs
boson~$H_{\mbox{\tiny SM}}$~($V=W,Z$):
\begin{subequations}
\begin{align}
  \Gamma{\left[h_i\to VV\right]} &=
  \Gamma^{\mbox{\tiny SM}}{\left[H_{\mbox{\tiny SM}}(M_{h_i})\to VV\right]}
  \left|\mathcal{R}_{ij}\cdot
  \frac{g_{h_jVV}^{\mbox{\tiny NMSSM}}}{g_{HVV}^{\mbox{\tiny SM}}}\right|^2,
  \label{ratio_NMSSM_SM}\\
  \frac{g_{h_jVV}^{\mbox{\tiny NMSSM}}}{g_{HVV}^{\mbox{\tiny SM}}} &\equiv
  \cos\beta\,\Un{j1}+\sin\beta\,\Un{j2}\,,
\end{align}
\end{subequations}
where~\mbox{$\Gamma{\left[h_i\to VV\right]}$} represents the decay
width of the physical Higgs state~$h_i$ in the~NMSSM,
while~\mbox{$\Gamma^{\mbox{\tiny SM}}{\left[H_{\mbox{\tiny
          SM}}{\left(M_{h_i}\right)}\to VV\right]}$} denotes the decay
width of an~SM-Higgs boson with the mass~$M_{h_i}$. The matrix
elements~$\mathcal{R}_{ij}$ reflect the connection between the
tree-level Higgs states and the physical states. This role is similar
to~$\mathbf{Z}^{\mbox{\tiny mix}}$. However, decoupling in
the~SM~limit of the model yields the additional condition that the
ratio in \refeq{ratio_NMSSM_SM} reduces to~$1$ in this limit for
the~SM-like Higgs boson of the~NMSSM. For this reason,
\texttt{FeynHiggs} employs the matrix~$\mathbf{U}^m$
(or~$\mathbf{U}^0$) as a unitary approximation
of~$\mathbf{Z}^{\mbox{\tiny mix}}$---see
Sect.\,\href{https://arxiv.org/pdf/1706.00437.pdf#subsection.2.6}{$2.6$}
of \citere{Domingo:2017rhb}. An alternative choice consists in
using~\mbox{$X_{ij}\equiv \left.Z^{\mbox{\tiny
      mix}}_{ij}\middle/\sqrt{\sum_k |Z^{\mbox{\tiny
        mix}}_{ik}|^2}\right.$}. However, the difference of the widths
when employing~$\mathbf{U}^0$, $\mathbf{U}^m$,
$\mathbf{Z}^{\mbox{\tiny mix}}$ or~$\mathbf{X} \equiv
\left(X_{ij}\right)$ corresponds to effects of higher order, which
should be regarded as part of the higher-order uncertainty. The
rescaling of the one-loop~SM~width should only be applied for
the~SM-like Higgs of the~NMSSM, where this implementation of
the~$h_i\to VV$~widths is expected to provide an approximation that is
relatively close to a full one-loop result incorporating all~NMSSM
contributions. However, for the other Higgs states of the~NMSSM
one-loop contributions beyond the~SM may well be dominant. Actually,
the farther the
quantity~\mbox{$[\mathcal{R}_{ij}\cdot\Un{j2}]\big/[\mathcal{R}_{ij}\cdot\Un{j1}]$}
departs from~$\tan\beta$, the more inaccurate the prediction based
on~SM-like radiative corrections becomes.

Our second approach consists in a one-loop calculation of the Higgs
decay widths into on-shell gauge bosons (see \citere{Gonzalez:2012mq}
for the~MSSM~case), including tree-level off-shell effects. This
evaluation is meant to address the case of heavy Higgs bosons at the
full one-loop order. The restriction to on-shell kinematics is
justified above the threshold for electroweak gauge-boson production
(off-shell effects at the one-loop level could be included via a
numerical integration over the squared momenta of the gauge bosons in
the final state---see \citeres{Hollik:2010ji,Hollik:2011xd} for a
discussion in the~MSSM). Our implementation largely follows the lines
described in Sect.\,\ref{sec:general}, with the noteworthy feature
that contributions from Higgs--electroweak mixing~$\Amp{1L}{$G/Z$}{}$
vanish. In the case of the~$W^+W^-$~final state,
the~QED~IR-divergences are regularized with a photon~mass and cancel
with bremsstrahlung corrections: soft and hard bremsstrahlung are
included according to \citeres{Kniehl:1991xe,Kniehl:1993ay} (see
also\,\cite{Gonzalez:2012mq}). We stress that the exact cancellation
of the~IR-divergences is only achieved through the replacement of
the~$h_iG^+G^-$~coupling by the expression in terms of the kinematical
Higgs mass, as discussed in Sect.\,\ref{sec:general}. This fact had
already been observed by \citere{Gonzalez:2012mq}. In order to extend
the validity of the calculation below threshold, we process the
Born-order term separately, applying an off-shell kinematic
integration over the squared external momentum of the gauge
bosons---see
\EG~Eq.\,(\href{https://arxiv.org/pdf/1612.07651.pdf#equation.2.37}{37})
in \citere{Spira:2016ztx}. Thus, this evaluation is performed at tree
level below threshold and at full one-loop order (for the on-shell
case) above threshold. The vanishing on-shell kinematical factor
multiplying the contributions of one-loop order ensures the continuity
of the prediction at threshold. Finally, we include the one-loop
squared term in the calculation. Indeed, as we will discuss later on,
the tree-level contribution vanishes for a decoupling doublet, meaning
that the Higgs decays to~$WW/ZZ$ can be dominated by one-loop
effects. To this end, the infrared divergences of two-loop order are
regularized in an ad-hoc fashion---which appears compulsory as long as
the two-loop order is incomplete---making use of the one-loop real
radiation and estimating the logarithmic term in the imaginary part of
the one-loop amplitude.

\paragraph{Radiative decays into gauge bosons}

Higgs decays into photon pairs, gluon pairs or~$\gamma Z$ appear at
the one-loop level---\IE~\mbox{$\Amp{tree}{}{}=0$} for all these
channels. We compute the one-loop order using the~\texttt{FeynArts}
modelfile, although the results are well-known analytically in the
literature---see \EG~\citere{Benbrik:2012rm} or
Sect.\,\href{https://arxiv.org/pdf/1402.3522.pdf#section*.3}{III} of
\citere{Belanger:2014roa} (\cite{Spira:2016ztx} for the~MSSM). The
electromagnetic coupling in these channels is set to the
value~$\alpha(0)$ corresponding to the Thomson~limit.

The use of tree-level Higgs--Goldstone couplings together with
loop-corrected kinematic Higgs masses~$M_{h_i}$ in our calculation
would induce an effective violation of Ward~identities by two-loop
order terms in the amplitude: as explained in
Sect.\,\ref{sec:general}, we choose to restore the proper gauge
structure by re-defining the Higgs--Goldstone couplings in terms of
the kinematic Higgs mass~$M_{h_i}$. Since our calculation is
restricted to the leading---here, one-loop---order, the transition of
the amplitude from tree-level to physical Higgs states is performed
via~$\mathbf{U}^m$ or~$\mathbf{X}$ instead of~$\mathbf{Z}^{\mbox{\tiny
    mix}}$ in order to ensure the appropriate behavior in the
decoupling limit.

Leading~QCD~corrections to the diphoton Higgs decays have received
substantial attention in the literature. A frequently used
approximation for this channel consists in multiplying the amplitudes
driven by quark and squark loops by the
factors~$\left[1-\alpha_s(M_{h_i})/\pi\right]$
and~$\left[1+8\,\alpha_s(M_{h_i})/(3\,\pi)\right]$, respectively---see
\EG~\citere{Lee:2003nta}. However, these simple factors are only valid
in the limit of heavy quarks and squarks (compared to the mass of the
decaying Higgs boson). More general analytical expressions can be
found in \EG~\citere{Aglietti:2006tp}. In our calculation, we apply
the correction
factors~$\left[1+C^S{(\tau_q)}\,\alpha_s(M_{h_i})/\pi\right]$
and~$\left[1+C^P{(\tau_q)}\,\alpha_s(M_{h_i})/\pi\right]$ to the
contributions of the quark~$q$ to the~\cp-even and
the~\cp-odd~$h_i\gamma\gamma$ operators, respectively,
and~$\left[1+C{(\tau_{\tilde{Q}})}\,\alpha_s(M_{h_i})/\pi\right]$ to
the contributions of the squark~$\tilde{Q}$ (to the~\cp-even
operator).  Here,~$\tau_X$ denotes the
ratio~$\left[4\,m^2_X{(M_{h_i}/2)}/M^2_{h_i}\right]$. The
coefficients~$C^{S,P}$ and~$C$ are extracted from
\citere{Spira:1995rr} and \citere{Muhlleitner:2006wx}. In order to
obtain a consistent inclusion of
the~$\mathcal{O}{(\alpha_s)}$~corrections, the quark and squark
masses~$m_X$ entering the one-loop amplitudes or the correction
factors are chosen as defined in
Eq.\,(\href{https://arxiv.org/pdf/hep-ph/9504378.pdf#page=7}{5}) of
\citere{Spira:1995rr} and in
Eq.\,(\href{https://arxiv.org/pdf/hep-ph/0612254.pdf#equation.2.12}{12})
of \citere{Muhlleitner:2006wx} (rather than~\MSbar~running masses).

The~QCD~corrections to the digluon decays include virtual corrections
but also gluon and light-quark radiation. They are thus technically
defined at the level of the squared amplitudes. In the limit of heavy
quarks and squarks, the corrections are known beyond~NLO---see the
discussion in \citere{Spira:2016ztx} for a list of references. The
full dependence in mass was derived at~NLO in
\citeres{Spira:1995rr,Muhlleitner:2006wx}, for both quark and squark
loops. In our implementation, we follow the prescriptions of
Eqs.\,(\href{https://arxiv.org/pdf/1612.07651.pdf#equation.2.51}{51}),
(\href{https://arxiv.org/pdf/1612.07651.pdf#equation.2.63}{63})
and~(\href{https://arxiv.org/pdf/1612.07651.pdf#equation.2.67}{67}) of
\citere{Spira:2016ztx} in the limit of light radiated quarks and heavy
particles in the loop. For consistency, the masses of the particles in
the one-loop amplitude are taken as pole masses. Effects beyond this
approximation can be sizable, as evidenced by
Fig.\,\href{https://arxiv.org/PS_cache/hep-ph/ps/9504/9504378v1.fig1-22.png}{20}
of \citere{Spira:1995rr} and
Fig.\,\href{https://arxiv.org/pdf/hep-ph/0612254.pdf#page.21}{12} of
\citere{Muhlleitner:2006wx}. As the~\cp-even and~\cp-odd
Higgs--$gg$~operators do not interfere, it is straightforward to
include both correction factors in the~\cp-violating case. Finally, we
note that parts of the leading~QCD~corrections to~$h_i\to gg$ are
induced by the real radiation of quark--antiquark pairs. In the case
of the heavier quark flavors (top, bottom and possibly charm), the
channels are experimentally well-distinguishable from gluonic
decays. Therefore, the partial widths related to these corrections
could be attached to the Higgs decays into quarks
instead\,\cite{Djouadi:1995gt}. The resolution of this ambiguity would
involve a dedicated experimental analysis of the kinematics of the
gluon radiation in~$h_i\to gq\bar{q}$ (collinear or back-to-back
emission). In the following section, we choose to present our results
for the~$h_i\to gg$~decay in the three-radiated-flavor approach, while
the contributions from the heavier quark~flavors are distributed among
the~$h_i\to c\bar{c}$, $b\bar{b}$, and~$t\bar{t}$~widths (provided the
Higgs state is above threshold). These contributions to the fermionic
Higgs decays are of~$\mathcal{O}{(\alpha_s^2)}$.

The~QCD~corrections to the quark loops of an~SM-Higgs decay
into~$\gamma Z$ have been studied in
\citeres{Spira:1991tj,Bonciani:2015eua,Gehrmann:2015dua}, but we do
not consider them here.

%% file: 03_NumericalAnalysis.tex
\section[Numerical Analysis]{\tocref{\label{sec:numerics}Numerical Analysis}}

In this section, we present our results for the decay widths of the
neutral Higgs bosons into~SM~particles in several scenarios and
compare them with the predictions of existing codes.  While a detailed
estimate of the uncertainty associated to missing higher-order
corrections goes beyond the scope of our analysis, we will provide
some discussion at the end of this section, based on our observations
and comparisons.

Throughout this section, the top-quark pole mass is chosen as~$m_t =
173.2$\,GeV. Moreover, all~\DRbar~parameters are defined at the
scale~$m_t$, and all stop~parameters are treated as on-shell
parameters. Concerning the Higgs phenomenology, we test the scenarios
presented in this section with the full set of experimental
constraints and signals implemented in the public
tools~\texttt{HiggsBounds-4.3.1}
(and~\texttt{5.1.1beta})\,\cite{Bechtle:2008jh, Bechtle:2011sb,
  Bechtle:2013gu, Bechtle:2013wla, Bechtle:2015pma, HB-www}
and~\texttt{HiggsSignals-1.3.1}
(and~\texttt{2.1.0beta})\,\cite{Bechtle:2013xfa, Bechtle:2014ewa,
  HB-www}. We refer the reader to the corresponding publications for a
detailed list of experimental references. The input parameters
employed in our scenarios are summarized in \refta{tab:scenarios}.

\begin{table}[h!]
\footnotesize
\centering
\begin{tabular}{||c|c||c|c|c|c||c}
\hhline{~~|t:====:t:=}
\multicolumn{2}{c||}{} & \multicolumn{4}{c||}{Sect.\,\ref{sec:feynhiggs}: comparison with \texttt{FeynHiggs}} & \\
\hhline{~~||----||-}
\multicolumn{2}{c||}{} & \reffi{fig:MSSM_hff} & \reffi{fig:MSSM_hVV} & \reffi{fig:MSSM_hgg} & \reffi{fig:MSSM_CPV} & \\
\hhline{|t:==::====::=}
\#1 & $\lambda$ & $1\cdot10^{-5}$ & $1\cdot10^{-5}$ & $1\cdot10^{-5}$ & $1\cdot10^{-5}$ & \\
\hhline{||--||----||-}
\#2 & $|\kappa|$ & $1\cdot10^{-5}$ & $1\cdot10^{-5}$ & $1\cdot10^{-5}$ & $1\cdot10^{-5}$ & \\
\hhline{||--||----||-}
\#3 & $\phi_{\kappa}$ & $0$ & $0$ & $0$ & $0$ & \\
\hhline{||--||----||-}
\#4 & $\tan\beta$ & $10$ & $10$ & $1+39 \cdot x$ & $10$ & \\
\hhline{||--||----||-}
\#5 & $\mu_{\mbox{\tiny eff}}$ (GeV) & $250$ & $250$ & $250$ & $250$ & \\
\hhline{||--||----||-}
\#6 & $m_{H^{\pm}}$ (TeV) & $1$ & $0.15+1.85\cdot x$ & $1$ & $0.5$ & \\
\hhline{||--||----||-}
\#7 & $A_{\kappa}$ (GeV) & $-100$ & $-100$ & $-100$ & $-100$ & \\
\hhline{||--||----||-}
\#8 & $m_{\tilde{Q}}$ (TeV) & $0.7+1.3\cdot x$ & $1.5$ & $1.5$ & $1.5$ & \\
\hhline{||--||----||-}
\#9 & $|A_{t}|$ (TeV) & $1.4+1.6\cdot x$ & $2.3$ & $2.3$ & $2.5$ & \\
\hhline{||--||----||-}
\#10 & $\phi_{A_{t}}$ & $0$ & $0$ & $0$ & $\pi(2\cdot x-1)$ & \\
\hhline{||--||----||-}
\#11 & $A_{b}$ (TeV) & $1.4+1.6\cdot x$ & $2.3$ & $2.3$ & $2.5$ & \\
\hhline{|b:==:b:====:b:=}
\end{tabular}\\[1ex]
\begin{tabular}{c||c|c|c|c||c|c|c||}
\hhline{=:t:====:t:===:t|}
& \multicolumn{4}{c||}{Sect.\,\ref{sec:nmssmcalc}: comparison with \texttt{NMSSMCALC}} & \multicolumn{3}{c||}{Sect.\,\ref{sec:singlet}: singlet Higgs at $\lsimord 100$\,GeV}\\
\hhline{-||----||---||}
& \reffi{fig:NMSSM_hff}& \reffi{fig:NMSSM_hVV} & \reffi{fig:NMSSM_hgg} & \reffi{fig:NMSSM_CPV} & \reffi{fig:hS_100GeV_TB12} & \reffi{fig:hS_100GeV_TB2} & \reffi{fig:hS_100GeV_CPV}\\
\hhline{=::====::===:|}
\#1& $0.3$ & $0.3$ & $0.3$ & $0.2$ & $0.1$ & $0.6$ & $0.7$\\
\hhline{-||----||---||}
\#2& $0.4$& $0.4$ & $0.4$ & $0.6$ & $0.15$ & $0.035$ & $0.1$\\
\hhline{-||----||---||}
\#3& $0$ & $0$ & $0$ & $\pi(2\cdot x-1)$ & $0$ & $0$ & $\frac{\pi}{8}(2\cdot x-1)$\\
\hhline{-||----||---||}
\#4&  $10$& $10$ & $1+39 \cdot x$ & $25$ & $12$ & $2$ & $2$\\
\hhline{-||----||---||}
\#5& $250$& $250$ & $250$ & $200$ & $140$ & $397+15\cdot x$ & $500$\\
\hhline{-||----||---||}
\#6& $1$ & $0.15+1.85\cdot x$ & $1$ & $1$ & $1.4$ & $1$ & $1.175$\\
\hhline{-||----||---||}
\#7& $-100$ & $-100$ & $-100$ & $-750$ & $-830+150\cdot x$ & $-325$ & $-70$\\
\hhline{-||----||---||}
\#8& $0.7+1.3\cdot x$ & $1.5$ & $1.5$ & $1.5$ & $1.5$ & $1$ & $0.5$\\
\hhline{-||----||---||}
\#9& $1.4+2.6\cdot x$ & $3$ & $3$ & $2.5$ & $2.5$ & $0$ & $0.1$\\
\hhline{-||----||---||}
\#10& $0$ & $0$ & $0$ & $\pi$ & $\pi$ & $0$ & $0$\\
\hhline{-||----||---||}
\#11& $1.4+2.6\cdot x$ & $3$ & $3$ & $-2.5$ & $0.5$ & $0$ & $0.1$\\
\hhline{=:b:====:b:===:b|}
\end{tabular}
\caption{\label{tab:scenarios}Input parameters for the scenarios
  considered in Sect.\,\ref{sec:numerics}. The bilinear soft
  SUSY-breaking parameter of all the squarks of the third generation
  is denoted by~$m_{\tilde{Q}}$. Moreover, $2\,M_1=M_2=M_3/5=500$\,GeV
  and~$m_{\tilde{F}}=1.5$\,TeV, where~$\tilde{F}$ represents any
  sfermion of the first two generations or sleptons of the third
  generation. We vary~$x$ in the interval~$[0,1]$.
  \label{inputtable}}
\end{table}


\subsection[Comparison with \texttt{FeynHiggs} in the~MSSM-limit]{\tocref{Comparison with \texttt{FeynHiggs} in the~MSSM-limit\label{sec:feynhiggs}}}

The~MSSM~limit of the~NMSSM is obtained at vanishingly small values
of~$\lambda$ and~$\kappa$: the singlet superfield then decouples from
the~MSSM~sector but the~$\mueff$~term remains relevant as long
as~$\kappa\sim\lambda$. It is then possible to compare our results for
the Higgs decays to the corresponding predictions
of~\verb|FeynHiggs-2.13.0|.  The settings of~\verb|FeynHiggs| are thus
adjusted in order to match the level of higher-order contributions and
renormalization conditions of our~NMSSM mass calculation: the
corresponding~\texttt{FeynHiggs} input flags read
\texttt{FHSetFlags[4,0,0,3,0,2,0,0,1,1]}. We will denote
the~MSSM(-like) Higgs bosons as~$h$ and~$H$, for the~\cp-even states,
and~$A$ for the~\cp-odd one.

\medskip

\begin{figure}[bp!]
  \centering
  \includegraphics[width=\textwidth]{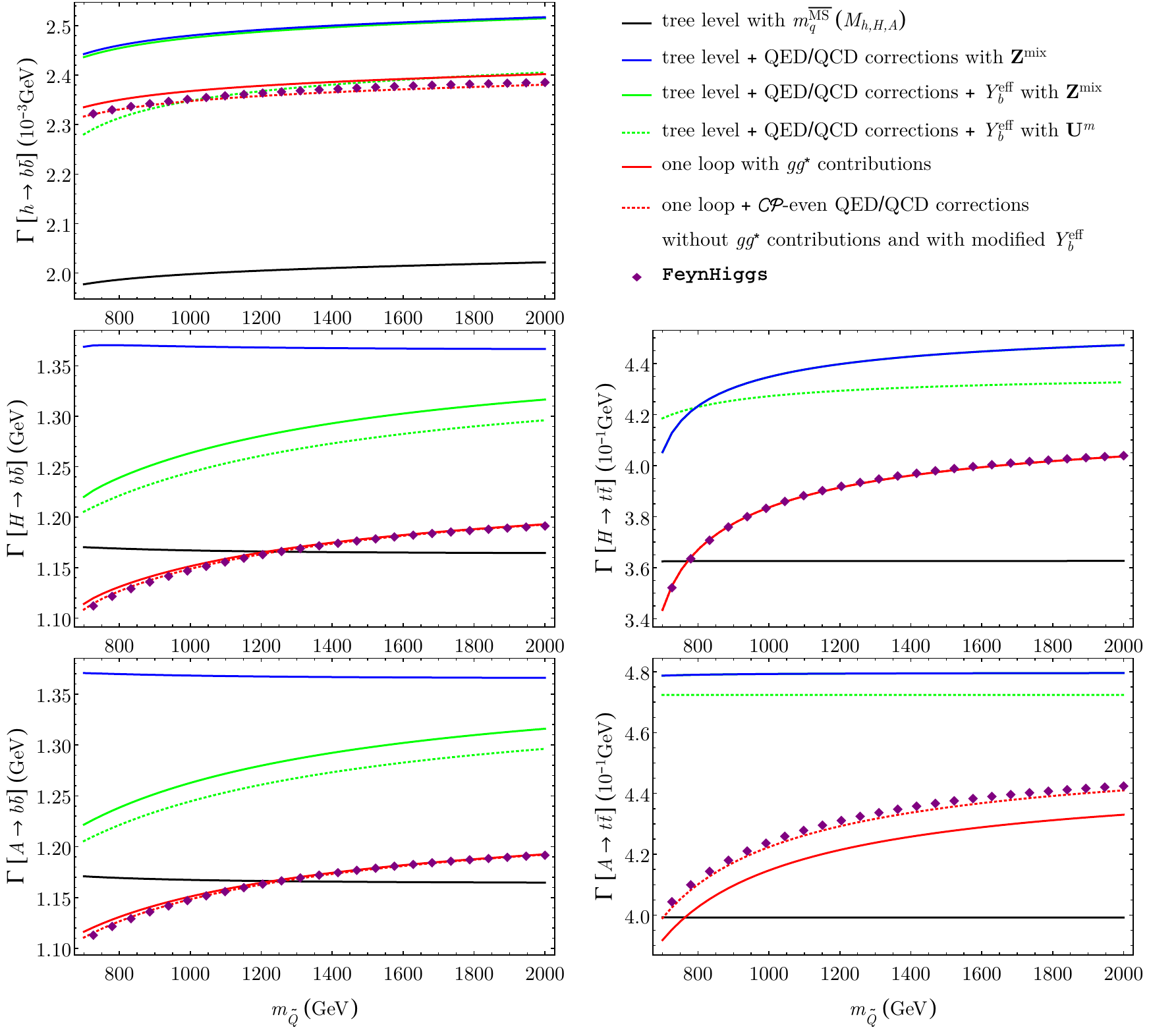}
  \caption{\label{fig:MSSM_hff} Comparison with \texttt{FeynHiggs} of
    the decay widths of the doublet Higgs states into~$b\bar{b}$
    and~$t\bar{t}$ in the~MSSM-limit of the~NMSSM. The input
    parameters are provided in \refta{tab:scenarios}. The black line
    shows the pure tree-level result (with~\MSbar~Yukawa couplings);
    for the blue line, QCD~and~QED~corrections are included, and the
    loop-corrected Higgs mass eigenstate is obtained
    using~$\mathbf{Z}^{\mbox{\tiny mix}}$; for the solid green line
    furthermore the SUSY corrections to the
    Higgs--$b\bar{b}$~couplings (in the case of the~$b\bar{b}$~final
    state) are included; the dashed green line shows the corresponding
    result where~$\mathbf{U}^m$ has been used instead
    of~$\mathbf{Z}^{\mbox{\tiny mix}}$; the red curve corresponds to
    our prediction at the full one-loop level (with higher-order
    improvements). The contribution from~$h_i\to g(g^*\to q\bar{q})$
    is included within the results in the solid red curves. The purple
    diamonds mark the corresponding evaluation by
    \texttt{FeynHiggs}. We also consider the widths without the
    contribution from~$gg^*$, as well as using~\cp-even
    QCD/QED-correction factors in the case of the~$A$~decays (in order
    to match \texttt{FeynHiggs}) as a dashed red line (which is not
    visible in the case~$H\to t\bar{t}$).}
\end{figure}

First, we consider the Higgs decays into~SM~fermions. We turn to a
region of the parameter space of the~\cp-conserving~NMSSM
characterized by the input provided in the column
`\reffi{fig:MSSM_hff}' of \refta{tab:scenarios}, where we vary the
masses and trilinear couplings associated with the squarks of the
third generation. In this setup, the lightest Higgs state is~SM-like,
with a mass in the range~$[124,126.5]$\,GeV, while the heavy doublet
states both have masses of about~$997$\,GeV in this scenario. The full
range under study is found to be in agreement with constraints in the
Higgs sector, as implemented in \texttt{HiggsBounds} and
\texttt{HiggsSignals}.

\medskip
In \reffi{fig:MSSM_hff}, we show the variations of the Higgs decay
widths into~$b\bar{b}$ and~$t\bar{t}$ for the doublet states (when
kinematically allowed). The solid lines correspond to our predictions
in several approximations: at the `tree~level' with Yukawa couplings
defined in terms of running~\MSbar~quark masses at the scale of the
physical Higgs mass~(black); including QCD~and~QED~corrections (but
without~SQCD~contributions) as well as the transition to the physical
Higgs state via~$\mathbf{Z}^{\mbox{\tiny mix}}$~(blue); replacing also
the Higgs couplings to the down-type quarks by their effective form as
expressed in Eq.\,\eqref{eq:effHbb}~(green); at full one loop, with
higher-order improvements as described above~(red). The green~dotted
line is similar to the solid green line, up to the replacement
of~$\mathbf{Z}^{\mbox{\tiny mix}}$ by its unitary
approximation~$\mathbf{U}^m$. The purple diamonds are obtained with
\texttt{FeynHiggs}. As the Higgs masses vary little over the range of
the scan, the modification of the decay widths is essentially driven
by radiative effects: this explains the relatively flat behavior of
the tree-level results---at least in the case of the heavy states; for
the light state, the mass and decay widths vary somewhat more. The
same remark applies to the~QCD/QED-corrected widths
with~$\mathbf{Z}^{\mbox{\tiny mix}}$, although the~$H\to t\bar{t}$
decay width displays a more pronounced variation due to
the~$h^0$--$H^0$~mixing at the loop level. The one-loop corrections to
the~$b\bar{b}$~decay width of the~SM-like state shift this quantity
upwards by~$\simord 20\%$. The bulk of this effect, beyond
the~QCD~corrections included in the running~$b$~quark mass, is already
contained within the~QCD/QED-corrected width. For the heavy-doublet
states, however,~QCD~and~QED~corrections (beyond the effect encoded
within the running Yukawa coupling) do not lead to a significant
improvement of the prediction of the decay widths, as the `tree-level'
result often appears closer to the full one-loop widths than the blue
curve---for the~$b\bar{b}$~final state, this deviation is only
partially explained by the radiative corrections that can be resummed
within the effective Higgs couplings to the bottom quark (solid green
curve).

In the case of the~$t\bar{t}$~final state, the solid green curve and
the blue curve are essentially identical as the effective couplings to
the down-type quarks only play a secondary part. The difference
between the solid and dotted green lines originates from the treatment
regarding the transition matrix employed for the description of the
external Higgs leg---$\mathbf{Z}^{\mbox{\tiny mix}}$ or the
approximation~$\mathbf{U}^m$: the consequences for the predicted width
reach~$\mathcal{O}(5\%)$. We note that, with the exception of \FH,
public tools usually neglect the effects associated with the momentum
dependence of the Higgs self-energies (only properly encoded
within~$\mathbf{Z}^{\mbox{\tiny mix}}$). On the other hand, the
estimate using~$\mathbf{U}^m$ is somewhat closer to the full one-loop
result (using~$\mathbf{Z}^{\mbox{\tiny mix}}$), which means that, as
long as one restricts to an `improved tree-level approximation', the
choice of a unitary transition matrix might provide slightly more
reliable results than the same level of approximation
with~$\mathbf{Z}^{\mbox{\tiny mix}}$. As we mentioned earlier,
the~$\simord10\%$~difference between the red and green curves in the
case of the heavy doublets hints at sizable electroweak effects of
one-loop order. This can be understood in terms of large electroweak
Sudakov logarithms for the heavy Higgs bosons. The impact of the
sfermion spectrum on the decay widths into~SM~fermions consists in a
suppression in the presence of light stops and sbottoms. This effect
is of order~$10\%$ over the considered range of squark
masses. Concerning the comparison with \texttt{FeynHiggs}, we observe
a very good agreement of the predicted one-loop widths for
the~\cp-even states: this is expected since we essentially apply the
same processing of the parameters. However, a small discrepancy in
the~$h\to b\bar{b}$~width is noticeable: it is related to our
inclusion of the contribution from~\mbox{$h\to g(g^*\to b\bar{b})$} to
the decay width. Subtracting this contribution (dashed red curve), we
recover the \texttt{FeynHiggs} prediction. This effect is negligible
for the other Higgs states.\footnote{The option for a similar
  processing of the~\mbox{$h\to g(g^*\to q\bar{q})$}~contributions
  will be integrated in an upcoming \FH~version.} Furthermore, for
the~\cp-odd state, we find a deviation of a few percent in
the~$t\bar{t}$~channel: this difference is due to \texttt{FeynHiggs}
employing~\cp-even~QCD/QED-correction factors instead of the~\cp-odd
ones. The agreement is restored if we adopt the same approximation
(dashed red curve). No such discrepancy appears for
the~$b\bar{b}$~final state, as the~\cp-even
and~\cp-odd~QCD/QED-correction factors converge in the limit of very
light fermions (as compared to the mass of the Higgs
state). Furthermore, the processing of the effective Higgs couplings
to down-type quarks differs between \FH~and our implementation,
leading to small numerical effects (below~$1\%$) for
light~SUSY~spectra: we evaluate~$\Delta_b$ at the scale defined by the
arithmetic mean of the~SUSY~masses involved, while \FH~employs a
geometric mean (`modified~$Y_b^{\text{eff}}$'). Another (numerically
minor) difference with \texttt{FeynHiggs} is the slightly different
treatment of the Goldstone--Higgs~couplings regarding the restoration
of gauge~invariance of the result---see Sect.\,\ref{sec:general}.

\begin{figure}[tp!]
  \centering
  \includegraphics[width=.5\textwidth]{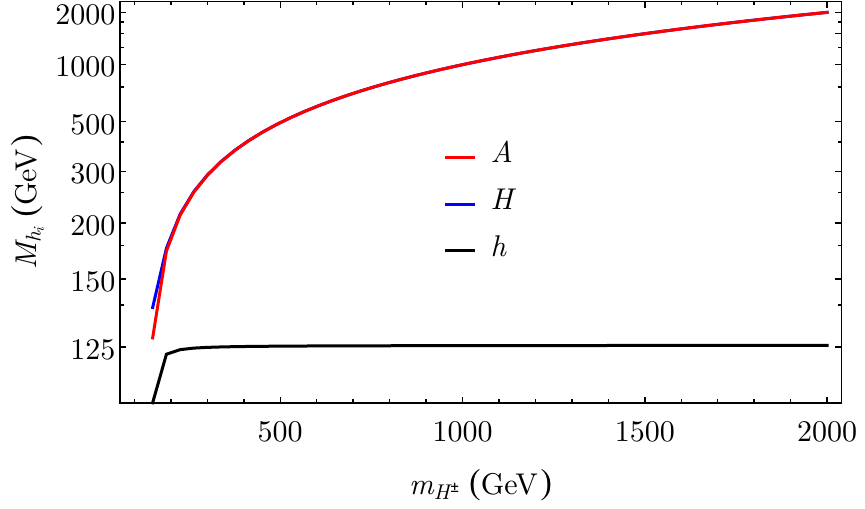}
  \vspace{-1ex}
  \caption{\label{fig:MSSM_hVV_mass} Higgs masses in the scenario of \reffi{fig:MSSM_hVV}.}
  \vspace*{3ex}
  \capstart
  \includegraphics[width=\textwidth]{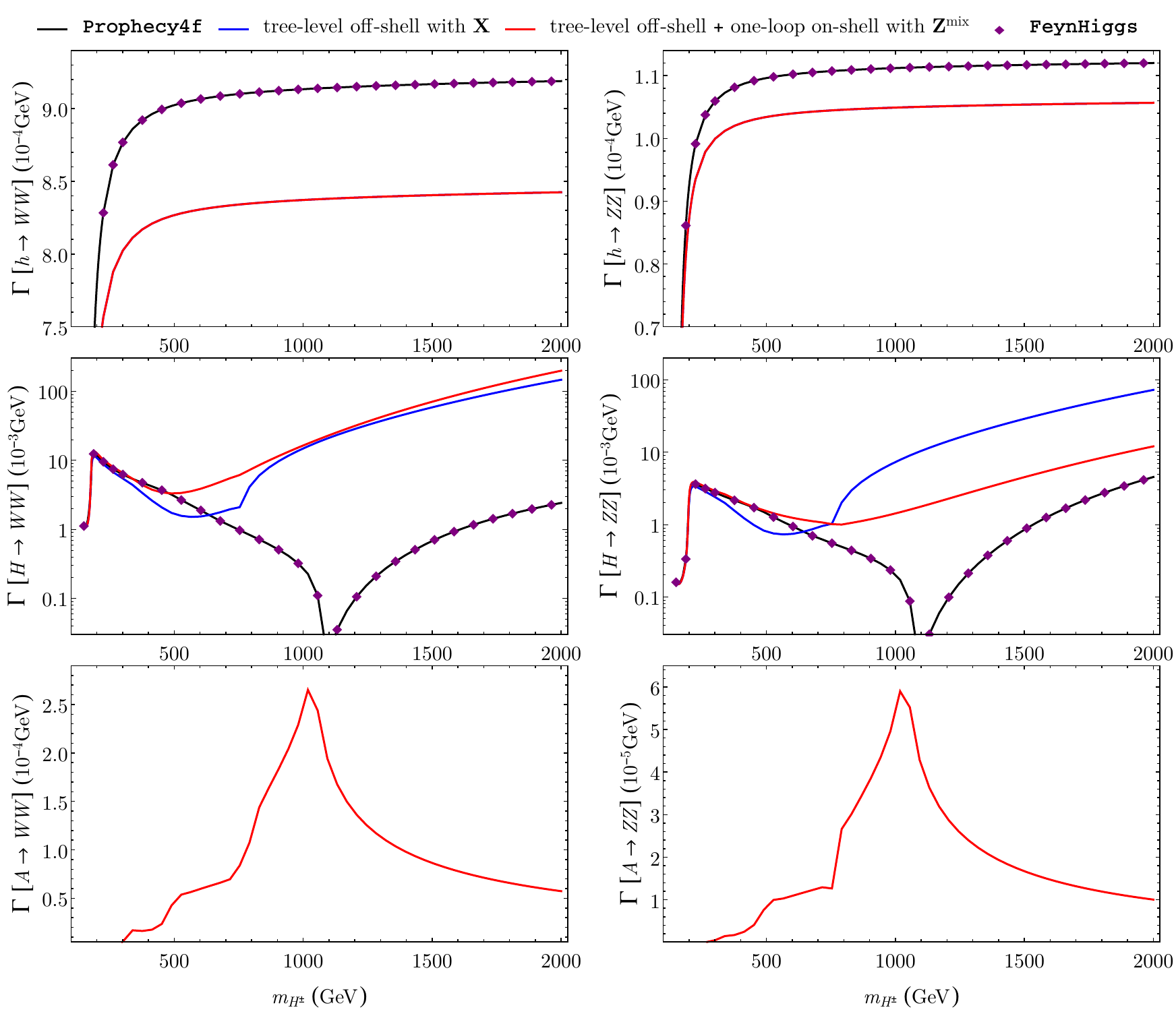}
  \vspace{-4ex}
  \caption{\label{fig:MSSM_hVV} Comparison with \texttt{FeynHiggs} of
    the decay widths of the doublet Higgs states into~$WW$ and~$ZZ$ in
    the~MSSM-limit of the~NMSSM. The input parameters are provided in
    \refta{tab:scenarios}. The black lines correspond to the widths
    estimated by rescaling the~SM~result of \texttt{Prophecy4f} by the
    relative tree-level coupling of the Higgs state in a unitary
    approximation. The blue line corresponds to a tree-level off-shell
    evaluation, including a transformation of the Higgs states
    by~$\mathbf{Z}^{\mbox{\tiny mix}}$. The red curve depicts our
    prediction for the full one-loop on-shell decay width (with
    higher-order improvements), including tree-level off-shell
    effects. The purple diamonds mark the corresponding evaluation by
    \texttt{FeynHiggs}.}
\end{figure}

\medskip
Then, we consider the Higgs decays into electroweak gauge bosons in
the~MSSM-limit. To this end, we perform a scan over the charged-Higgs
mass in the range~$[150,2000]$\,GeV; the rest of the parameters is set
as in \reffi{fig:MSSM_hff}, but the stop and sbottom soft masses and
trilinear couplings are frozen to~$1.5$\,TeV and~$2.3$\,TeV,
respectively. Correspondingly, the lightest~\cp-even Higgs is~SM-like
with a mass of~$\simeqord 124$--$125$\,GeV as soon
as~$m_{H^{\pm}}\gtrsim 350$\,GeV. The mass of the heavy doublet states
varies over the scan and is comparable to~$m_{H^{\pm}}$. For clarity,
the masses are shown in \reffi{fig:MSSM_hVV_mass}. Except
for~\mbox{$m_{H^{\pm}}\lesssim300$}\,GeV, where the~SM-like Higgs is
not in the desired experimental window, this scenario is consistent
with experimental bounds implemented in \texttt{HiggsBounds} and
\texttt{HiggsSignals}.

In \reffi{fig:MSSM_hVV}, we show our results for the decay widths of
the neutral Higgs states into~$WW$ and~$ZZ$. The black curves
correspond to the prediction obtained by rescaling the~SM~width of
\texttt{Prophecy4f} by the tree-level
Higgs--$WW/ZZ$~couplings~(relative to the~SM). The rotation to the
loop-corrected Higgs state is then described in the unitary
approximation~$\mathbf{U}^m$. This is the approach employed by
\texttt{FeynHiggs}~(purple diamonds), leading to a good agreement. In
the case of the~SM-like state (upper plots), the decays are off-shell
and both the (superposed) blue and red curves correspond to the
tree-level results with off-shell kinematics but
with~$\mathbf{Z}^{\mbox{\tiny mix}}$ applied to the external Higgs
leg. Thus, the difference between the red and black lines must be
interpreted as the magnitude of the~SM~radiative corrections
implemented in \texttt{Prophecy4f} (which are not included in our
result displayed by the red curve) and amounts to somewhat less
than~$10\%$. For the heavy~\cp-even Higgs (plots in the middle), the
red and blue curves are differentiated by the inclusion of on-shell
one-loop contributions to the widths (red curves). While the results
essentially agree with the~`\texttt{Prophecy4f}~approach' at low mass
(where the tree-level Higgs--$WW/ZZ$~coupling remains sizable due to a
substantial mixing of the heavy~\cp-even Higgs with the~SM-like
state), large deviations are observed
for~\mbox{$m_{H^{\pm}}\gtrsim500$}\,GeV. Indeed, for this state the
tree-level Higgs coupling to electroweak gauge bosons is very small
(as a consequence of the decoupling limit), so that the decay width is
largely dominated by one-loop effects arising both from the
contributions to the external Higgs leg~(via~$\mathbf{Z}^{\mbox{\tiny
    mix}}$) and in the vertex corrections. The dip of the
`\texttt{Prophecy4f}~prediction'~(black line)
at~\mbox{$m_{H^{\pm}}\simeq1.1$}\,TeV is due to an exactly vanishing
Higgs--gauge coupling (tree level, unitarily rotated) at this point in
parameter space. This does not happen when the couplings are
transformed to the physical Higgs state via~$\mathbf{Z}^{\mbox{\tiny
    mix}}$, due to the imaginary part of this transition matrix. On
the other hand,
the~`\mbox{$\text{tree-level}\cdot\mathbf{Z}^{\mbox{\tiny
      mix}}$}'~approach~(blue line) does not offer an accurate
estimate of the~\mbox{$\text{Higgs}\to WW/ZZ$}~decay widths either,
due to sizable vertex corrections. The `sudden drop' of the~$H$~decay
widths for low values of~$m_{H^{\pm}}$~(for all the curves) is
associated to the off-shell regime (the Higgs state has a mass below
threshold).  Finally, the~\cp-odd Higgs decays into~$WW$
and~$ZZ$~(lower plots) are generated at the radiative level. In this
case, all the approximations based on tree-level predictions
(`\texttt{Prophecy4f}', `$\mathbf{Z}^{\mbox{\tiny mix}}$',
\texttt{FeynHiggs}) vanish, and therefore only our one-loop on-shell
description~(red curve) is displayed in the plots. The peculiar shape
of the predicted decay widths---in particular the peak
at~\mbox{$m_{H^{\pm}}\simeq1$}\,TeV---is associated to various
thresholds---in particular the two-wino threshold
at~$\simeqord1$\,TeV. We stress that one-loop effects dominate the
decays of the heavy doublet Higgs states into~$WW/ZZ$, which means
that our results, albeit formally of next-to-leading order, come with
a large uncertainty due to~QCD~(two-loop)~corrections: in particular,
we observed that the use of pole instead of~\MSbar~quark masses within
the loop could shift the widths by~$\simord50\%$.

\begin{figure}[tp!]
  \centering
  \includegraphics[width=\textwidth]{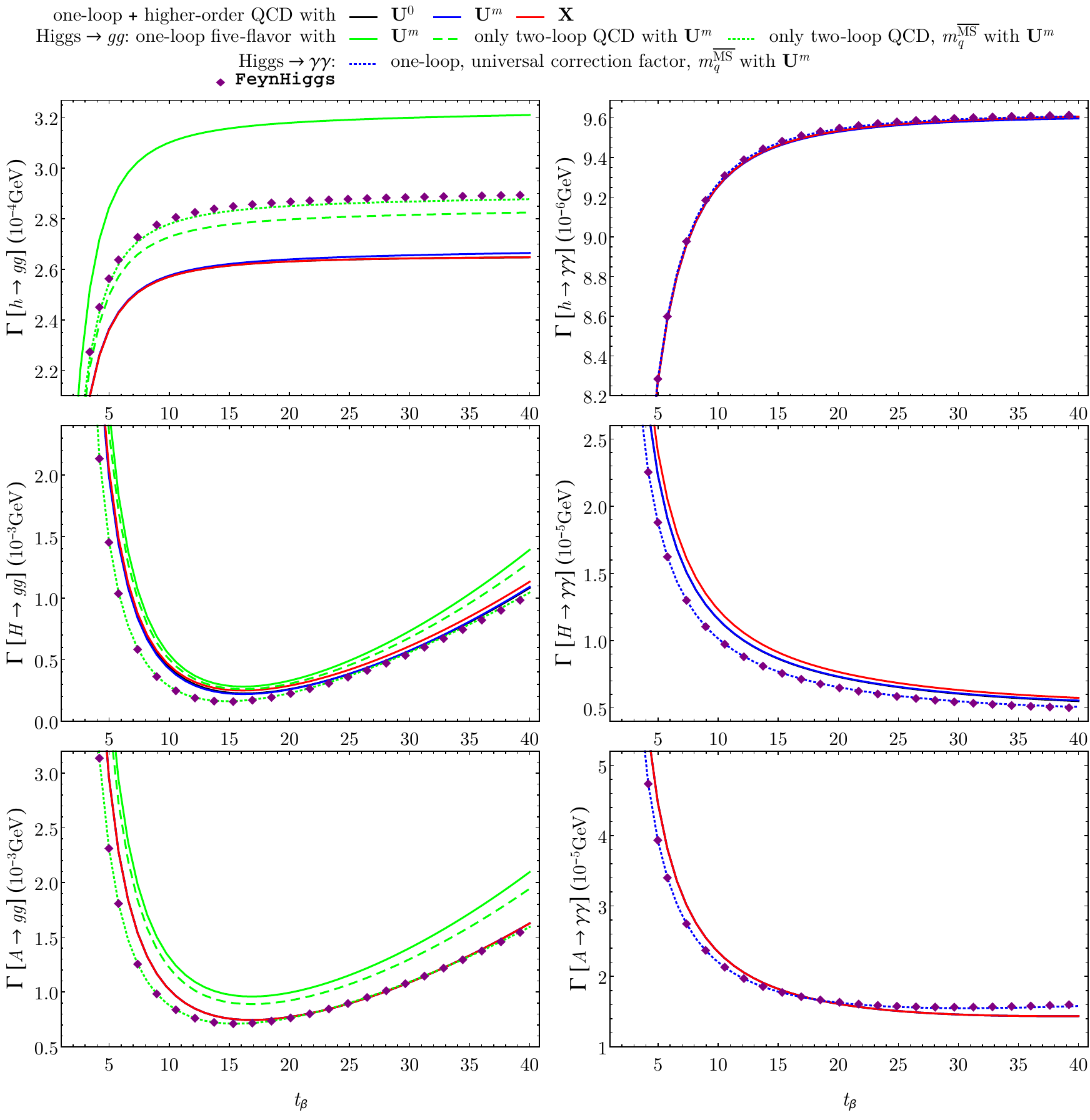}
  \vspace{-4ex}
  \caption{\label{fig:MSSM_hgg} Comparison with \texttt{FeynHiggs} of
    the decay widths of the doublet Higgs states into~$gg$
    and~$\gamma\gamma$ in the~MSSM-limit of the~NMSSM. The input
    parameters are provided in \refta{tab:scenarios}. The black, blue
    and red curves correspond to a rotation to the physical Higgs
    states via~$\mathbf{U}^0$, $\mathbf{U}^m$ and~$\mathbf{X}$
    respectively. In the case of the~$gg$ final state, these lines are
    obtained for three radiated quark flavors in
    the~QCD~corrections. The green curves are obtained for five
    radiated quark flavors, employing universal~QCD~corrections at
    NLO~(dashed) or beyond~(solid). Furthermore the dotted green line
    shows the widths for~NLO~QCD-correction factors and~\MSbar~quark
    masses, which is the current approach of
    \texttt{FeynHiggs}~(purple diamonds). In the case of the diphoton
    final state, the dashed lines are obtained with~QCD~corrections in
    the approximation of heavy quarks/squarks and~\MSbar~masses like
    in the current version of \FH.}
\end{figure}

\medskip
We choose to discuss the diphoton and digluon decay widths in
the~MSSM-limit in a scenario where~$\tan\beta$ scans the
range~$[1,40]$---$m_{H^{\pm}}$ is set to~$1$\,TeV again; the details
of the input are available in \refta{tab:scenarios}. The mass of
the~SM-like Higgs state is of order~$100$\,GeV
at~\mbox{$\tan\beta=1$}, but settles in the interval~$[123.5,126.5]$
when~\mbox{$\tan\beta\gtrsim7$}. Correspondingly, the
low-$\tan\beta$~limit is disfavored by \texttt{HiggsSignals} in this
scenario. The heavy doublet states have a mass of
about~$996$--$997$\,GeV. \texttt{HiggsBounds} excludes the
large-\mbox{$\tan\beta\sim40$} endpoints, due to constraints on
heavy-Higgs searches in the~$\tau\tau$~channel\,\cite{CMS:2015mca}.

We display the Higgs decay widths into~$gg$ and~$\gamma\gamma$ in
\reffi{fig:MSSM_hgg}. The transition to the physical Higgs states is
performed using various approximations:~$\mathbf{U}^0$ (black
curves),~$\mathbf{U}^m$ (blue curves) and~$\mathbf{X}$ (red
curves). These three descriptions agree rather well. The predictions
for the digluon final state employ the~QCD~corrections for three
radiated quark flavors---as radiated~$c\bar{c}$, $b\bar{b}$
or~$t\bar{t}$ are regarded as contributions to the fermionic Higgs
decays. Nevertheless, in order to compare with \texttt{FeynHiggs}, we
also show the results in the five-radiated-flavor~approach~(solid
green curves) and cutting off universal~QCD~corrections beyond
NLO~(dashed green curves). The resulting deviation from the
predictions of \texttt{FeynHiggs}~(purple diamonds) is resolved when
replacing the pole quark masses by~\MSbar~masses in the one-loop
amplitude~(dotted green curves). The appropriate choice for the
considered~QCD-correction factor is that of pole masses in the loop,
however. The difference between the widths depicted by the
black/blue/red lines and by the solid green lines is of order~$20\%$:
this enhancement is due to the larger number of radiated
flavors. Universal~QCD~corrections beyond~NLO (difference between the
solid and the dashed green curves) represent almost~$15\%$ of the
width. In the case of the diphoton widths, the results of
\texttt{FeynHiggs}~(purple diamonds) should be compared to our
predictions employing~$\mathbf{U}^m$~(blue curves): a deviation of
somewhat less than~$10\%$ is noticeable for the heavy states. We
checked that this discrepancy can be interpreted in terms of the
heavy-quark/squark approximation that \texttt{FeynHiggs} employs for
the~NLO~QCD~corrections as well as the use of~\MSbar~running masses
(instead of the running masses defined in
Eq.\,(\href{https://arxiv.org/pdf/hep-ph/9504378.pdf#page=7}{5}) of
\citere{Spira:1995rr}): simplifying our processing of the widths to
this approximation (dashed blue curves) yields a very good agreement
with the results of \texttt{FeynHiggs}.

\begin{figure}[tp!]
  \centering
  \includegraphics[width=\textwidth]{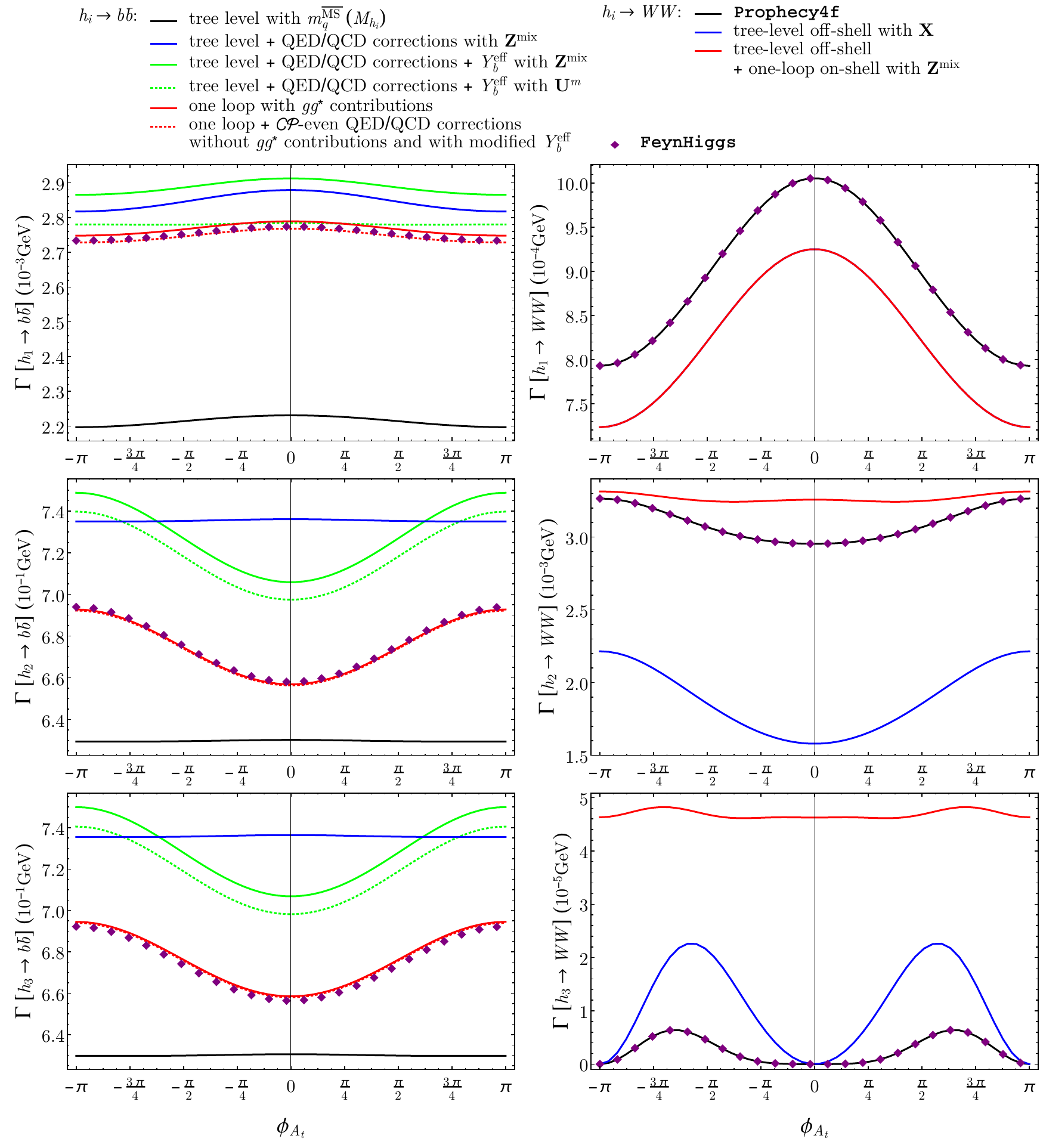}
  \vspace{-4ex}
  \caption{\label{fig:MSSM_CPV} Comparison with FeynHiggs of the decay
    widths of the doublet Higgs states into~$b\bar{b}$ and~$WW$ in
    the~\cp-violating~MSSM-limit of the~NMSSM. The input parameters
    are provided in \refta{tab:scenarios}. For the~$b\bar{b}$~final
    state, the widths shown in black include only tree-level effects;
    these are rescaled by the~QCD/QED~corrections and transformed into
    physical states for the blue curves; for the green curve, leading
    corrections to the bottom Yukawa couplings are added; the red
    curves correspond to the full one-loop results (with higher-order
    improvements). In the case of the~$WW$~final state, the black
    curves correspond to the approach where the~SM~widths of
    \texttt{Prophecy4f} are rescaled; the blue curves depict
    tree-level off-shell widths where~$\mathbf{Z}^{\mbox{\tiny mix}}$
    is applied; the red curves include on-shell one-loop corrections.
    The purple diamonds mark the predictions of \texttt{FeynHiggs}.}
\end{figure}

\medskip
Finally, we consider a~\cp-violating scenario in \reffi{fig:MSSM_CPV}:
the parameters are set as indicated in the
column~`\reffi{fig:MSSM_CPV}' of \refta{tab:scenarios}. We perform a
scan over~$\phi_{A_t}\in[-\pi,\pi]$. The doublet Higgs states have
masses of about~$\simord125$\,GeV,~$493$\,GeV and~$494$\,GeV. This
scenario is phenomenologically consistent with the limits on the Higgs
sector as implemented in \texttt{HiggsBounds} and
\texttt{HiggsSignals}. Ideally, limits from the measured electric
dipole moments~(EDMs) should be considered as well: in particular,
Barr--Zee contributions involving squark loops are sensitive to
variations of~$\phi_{A_t}$. On the other hand, such effects are
relatively suppressed given the high mass of the
stops~($\simord1.5$~TeV). In any case, we mainly consider this
scenario for the sake of comparison.

\reffi{fig:MSSM_CPV} displays the Higgs decay widths into~$b\bar{b}$
and~$WW$.  For the~$b\bar{b}$~final state, we consider the tree-level
widths (corresponding to an~\MSbar~running Yukawa coupling; black
lines), incorporate~QCD/QED~corrections and transform to the physical
Higgs state using~$\mathbf{Z}^{\mbox{\tiny mix}}$~(blue), subtract and
redistribute the~SUSY~corrections to the bottom-quark mass in the
definition of the Higgs-bottom couplings~(green) and finally evaluate
the widths at full one-loop order~(red, including two-loop pieces as
described in Sect.\,\ref{sec:theory}). The predictions of
\texttt{FeynHiggs}~(purple diamonds) are in good agreement with our
full one-loop results. While the inclusion of the corrections
associated to~QED/QCD~effects and the definition of effective Higgs
couplings to~$b\bar{b}$ notably improve the tree-level width of
the~SM-like state, as compared to the one-loop results---from a
discrepancy of~$\simord20\%$ to less than~$\simord4\%$---the
performance of these~`leading'~corrections is less convincing in the
case of the heavy doublet states: the deviations with respect to the
full one-loop results remain of order~\mbox{$5$--$10\%$}.

Turning to the~$WW$~final state, we expectedly recover the results of
\texttt{FeynHiggs} in the approximation with the rescaled~SM~widths of
\texttt{Prophecy4f}~(black lines). The impact of the one-loop
corrections~(red curves) on the width of the mostly~\cp-even heavy
doublet state~$h_2$ are rather mild, which we should put in
perspective with the fact that this state is comparatively
light. However, for the mostly~\cp-odd state~$h_3$, the tree-level
approximations sizably underestimate the one-loop widths.

\medskip
To summarize, in this comparison with \texttt{FeynHiggs} in
the~MSSM~limit of the~NMSSM, we were able to quantitatively recover
the widths predicted by \texttt{FeynHiggs} and interpret the origin of
the differences with our results. In the case of the Higgs decays into
electroweak gauge bosons, our one-loop approach goes beyond the
current approach of \texttt{FeynHiggs} and shows that the tree-level
approximation for the~SUSY~contributions---even though rescaled from
the~\texttt{Prophecy4f}~SM~widths---leads to significant deviations
for the heavy doublet states. In the case of the digluon decay,
universal~QCD~corrections beyond~NLO as well as the number of radiated
quark flavors have a sizable impact on the widths. Finally, we
observed that accounting for the mass dependence in
the~QCD~corrections to the diphoton widths has a mild effect on the
decay of the heavy states. It is planned to include all the
refinements that go beyond the current status of \texttt{FeynHiggs}
and that we have employed here into the predictions of the~MSSM~Higgs
decays of a future update of \texttt{FeynHiggs}.

\subsection[Comparison with \texttt{NMSSMCALC}]{\tocref{Comparison with \texttt{NMSSMCALC}\label{sec:nmssmcalc}}}

We now depart from the~MSSM-limit of the~NMSSM. We first investigate
the Higgs decays in scenarios that are similar to those that we
considered in the~MSSM-limit.

\begin{figure}[tp!]
  \centering
  \includegraphics[width=\textwidth]{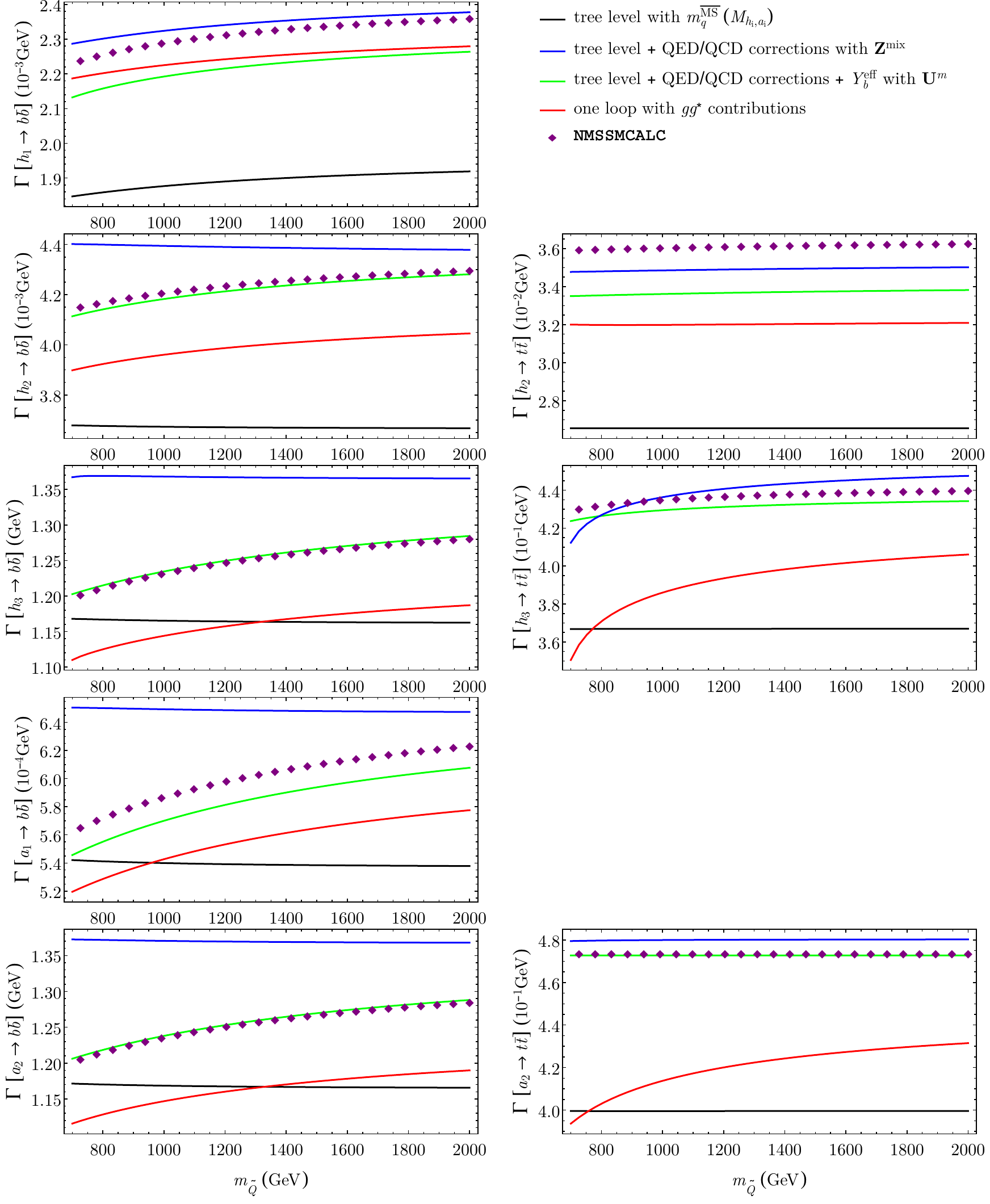}
  \vspace{-4ex}
  \caption{\label{fig:NMSSM_hff} Comparison with \texttt{NMSSMCALC} of
    the decay widths of the Higgs states into~$b\bar{b}$
    and~$t\bar{t}$. The input parameters are provided in
    \refta{tab:scenarios}. The black lines are pure tree-level
    results; for the blue lines,~QCD and~QED~corrections are included
    and the Higgs states are transformed to the loop-corrected
    eigenstates using~$\mathbf{Z}^{\mbox{\tiny mix}}$; the green
    curves employ~$\mathbf{U}^m$,~QCD/QED~corrections and an effective
    bottom Yukawa coupling; the red curve corresponds to our
    prediction at the one-loop level. The purple diamonds mark the
    corresponding evaluation by \texttt{NMSSMCALC}.}
\end{figure}

We will compare our estimates for the Higgs decay widths to the
predictions of \texttt{NMSSMCALC-2.1}\,\cite{Baglio:2013iia,
  NMSSMCALC-www}. In order to minimize the impact of the Higgs-mass
calculation, we bypass the mass-evaluation routines of
\texttt{NMSSMCALC}---we refer the reader to
\citeres{Drechsel:2016htw,Domingo:2017rhb} for a comparison of the
Higgs-mass predictions---and directly inject our Higgs spectrum
(two-loop masses and~$\mathbf{U}^m$ mixing matrix) in
the~SLHA\,\cite{Skands:2003cj,Allanach:2008qq}~file serving as an
interface between the mass-evaluation and the decay-evaluation
routines. We proceed similarly with the squark masses and mixing
angles. The subroutine of \texttt{NMSSMCALC} that evaluates the Higgs
decays is based on a generalization of
\texttt{HDECAY}\verb|_6.1|\,\cite{Djouadi:1997yw,Butterworth:2010ym}. The
corresponding widths include leading~NLO~effects
(\EG~QCD/QED~corrections, effective Higgs couplings to the bottom
quark, etc.) but do not represent a complete one-loop order
evaluation. Furthermore, internal parameters, renormalization scales
or~RGE~runnings are not strictly identical to our choices. For
instance, the decay widths predicted by \texttt{NMSSMCALC} are
systematically normalized to~$G_F$, while we employ other
parametrizations in terms of~$M_W$, $M_Z$ and \EG~$\alpha(M_Z)$: the
corresponding tree-level contributions numerically differ by a few
percents. In the case of \texttt{NMSSMCALC}, such effects are of
one-loop electroweak order, hence beyond the considered
approximation. In our calculation, the one-loop electroweak order is
consistently implemented with respect to our renormalization
scheme. Other small numerical differences appear at the level of
\EG~running quark masses. Therefore, the numerical comparison between
the two sets of results is expected to show a certain level of
deviations.

\medskip
We set~$\kappa=0.4$, $\lambda=0.3$ and $A_{\kappa}=-100$\,GeV. In
\reffi{fig:NMSSM_hff}, we consider the Higgs decay widths
into~$b\bar{b}$ and~$t\bar{t}$ in a scenario where the squark masses
of third generation vary between~$0.7$ and~$2$\,TeV while the
trilinear couplings are in the interval~$[1.4,4]$\,TeV (see
\refta{tab:scenarios} for details). Correspondingly, the
lightest~\cp-even Higgs state is~SM-like with a mass
of~$\simord120$--$126$\,GeV (the lowest range in~$m_{Q}$ is in tension
with the measured Higgs data); the second-lightest~\cp-even and the
lightest~\cp-odd Higgs states are singlet-like, with masses
of~$\simord643$\,GeV and~$\simord319$\,GeV respectively; the heavy
doublet states have masses of~$\simord999$\,GeV. This scenario
satisfies the tests of \texttt{HiggsBounds} and \texttt{HiggsSignals}
in most of the~$m_{\tilde{Q}}$~range. The decay widths are shown at
the tree level~(with~\MSbar~Yukawa couplings; black curves), in the
approximation of~QCD/QED~corrections and including the transition
by~$\mathbf{Z}^{\mbox{\tiny mix}}$~(blue curves), including the
leading~SUSY~corrections to the relation between the bottom-quark mass
and the bottom Yukawa coupling, and furthermore
substituting~$\mathbf{Z}^{\mbox{\tiny mix}}$ by~$\mathbf{U}^m$ (green
curves), and finally for our full one-loop calculation (red
curves). These various approximations perform in the same fashion as
what we observed in the~MSSM-limit: while the inclusion of~QCD/QED-
and Yukawa-driven leading corrections improve the tree-level-based
predictions for the width of the~SM-like state~$h_1$, the improvement
is less obvious for the other Higgs states, pointing at sizable
electroweak effects. The purple diamonds represent the predictions of
\texttt{NMSSMCALC}/\texttt{HDECAY}: they should correspond to the
approximation of our results shown in green. These results
qualitatively agree at the level of a few percent (as expected, given
\EG~the differing parametrizations at the tree level). These~`improved
tree-level'~predictions of the decay widths typically
remain~$5$--$10\%$ away from our full one-loop implementation.

\begin{figure}[bp!]
  \centering
  \includegraphics[width=.5\textwidth]{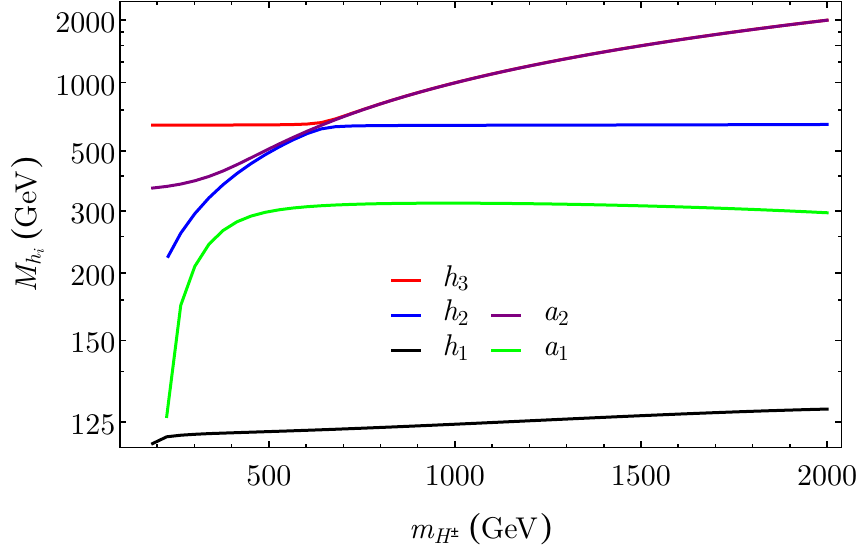}
  \vspace{-2ex}
  \caption{\label{fig:NMSSM_hVV_mass} Higgs masses in the scenario of
    \reffi{fig:NMSSM_hVV}.}
  \vspace{-3ex}
\end{figure}


\begin{figure}[tp!]
  \centering
  \includegraphics[width=\textwidth]{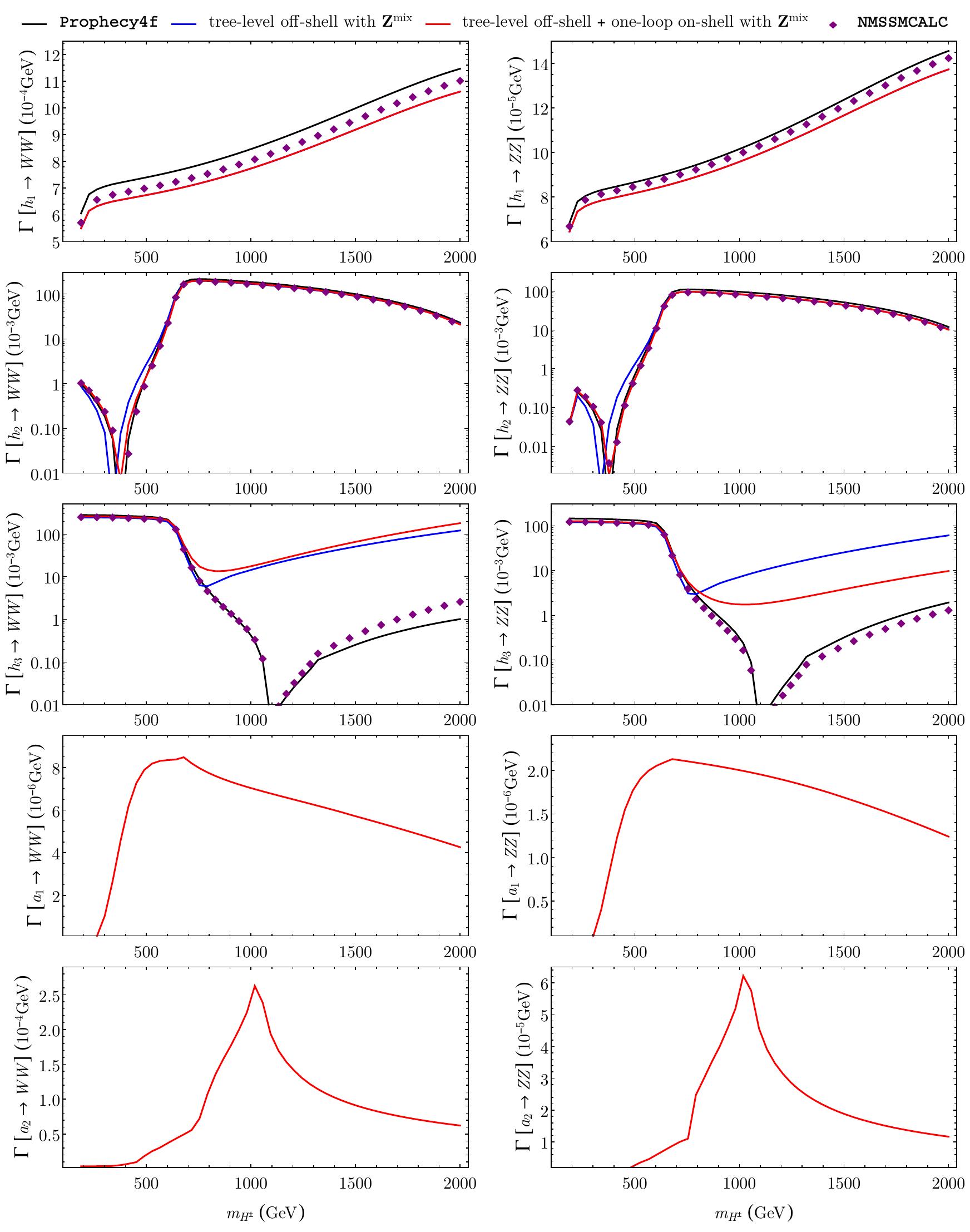}
  \vspace{-4ex}
  \caption{\label{fig:NMSSM_hVV} Comparison with \texttt{NMSSMCALC} of
    the decay widths of the Higgs states into~$WW$ and~$ZZ$. The input
    parameters are provided in \refta{tab:scenarios}. The black lines
    correspond to the widths estimated by rescaling the~SM~result of
    \texttt{Prophecy4f} by the relative tree-level coupling of the
    Higgs states in a unitary approximation. The blue lines correspond
    to a tree-level off-shell evaluation,
    including~$\mathbf{Z}^{\mbox{\tiny mix}}$. The red curves depict
    our prediction for the full one-loop on-shell decay widths,
    including tree-level off-shell effects. The purple diamonds mark
    the corresponding evaluation by \texttt{NMSSMCALC}.}
  \vspace{-4ex}
\end{figure}

\begin{figure}[tp!]
  \centering
  \includegraphics[width=\textwidth]{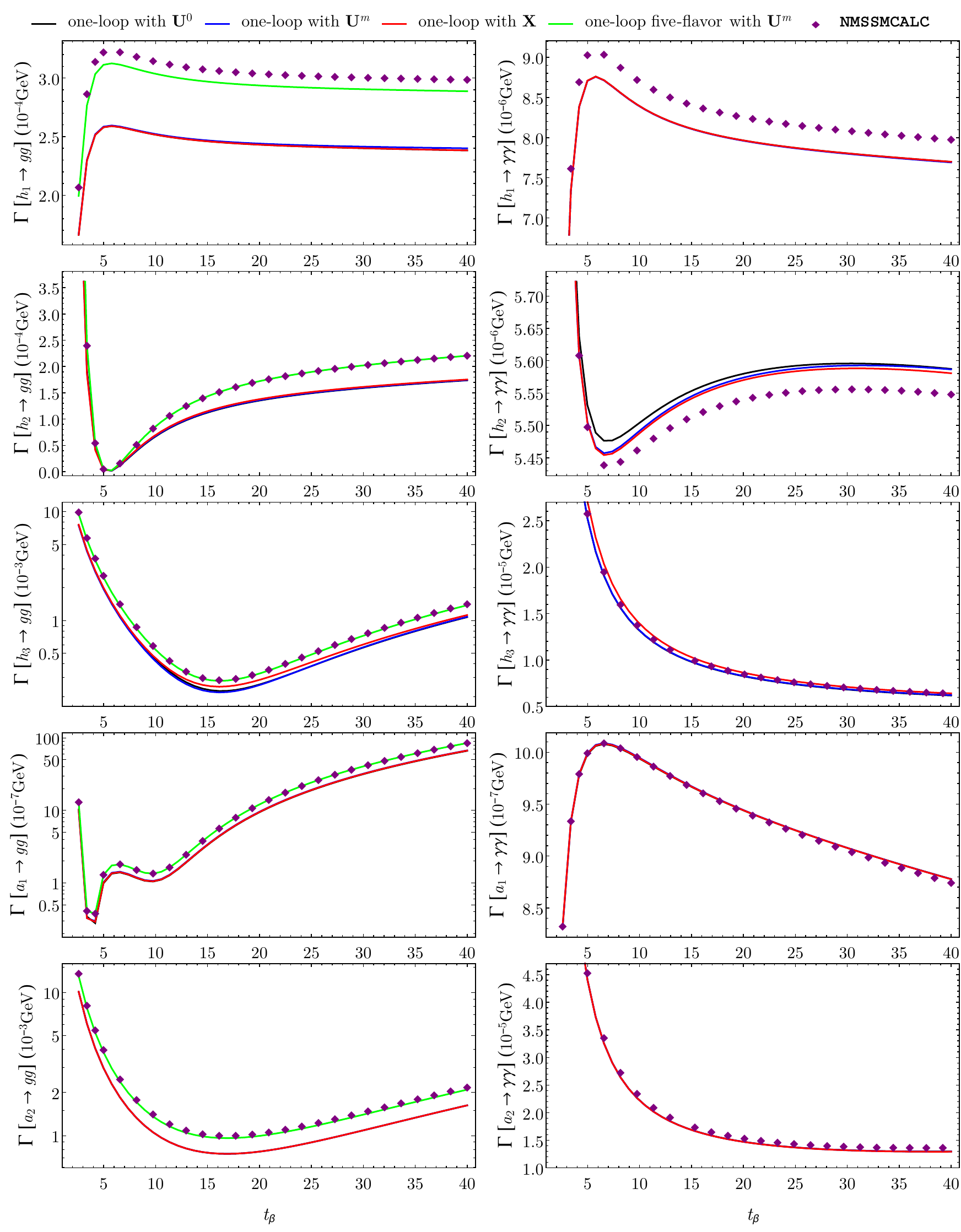}
  \vspace{-4ex}
  \caption{\label{fig:NMSSM_hgg} Comparison with \texttt{NMSSMCALC} of
    the decay widths of the Higgs states into~$gg$ and~$\gamma\gamma$.
    The input parameters are provided in \refta{tab:scenarios}. The
    black, blue and red curves correspond to a transition to the
    physical Higgs state via~$\mathbf{U}^0$, $\mathbf{U}^m$
    and~$\mathbf{X}$ respectively. In the case of the~$gg$~final
    state, the black/blue/red lines are obtained for three radiated
    quark flavors in the~QCD~corrections. The green lines are obtained
    for five radiated quark flavors, which agrees with the approach of
    \texttt{NMSSMCALC}~(purple diamonds).}
  \vspace{-5ex}
\end{figure}

\medskip
Then, we study the Higgs decays to electroweak gauge bosons as a
function of the charged-Higgs mass---the squark masses and trilinear
couplings of third generation are frozen to~$1.5$\,TeV and~$3$\,TeV
respectively. The lightest~\cp-even Higgs with
mass~$\simord122$--$127$\,GeV is~SM-like. The mostly
singlet-like~\cp-even and~\cp-odd states have masses of
order~$650$\,GeV and~$300$\,GeV, respectively.  With
growing~$m_{H^{\pm}}$, the masses of the heavy doublet states increase
in proportion and successively cross the singlet masses, leading to
strong mixing regimes. For clarity, the Higgs masses are plotted in
\reffi{fig:NMSSM_hVV_mass}. The decay widths into~$WW$ and~$ZZ$ are
displayed in \reffi{fig:NMSSM_hVV} in the approximation where
the~SM~widths of \texttt{Prophecy4f} are rescaled (black curves), in
the tree-level approximation using~$\mathbf{Z}^{\mbox{\tiny mix}}$
(blue curves), in our on-shell one-loop description (with off-shell
tree-level contributions; red curves). The predictions of
\texttt{NMSSMCALC} are in agreement with our result obtained by a
rescaling of the \texttt{Prophecy4f} widths. Again, we find that this
approximation is not appropriate for heavy doublet Higgs states
($h_3$~for~\mbox{$m_{H^{\pm}}\gtrsim650$\,GeV}), in which case
one-loop contributions dominate the decays into~$WW$
and~$ZZ$. Differences may reach a factor~$100$ in certain mass
ranges/scenarios. On the other hand, the various predictions for the
width of the~\cp-even singlet at~$\simord650$\,GeV ($h_3$~for
low~$m_{H^{\pm}}$, then~$h_2$) are all consistent with one another: at
least in the scenario under consideration, the tree-level contribution
remains dominant for this state. At~\mbox{$m_{H^{\pm}}\sim400$\,GeV},
we observe a suppression of the decay widths of the heavy doublet
state~$h_2$: once again, it is associated with a
vanishing~$h_2$--$WW/ZZ$~coupling. One-loop effects do not lead to
large deviations for this state. Furthermore, the `drop' at
low~$m_{H^{\pm}}$ in the~\mbox{$h_2\to ZZ$}~channel can be traced back
to the crossing of the kinematical threshold for on-shell~$Z$~bosons
(the width below threshold is suppressed). Finally, the decays of
the~\cp-odd Higgs are generated at the one-loop level: consequently,
they are non-trivial only in our one-loop approach.

\begin{figure}[b!]
  \centering
  \includegraphics[width=\textwidth]{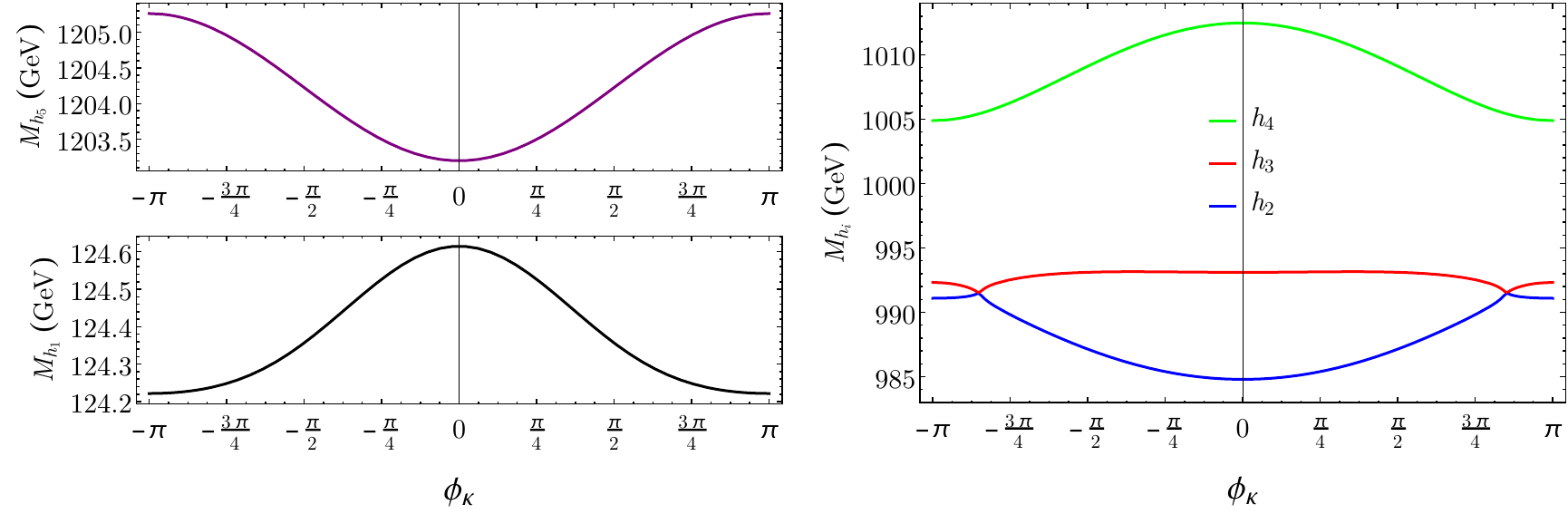}
  \vspace{-4ex}
  \caption{\label{fig:NMSSM_CPV_mass} Higgs masses in the scenario of
    \reffi{fig:NMSSM_CPV}.}
  \vspace{-1ex}
\end{figure}

\medskip
We investigate the Higgs decay widths into gluon and photon pairs as a
function of~$\tan\beta$. The results are displayed in
\reffi{fig:NMSSM_hgg}. The Higgs masses are of order~$\simord125$\,GeV
for the~SM-like state (except for~\mbox{$\tan\beta\lesssim3$}, in
which case the corresponding mass is too low to satisfy the observed
Higgs properties), $\simord1$\,TeV for the heavy-doublet states,
$\simord320$\,GeV and~$\simord645$\,GeV for the~\cp-odd and~\cp-even
mostly singlet-like states, respectively. The points
with~\mbox{$\tan\beta\gtrsim35$} appear to be in tension
with~LHC~searches for heavy Higgs bosons in the~$\tau^+\tau^-$~decay
channel. Also in this case, we consider several descriptions of the
transition to the physical Higgs state~(black, blue and red lines):
all the predictions agree to a good accuracy. In the case of the
digluon decays, we also show the widths obtained with five radiated
quarks~(green curves): this result is essentially in agreement with
the prediction of \texttt{NMSSMCALC}~(purple diamonds), using the same
approach. For the diphoton decays, \texttt{NMSSMCALC} also applies
full~NLO~QCD~corrections. Again, we observe a good agreement with our
results. We checked that the remaining discrepancies---at the percent
level---between the predictions from \texttt{NMSSMCALC} for the
digluon and diphoton widths and ours are largely accounted for by a
relative normalization
factor~$\left.\left(G_F\,M_W^2\,s_{\text{w}}^2\,\sqrt{2}\right)\middle/\left(\pi\,\alpha\right)\right.=\left[1+\mathcal{O}(\alpha)\right]$
and minor deviations in the running of~$\alpha_s$ and the quark
masses.

\begin{figure}[tp!]
  \vspace{-3ex}
  \centering
  \includegraphics[width=\textwidth]{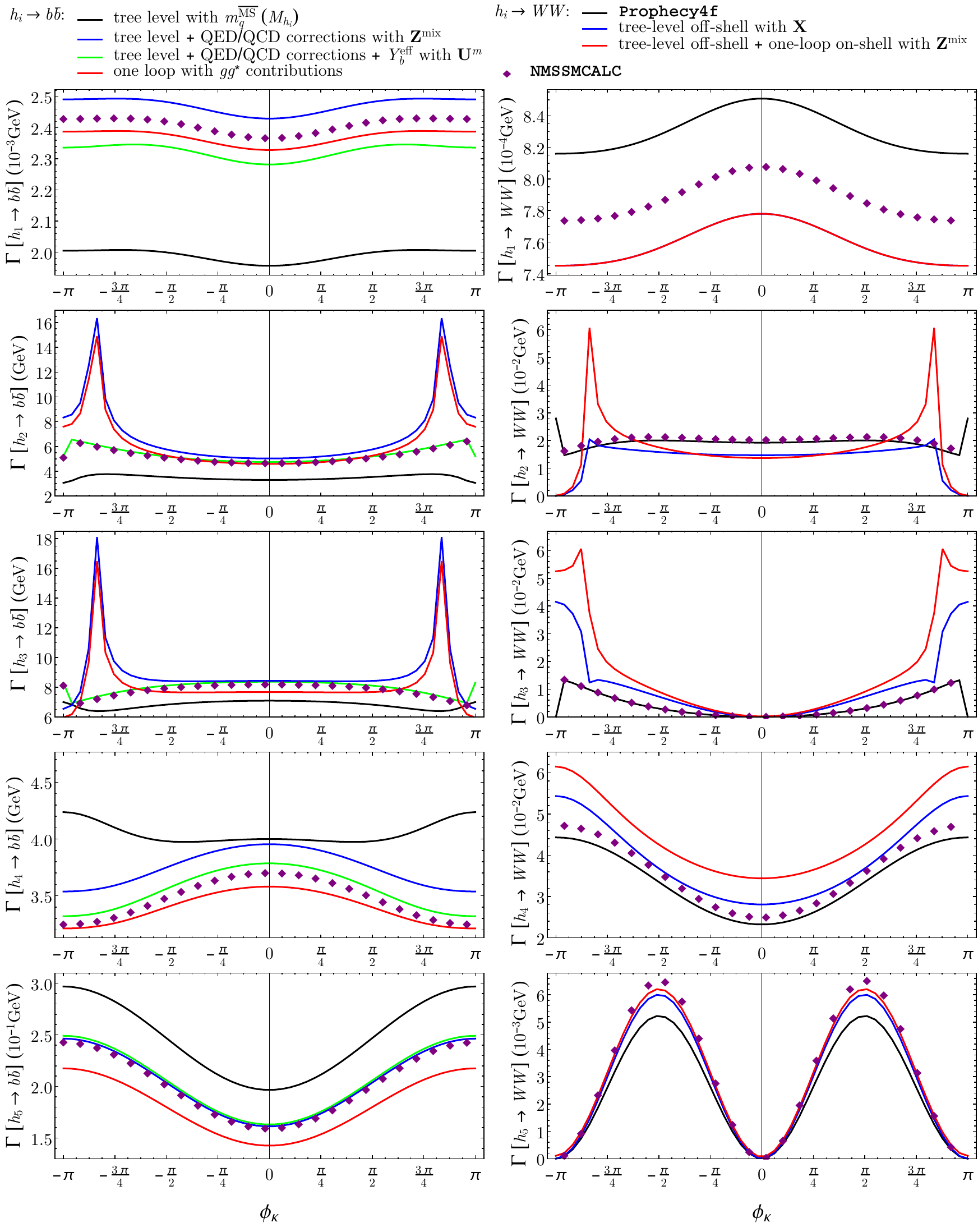}
  \vspace{-4ex}
  \caption{\label{fig:NMSSM_CPV} Comparison with \texttt{NMSSMCALC} of
    the decay widths of the Higgs states into~$b\bar{b}$ and~$WW$ in
    the~\cp-violating~NMSSM. The input parameters are provided in
    \refta{tab:scenarios}. For the~$b\bar{b}$~final state, the widths
    shown in black include only tree-level effects; they are rescaled
    by the~QCD/QED~corrections, and transformed into physical states
    for the blue curves; for the green curves, leading corrections to
    the bottom Yukawa couplings are added; the red curves correspond
    to the full one-loop results. In the case of the~$WW$~final state,
    the black curves correspond to the approach where the~SM~widths of
    \texttt{Prophecy4f} are rescaled; the blue curves depict
    tree-level off-shell widths using~$\mathbf{Z}^{\mbox{\tiny mix}}$;
    the red curves include on-shell one-loop corrections. The purple
    diamonds mark the predictions of \texttt{NMSSMCALC}.}
  \vspace{-5ex}
\end{figure}

\medskip
Finally, we consider a~\cp-violating scenario where we
set~$\lambda=0.2$, $\lvert\kappa\rvert=0.6$, $\tan\beta=25$\,GeV
and~$A_{\kappa}=-750$\,GeV. We scan
over~\mbox{$\phi_{\kappa}\in[-\pi,\pi]$}, which is phenomenologically
realistic in the sense that~EDMs in principle allow for large
variations of this phase\,\cite{Cheung:2011wn,King:2015oxa}; however,
scenarios with large~\cp-violating mixing of the Higgs states---as the
one we consider---tend to be constrained. The Higgs spectrum consists
of an~SM-like state with a mass of~$\simord124.5$\,GeV, a triplet
of~\mbox{\cp-even/\cp-odd} doublet and~\cp-even singlet states
near~$\simord1$\,TeV with large and fluctuating mixing depending
on~$\phi_{\kappa}$, as well as a mostly~\cp-odd singlet
at~$\simord1.2$\,TeV. The masses are depicted in
\reffi{fig:NMSSM_CPV_mass}. The Higgs decays into~$b\bar{b}$ and~$WW$
are plotted in \reffi{fig:NMSSM_CPV}. For the~$b\bar{b}$~final state
we find a relatively good agreement between the predictions of
\texttt{NMSSMCALC}~(purple diamonds) and our predictions employing the
same approximation (shown in green: transition via~$\mathbf{U}^m$,
QCD/QED~corrections and effective~SUSY-corrected
Higgs--$b\bar{b}$~couplings). A sizable discrepancy with our full
one-loop result appears
for~\mbox{$\lvert\phi_{\kappa}\rvert\simeq2.6$} in the case
of~$h_{2,3}$: this difference originates in large Higgs-mixing effects
encoded within~$\mathbf{Z}^{\mbox{\tiny mix}}$ that are not captured
by the~$\mathbf{U}^m$~approximation (see
Sect.\,\href{https://arxiv.org/pdf/1706.00437.pdf#subsection.3.3}{3.3}
and
Sect.\,\href{https://arxiv.org/pdf/1706.00437.pdf#subsection.3.4}{3.4}
of \citere{Domingo:2017rhb} for a detailed discussion). In the case of
the~$WW$~channel, deviations can be large when the tree-level
approximation fails to capture the leading contribution to the decay
width (as \EG~for~$h_{2,3,4}$): the \texttt{NMSSMCALC} predictions are
comparable to our rescaled widths from \texttt{Prophecy4f}.

\medskip
We have thus observed a qualitative agreement of our results with the
decay widths predicted by \texttt{NMSSMCALC}, when employing the same
approximations as this code. Our calculation goes beyond the approach
of \texttt{NMSSMCALC}/\texttt{HDECAY} at the level of the Higgs decays
into~SM~fermions or into~$WW/ZZ$, since we consider the full
one-loop order. Sizable differences may thus appear, \EG~in the
case of the decays into electroweak gauge bosons. Concerning the
decays into gluon or photon pairs, the two codes are considering the
decay widths at the same order, and the results are comparable.

\subsection[A singlet dominated state at 
\texorpdfstring{$\lsimord 100$\,}{\unicodelesssim 100 }GeV and possible 
explanation of slight excesses 
in the~CMS and~LEP data]{\tocref{\label{sec:singlet}A singlet 
dominated state at \boldmath{$\lsimord 100$}\,GeV and 
possible explanation of slight excesses in the~CMS and~LEP data}}

The possible presence of a singlet-dominated state with mass in the
ballpark of~$\simord100$\,GeV is a long-standing phenomenological
trademark of the~NMSSM---see
\EG~\citeres{Dermisek:2007ah,Belanger:2012tt}. One motivation for such
a scenario is for instance the~$2.3\,\sigma$~local excess observed in
Higgs searches at~LEP in the~\mbox{$e^+e^-\to Z(H\to
  b\bar{b})$}~channel\,\cite{Barate:2003sz}, which would be consistent
with a scalar mass of~$\simord98$\,GeV (but with a rather coarse mass
resolution). It would correspond to a signal strength with respect to
the~SM at the level of~$\simord 10\%$. A natural candidate to explain
this excess consists in a mostly singlet-like Higgs with a doublet
component of about~$10\%$~(mixing squared). Interestingly,
recent~LHC~Run\,II~results\,\cite{CMS:2017yta} for~CMS~Higgs searches
in the diphoton~final state show a local excess of~$\simord3\,\sigma$
in the vicinity of~$\simord 96$\,GeV, while a similar upward
fluctuation of~$2\,\sigma$ had been observed in the~CMS~Run\,I~data at
a comparable mass. A hypothetical signal of this kind would amount
to~\mbox{$60\% \pm 20\%$} of that of an~SM~Higgs boson at the same
mass. In the~NMSSM, relatively large Higgs branching fractions
into~$\gamma\gamma$ are possible due to the three-state mixing, in
particular when the effective Higgs coupling to~$b\bar{b}$ becomes
small---see \EG~\citeres{Ellwanger:2010nf,Benbrik:2012rm}. However, it
then appears more difficult to interpret the~LEP~excess
simultaneously.\footnote{For instance, the authors of
  \citere{Cao:2016uwt} came to a negative answer when considering
  the~Run\,I~`excess' together with Dark~Matter constraints in a
  specific region of the parameter space.} Below, we consider the
decays of a light singlet-like Higgs in this regime employing our
calculation, to show that it is indeed possible to describe both
`excesses' simultaneously (without exploiting all possibilities within
the NMSSM to desribe these effects).

\begin{figure}[tp!]
  \centering
  \includegraphics[width=.5\textwidth]{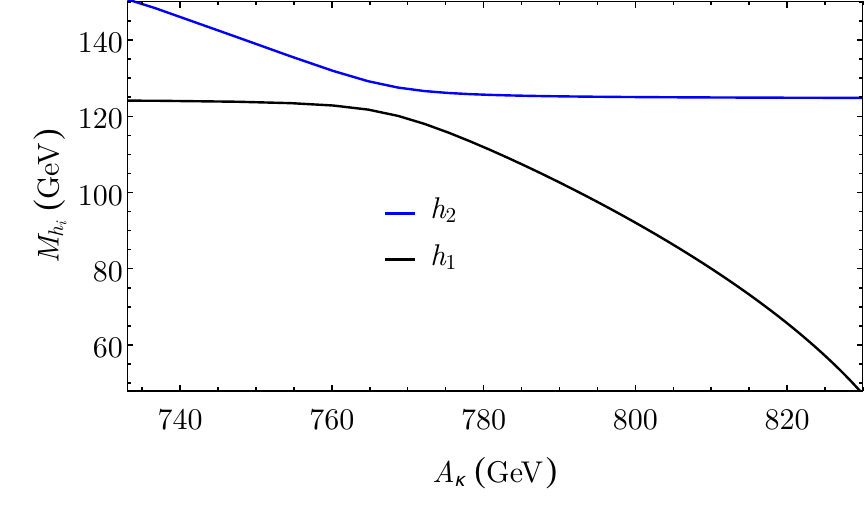}
  \vspace{-2ex}
  \caption{\label{fig:hS_100GeV_TB12_mass} Higgs masses in the
    scenario of \reffi{fig:hS_100GeV_TB12}.}
  \vspace{4ex}
  \capstart
  \includegraphics[width=\textwidth]{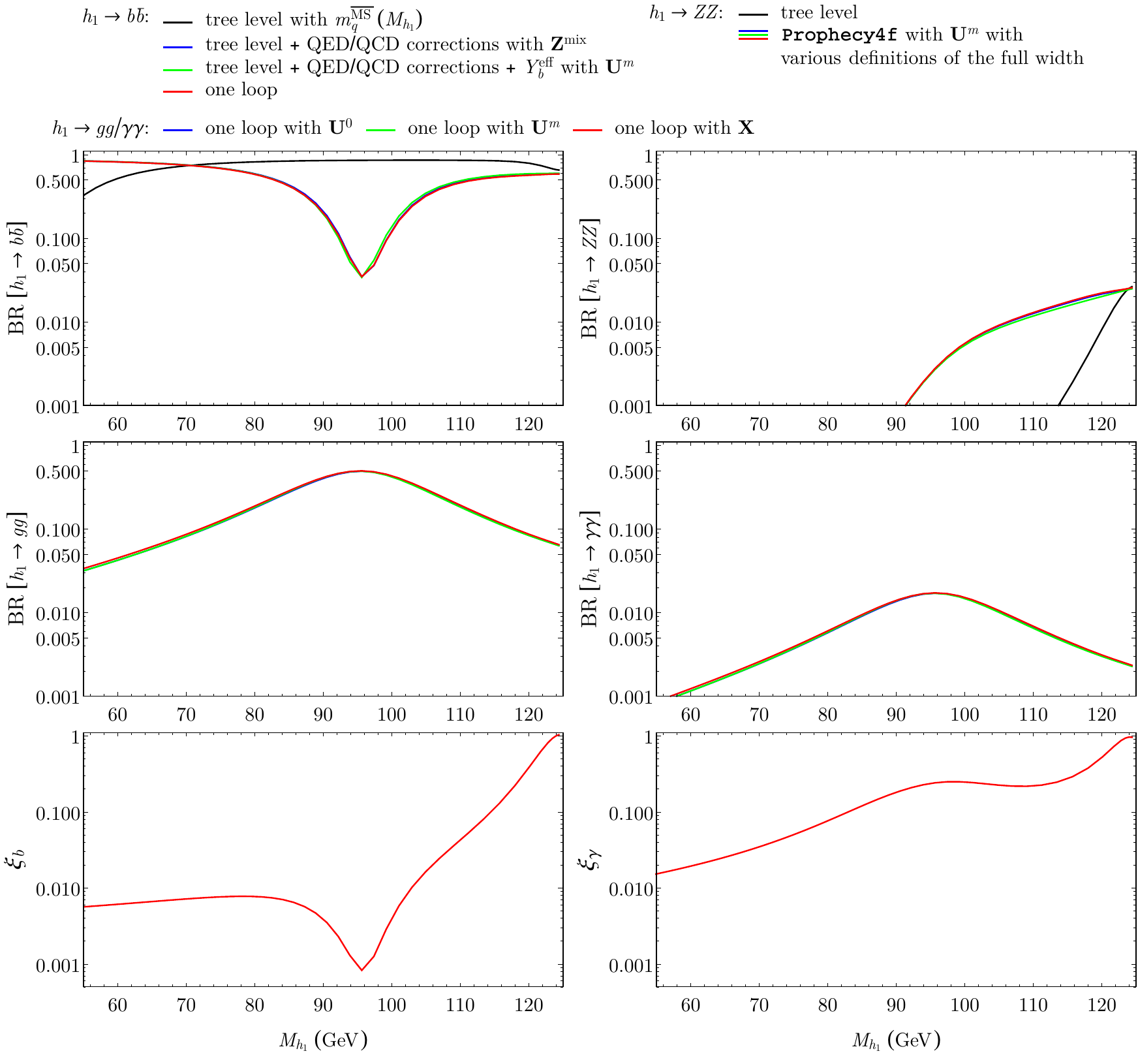}
  \vspace{-4ex}
  \caption{\label{fig:hS_100GeV_TB12} The decay properties of the
    mostly singlet~\cp-even Higgs state as a funcion of its mass are
    shown. The input parameters are provided in
    \refta{tab:scenarios}. The branching ratios into~$b\bar{b}$, $ZZ$,
    $gg$ and~$\gamma\gamma$ are shown at the tree level~(black), under
    partial one-loop approximations~(blue, green) or at the full
    one-loop order~(red). The plots at the bottom show the
    quantities~$\xi_b$ and~$\xi_{\gamma}$ of Eqs.\,\eqref{eq:xi},
    estimating the signals associated with~$h_1$ in
    the~$b\bar{b}$~channel at~LEP and in the~$\gamma\gamma$~channel at
    the~LHC, as compared to an~SM~Higgs at the same mass. Explicit
    values at~\mbox{$M_{h_1}=95.4$}\,GeV are given in
    \refta{tab:examples}.}
\end{figure}

\medskip
We first consider a region of the~NMSSM parameter space characterized
by the input displayed in the column `\reffi{fig:hS_100GeV_TB12}' of
\refta{tab:scenarios}: we scan over~$A_{\kappa}$
with~\mbox{$\tan\beta=12$}. In this regime, the lightest~\cp-even
state is singlet-like (except at the upper boundary
in~$A_{\kappa}$). Its mass varies within the
range~$[50,125]$\,GeV. The second-lightest Higgs has~SM-like
properties with a mass of~$\simord125$\,GeV. The variations of these
masses as functions of~$A_{\kappa}$ are shown in
\reffi{fig:hS_100GeV_TB12_mass}. We do not discuss the~\cp-odd singlet
at~$\simord700$\,GeV and the heavy doublet states at~$\simord1.4$\,TeV
in this context. The experimental constraints from Higgs searches, as
summarized in \texttt{HiggsBounds} and \texttt{HiggsSignals}, are
satisfied over the whole interval. In \reffi{fig:hS_100GeV_TB12}, we
display the branching ratios of the lightest~\cp-even Higgs state
into~$b\bar{b}$, $ZZ$, $gg$ and $\gamma\gamma$---the total width is
calculated as the sum of the decay widths in the~$b\bar{b}$,
$\tau^+\tau^-$, $c\bar{c}$, $WW$, $ZZ$, $gg$
and~$\gamma\gamma$~channels. The tree-level results (including a
tree-level evaluation of the full width, \EG~excluding the~$gg$
or~$\gamma\gamma$~channels) are shown in black while the full one-loop
results appear in red. Various approximations of the one-loop
branching ratios are shown in blue and green: they are in good
agreement with the full one-loop prediction for all the considered
channels. We observe sizable deviations for the tree-level and
one-loop predictions of the branching fraction into~$b\bar{b}$: this
is due to the existence of a region with
suppressed~$h_1$--$b\bar{b}$~coupling arising as a consequence of the
mixing between the three neutral~\cp-even Higgs states. This
suppressed region occurs below the displayed range of~$M_{h_1}$ at the
tree level, but is shifted to~\mbox{$M_{h_1}\sim90$--$100$}\,GeV at
the one-loop order. At its minimum for~\mbox{$M_{h_1}\simeq95$}\,GeV,
the~\mbox{$h_1\to b\bar{b}$}~decay is in fact dominated by
its~$\mathcal{O}(\alpha_s^2)$~contribution from~\mbox{$h_1\to g
  (g^*\to b\bar{b})$}. Concerning the~$ZZ$~final state, the difference
between tree-level and one-loop branching fractions is largely driven
by the magnitude of the full width: the large~\mbox{$h_1\to gg$}
contributions are absent at the tree level, and the
large~\mbox{$h_1\to b\bar{b}$}~contributions differ sizably. The
branching fractions into~$gg$ and~$\gamma\gamma$ exist only at the
loop level. They show a peak at~\mbox{$M_{h_1}\sim90$--$100$}\,GeV,
which mostly originates from the suppression of the~\mbox{$h_1\to
  b\bar{b}$}~decay in this range. In particular, the~$gg$~final state
contributes about~$40\%$ to the total width. Finally, we display the
quantities~$\xi_b$ and~$\xi_{\gamma}$, defined as follows:
\begin{subequations}\label{eq:xi}
\begin{align}
& \xi_b\equiv\frac{\Gamma[h_1\to ZZ]\cdot \mbox{BR}[h_1\to b\bar{b}]}{\Gamma[H_{\mbox{\tiny SM}}(M_{h_1})\to ZZ]\cdot \mbox{BR}[H_{\mbox{\tiny SM}}(M_{h_1})\to b\bar{b}]}\sim\frac{\sigma[e^+e^-\to Z (h_1\to b\bar{b})]}{\sigma[e^+e^-\to Z (H_{\mbox{\tiny SM}}(M_{h_1})\to b\bar{b})]}\\
& \xi_{\gamma}\equiv\frac{\Gamma[h_1\to gg]\cdot \mbox{BR}[h_1\to \gamma\gamma]}{\Gamma[H_{\mbox{\tiny SM}}(M_{h_1})\to gg]\cdot \mbox{BR}[H_{\mbox{\tiny SM}}(M_{h_1})\to \gamma\gamma]}\sim\frac{\sigma[gg\to h_1\to \gamma\gamma]}{\sigma[gg\to H_{\mbox{\tiny SM}}(M_{h_1})\to\gamma\gamma]}\ .
\end{align}
\end{subequations}
These definitions of~$\xi_{b,\gamma}$ give loose estimates of the
signals that~$h_1$ would generate in the~LEP~searches
for~\mbox{$e^+e^-\to Z (H\to b\bar{b})$} and the~LHC~searches
for~\mbox{$pp\to H\to\gamma\gamma$}, normalized to
the~SM~cross-sections. The decay widths of an~SM-Higgs boson are
evaluated with our code, taking the~SM-limit of the~NMSSM. In our
example, the signal in the~$b\bar{b}$~searches would thus reach a few
percent of an~SM~signal for~\mbox{$M_{h_1}\sim100$--$110$\,GeV}, but
would be very suppressed in the
range~\mbox{$M_{h_1}\sim90$--$100$\,GeV}, due to the small branching
fraction into~$b\bar{b}$. On the other hand, the signal in the
diphoton~channel would amount to~$\simord20\%$ of the
corresponding~SM~cross-section
for~\mbox{$M_{h_1}\sim90$--$100$\,GeV}. The Higgs properties for a
specific point in this scenario are provided in the column
`\reffi{fig:hS_100GeV_TB12}' of \refta{tab:examples}. The production
cross-sections, approximated by~$\widehat{\Gamma}_{ZZ}$ for~LEP
and~$\widehat{\Gamma}_{gg}$ for the~LHC (in gluon--gluon fusion),
amount to a comparable fraction---a few percent---of the
corresponding~SM-Higgs cross-sections. The branching ratio
into~$b\bar{b}$ is considerably reduced, as compared to an~SM-Higgs
state at the same mass, whereas the diphoton~branching ratio is
enhanced by up to a factor~$\simord10$. This is caused by the mixing
between the three neutral Higgs states, which in this parameter region
gives rise to a very small~$H_d^0$~component of the mostly
singlet-like state implying suppressed couplings to down-type quarks
(and leptons). Thus, the mechanism leading to a large diphoton~signal
in this scenario is purely driven by the decay properties of the light
singlet state. This scenario would therefore give rise to
a~$\gamma\gamma$~signal at the~LHC without a
corresponding~$b\bar{b}$~excess at~LEP. In fact, the reverse is also
possible: an~NMSSM-Higgs singlet could cause an excess in
the~\mbox{$e^+e^-\to Z(H\to b\bar{b})$}~channel without generating a
significant signal in~\mbox{$pp\to H\to\gamma\gamma$}. A relatively
large~$b\bar{b}$~decay rate can naturally occur for a Higgs state in
the~$100$\,GeV mass range, in which case the diphoton~branching
fraction remains small.

\begin{table}[t]
\footnotesize
\newcommand{\fancyhline}{\hhline{||-||->{\arrayrulecolor{lightgray}}->{\arrayrulecolor{black}}||>{\arrayrulecolor{black}}->{\arrayrulecolor{lightgray}}->{\arrayrulecolor{black}}||>{\arrayrulecolor{black}}->{\arrayrulecolor{lightgray}}->{\arrayrulecolor{black}}||}}
\newcommand{\fancydiag}[1]{\color{lightgray}{\diagbox[dir=SW,height=#1,width=53pt]{}{}}}
\centering
\begin{tabular}{||c||c|c||c|c||c|c||}
\hhline{~|t:==:t:==:t:==:t|}
\multicolumn{1}{c||}{} & \multicolumn{2}{c||}{\reffi{fig:hS_100GeV_TB12}} & \multicolumn{2}{c||}{\reffi{fig:hS_100GeV_TB2}} & \multicolumn{2}{c||}{\reffi{fig:hS_100GeV_CPV}}\\
\hhline{~||--||--||--||}
\multicolumn{1}{c||}{} & $h_1$ & $h_2$ & $h_1$ & $h_2$ & $h_1$ & $h_2$ \\
\hhline{|t:=::==::==::==:|}
$M_{h_i}$ & $95.4$ & $125.1$ & $95.0$ & $125.5$ & $95.1$ & $127.0$ \\
\hhline{|:=::==::==::==:|}
$\Gamma_{b\bar{b}}$ & $1.9\cdot10^{-7}$ & $2.4\cdot10^{-3}$ & $1.4\cdot10^{-4}$ & $2.1\cdot10^{-3}$ & $4.0\cdot10^{-4}$ & $1.6\cdot10^{-3}$ \\
\hhline{||-||--||--||--||}
$\Gamma_{\tau\tau}$ & $3.8\cdot10^{-9}$ & $2.7\cdot10^{-4}$ & $1.5\cdot10^{-5}$ & $2.4\cdot10^{-4}$ & $4.3\cdot10^{-5}$ & $1.8\cdot10^{-4}$ \\
\hhline{||-||--||--||--||}
$\Gamma_{c\bar{c}}$ & $2.2\cdot10^{-6}$ & $1.3\cdot10^{-4}$ & $1.8\cdot10^{-5}$ & $1.2\cdot10^{-4}$ & $2.3\cdot10^{-5}$ & $1.1\cdot10^{-4}$ \\
\hhline{||-||--||--||--||}
$\Gamma_{WW}$ & $1.2\cdot10^{-7}$ & $8.8\cdot10^{-4}$ & $8.4\cdot10^{-6}$ & $7.9\cdot10^{-4}$ & $1.2\cdot10^{-6}$ & $7.9\cdot10^{-4}$ \\
\hhline{||-||--||--||--||}
$\Gamma_{ZZ}$ & $1.4\cdot10^{-8}$ & $1.1\cdot10^{-4}$ & $1.0\cdot10^{-7}$ & $9.7\cdot10^{-5}$ & $1.5\cdot10^{-7}$ & $9.9\cdot10^{-5}$ \\
\hhline{||-||--||--||--||}
$\Gamma_{gg}$ & $2.6\cdot10^{-6}$ & $2.5\cdot10^{-4}$ & $2.2\cdot10^{-5}$ & $2.2\cdot10^{-4}$ & $2.8\cdot10^{-5}$ & $2.1\cdot10^{-4}$ \\
\hhline{||-||--||--||--||}
$\Gamma_{\gamma\gamma}$ & $9.0\cdot10^{-8}$ & $9.0\cdot10^{-6}$ & $6.3\cdot10^{-7}$ & $7.6\cdot10^{-6}$ & $7.2\cdot10^{-7}$ & $7.1\cdot10^{-6}$
\\
\hhline{|:=::==::==::==:|}
\rule{0pt}{12pt}$\widehat{\Gamma}_{ZZ}$ & $0.02$ & \fancydiag{19pt} & $0.15$ & \fancydiag{19pt} & $0.22$ & \fancydiag{19pt}\\
\fancyhline
\rule{0pt}{12pt}$\widehat{\Gamma}_{gg}$ & $0.02$ & \fancydiag{19pt} & $0.18$ & \fancydiag{19pt} & $0.23$ & \fancydiag{19pt}\\
\fancyhline
\rule{0pt}{12pt}$\widehat{\text{BR}}_{b\bar{b}}$ & $0.04$ & \fancydiag{19pt} & $0.88$ & \fancydiag{19pt} & $1.0$ & \fancydiag{19pt}\\
\fancyhline
\rule{0pt}{12pt}$\widehat{\text{BR}}_{\gamma\gamma}$ & $11.8$ & \fancydiag{19pt} & $2.2$ & \fancydiag{19pt} & $1.0$ & \fancydiag{19pt}\\
\fancyhline
$\xi_b$ & $8.4\cdot10^{-4}$ & \fancydiag{14pt} & $0.13$ & \fancydiag{14pt} & $0.22$ & \fancydiag{14pt}\\
\fancyhline
$\xi_{\gamma}$ & $0.26$ & \fancydiag{14pt} & $0.41$ & \fancydiag{14pt} & $0.23$ & \fancydiag{14pt}\\
\hhline{|b:=:b:==:b:==:b:==:b|}
\end{tabular}
\caption{\label{tab:examples} The Higgs properties for a few example
  points in Sect.\,\ref{sec:singlet} are shown. The Higgs width
  into~$xx'$ is denoted by~$\Gamma_{xx'}$, and the width normalized to
  the~SM~width at the same mass is represented
  by~$\widehat{\Gamma}_{xx'}$. The symbol~$\widehat{\text{BR}}_{xx'}$
  represents the Higgs branching ratio into~$xx'$, normalized to
  the~SM~branching ratio at the same mass.  The mass values and the
  partial decay widths are are given in~GeV. The variation in the
  values of~$M_{h_1}$ is an artefact of the scans performed for the
  three scenarios.}
\end{table}

\begin{figure}[tp!]
  \centering
  \includegraphics[width=.5\textwidth]{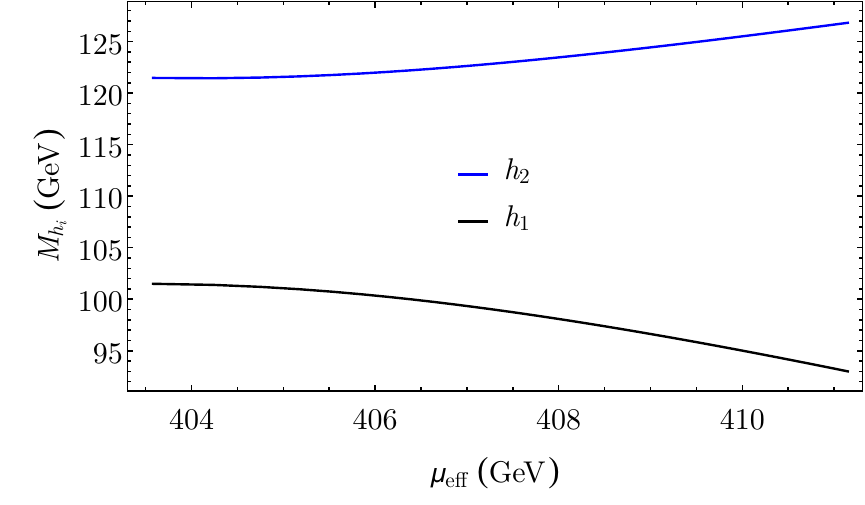}
  \vspace{-2ex}
  \caption{\label{fig:hS_100GeV_TB2_mass} Higgs masses in the scenario
    of \reffi{fig:hS_100GeV_TB2}.}
  \vspace{4ex}
  \capstart
  \includegraphics[width=\textwidth]{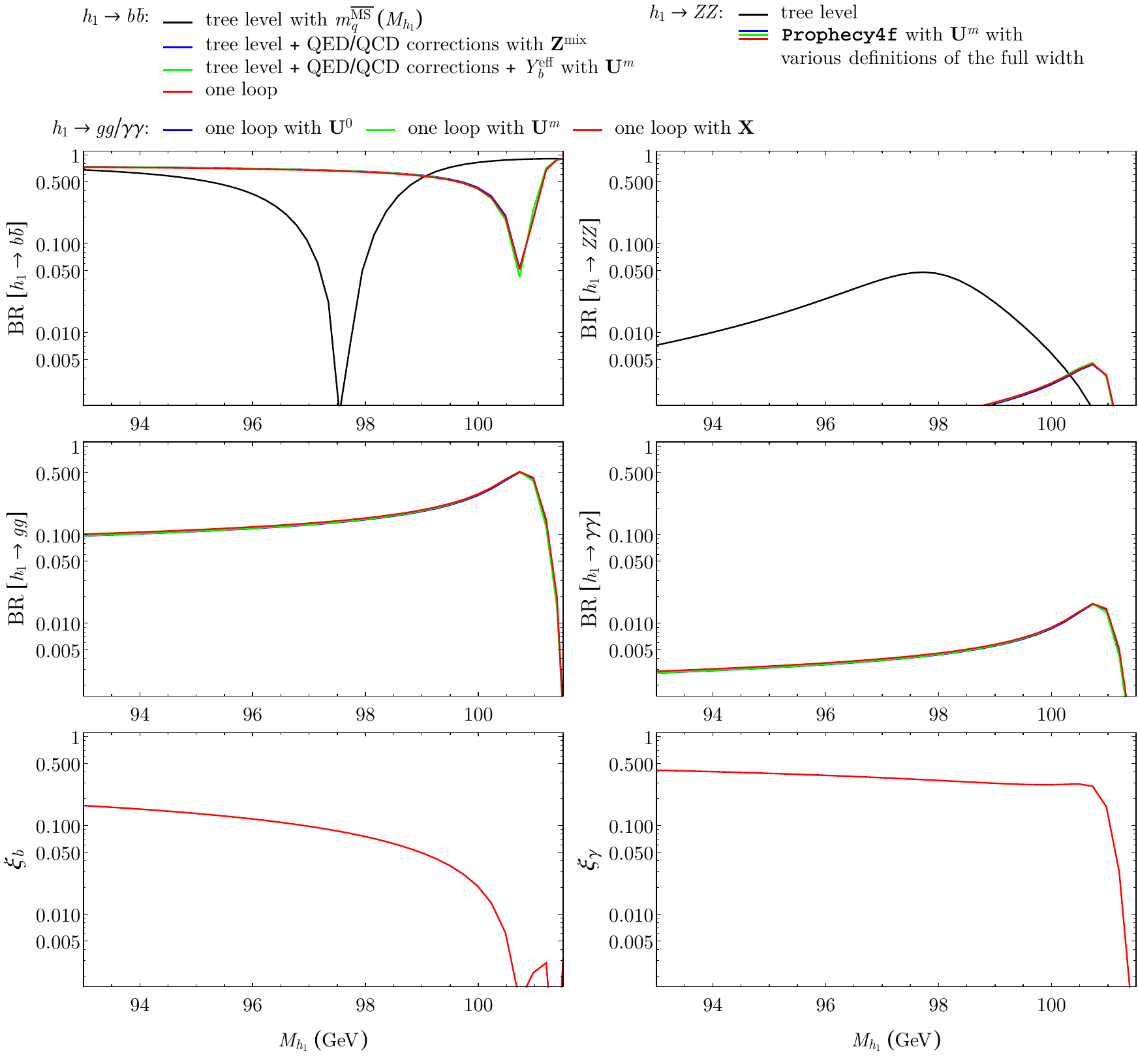}
  \vspace{-4ex}
  \caption{\label{fig:hS_100GeV_TB2} The decay properties of the
    mostly singlet-like~\cp-even Higgs state are depicted as a
    function of its mass. The input parameters are provided in
    \refta{tab:scenarios}. The branching ratios into~$b\bar{b}$, $ZZ$,
    $gg$ and~$\gamma\gamma$ are shown at the tree level~(black), under
    partial one-loop approximations~(blue, green) and at the full
    one-loop order~(red). The plots at the bottom show the
    quantities~$\xi_b$ and~$\xi_{\gamma}$ of Eqs.\,\eqref{eq:xi},
    estimating the signals associated with~$h_1$ in
    the~$b\bar{b}$~channel at~LEP and in the~$\gamma\gamma$~channel at
    the~LHC, as compared to an~SM~Higgs at the same mass. Explicit
    values at~\mbox{$M_{h_1}=95.0$}\,GeV are given in
    \refta{tab:examples}.}
\end{figure}

\medskip
We now turn to another scenario in the low-$\tan\beta$~regime with
large $\lambda$: the chosen input is provided in
column~`\reffi{fig:hS_100GeV_TB2}' of \refta{tab:scenarios}. We
vary~$\mu_{\mbox{\tiny eff}}$ in a narrow interval. The masses of the
two lightest Higgs states are shown in
\reffi{fig:hS_100GeV_TB2_mass}. In the lower range of values
for~$\mu_{\mbox{\tiny eff}}$, the mixing between the light singlet
at~$\simord101$\,GeV and the~SM-like state at~$\simord121$\,GeV almost
vanishes. These points are in tension with the measured properties of
the~SM-like state, as tested with \texttt{HiggsSignals}, because of
the relatively low mass of the~SM-like state. However, with
growing~$\mu_{\mbox{\tiny eff}}$, the mixing between the two
light~\cp-even states increases, eventually pushing the singlet mass
down to~$\simord90$\,GeV and the mass of the~SM-like state up
to~$\simord128$\,GeV. Consistency with the experimental results
obtained on the observed state at~$125$\,GeV is achieved for a mass of
the~SM-like state that is compatible with the~LHC~discovery within
experimental and theoretical uncertainties. The~\cp-odd singlet has a
mass of~$\simord150$\,GeV, while the heavy doublet states are
at~$\simord1$\,TeV in this scenario. The decay properties of~$h_1$ are
documented in \reffi{fig:hS_100GeV_TB2}, as well as in the
column~`\reffi{fig:hS_100GeV_TB2}' of \refta{tab:examples} (for a
specific point). The branching ratio into~$b\bar{b}$ changes very
significantly between the tree-level and the one-loop approach: again,
the point of vanishing~$H_d^0$~component in~$h_1$ is shifted in
parameter space from~\mbox{$M_{h_1}\sim98$}\,GeV
to~\mbox{$M_{h_1}\sim101$}\,GeV. On the other hand,
the~$H_u^0$~component in~$h_1$ vanishes
at~\mbox{$M_{h_1}\sim101.5$}\,GeV, leading to a suppression of all the
decay widths into gauge bosons at this mass. The magnitude of the
estimated~\mbox{$e^+e^-\to Z(h_1\to b\bar{b})$}~signal
reaches~$\simord13\%$ of that of an~SM~Higgs
at~\mbox{$M_{h_1}\sim95$}\,GeV, while~\mbox{$pp\to
  h_1\to\gamma\gamma$} corresponds to more than~$40\%$ of an~SM~signal
in the same mass range. In this
example,~\mbox{$\text{BR}[h_1\to\gamma\gamma]$}
(or~\mbox{$\text{BR}[h_1\to gg]$}) is only moderately enhanced with
respect to the~SM~branching fraction due to an~$H_u^0$-dominated
doublet composition of~$h_1$, while~\mbox{$\text{BR}[h_1\to
    b\bar{b}]$} remains dominant, albeit slightly suppressed. This
scenario would thus simultaneously address the~LEP and
the~CMS~excesses in a phenomenologically consistent manner.

\begin{figure}[tp!]
  \centering
  \includegraphics[width=.5\textwidth]{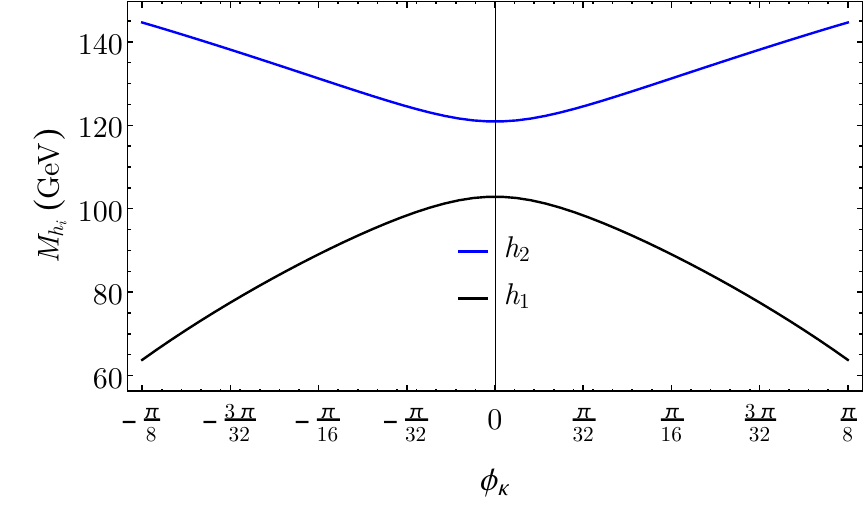}
  \vspace{-2ex}
  \caption{\label{fig:hS_100GeV_CPV_mass} Higgs masses in the scenario
    of \reffi{fig:hS_100GeV_CPV}.}
  \vspace{4ex}
  \capstart
  \includegraphics[width=\textwidth]{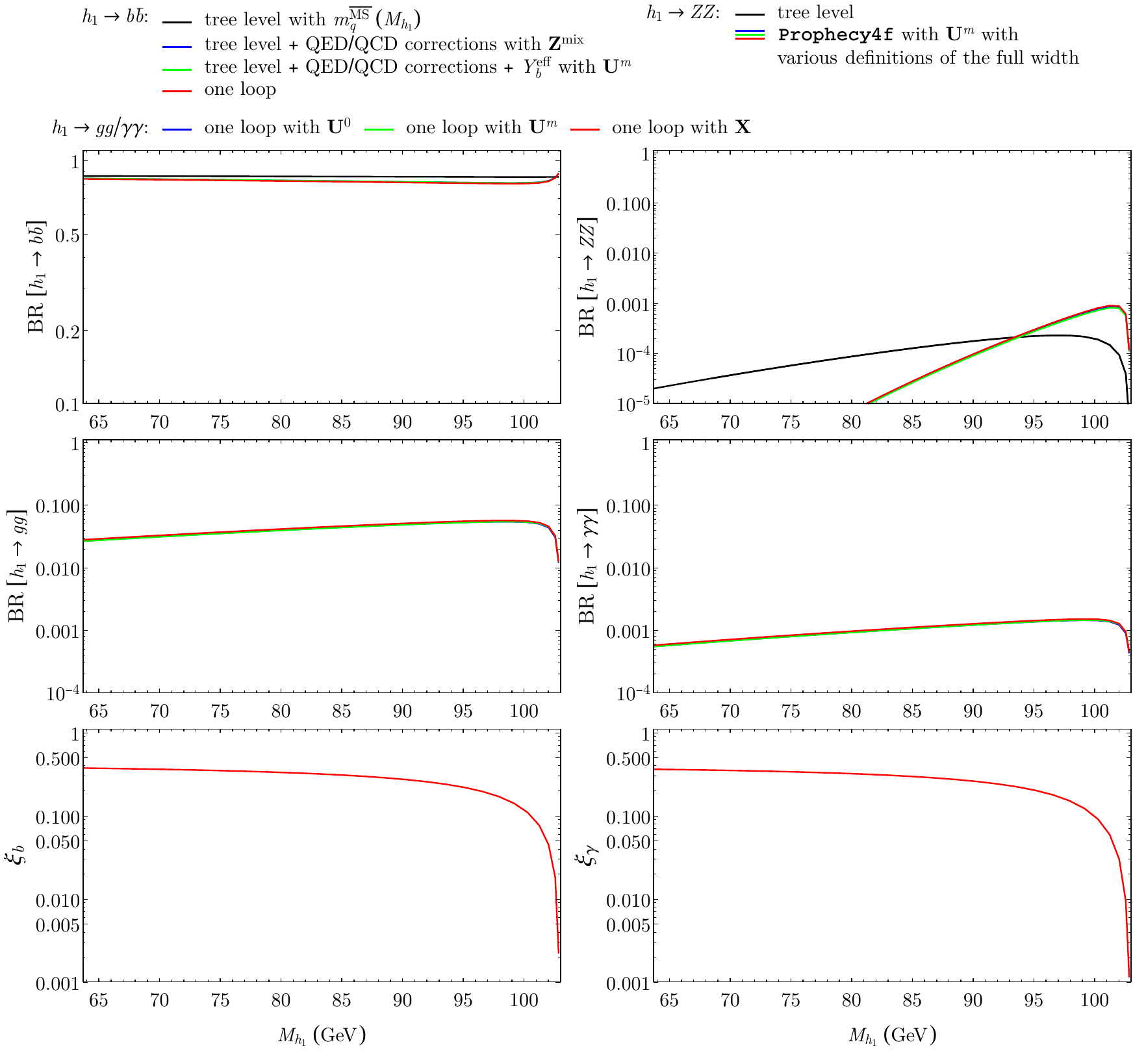}
  \vspace{-4ex}
  \caption{\label{fig:hS_100GeV_CPV} The decay properties of the
    mostly singlet-like~\cp-even Higgs state are depicted as a
    function of its mass. The input parameters are provided in
    \refta{tab:scenarios}. The branching ratios into~$b\bar{b}$, $ZZ$,
    $gg$ and~$\gamma\gamma$ are shown at the tree level~(black), under
    partial one-loop approximations~(blue, green) and at the full
    one-loop order~(red). The plots at the bottom show the
    quantities~$\xi_b$ and~$\xi_{\gamma}$ of Eqs.\,\eqref{eq:xi},
    estimating the signals associated with~$h_1$ in
    the~$b\bar{b}$~channel at~LEP and in the~$\gamma\gamma$~channel at
    the~LHC, as compared to an~SM~Higgs at the same mass. Explicit
    values at~\mbox{$M_{h_1}=95.1$}\,GeV are given in
    \refta{tab:examples}.}
\end{figure}

\medskip
Finally, we consider a~\cp-violating scenario, still in the
large-$\lambda$, low-$\tan\beta$~regime. We scan
over~\mbox{$\phi_{\kappa}\in\left[-\frac{\pi}{8},\frac{\pi}{8}\right]$}. The
lightest Higgs state is dominantly~\cp-odd, with a mass
of~$\simord100$\,GeV in the~\cp-conserving limit. The second-lightest
Higgs is~SM-like, with a mass of~$\simord120$\,GeV
for~\mbox{$\phi_{\kappa}=0$}. With
increasing~$\lvert\phi_{\kappa}\rvert$, these states mix and the
masses draw apart, reaching~$\simord60$\,GeV and~$\simord150$\,GeV
at~\mbox{$\lvert\phi_{\kappa}\rvert\sim\frac{\pi}{8}$}, which can be
seen in \reffi{fig:hS_100GeV_CPV_mass}. Correspondingly, appropriate
Higgs properties, as tested with \texttt{HiggsBounds} and
\texttt{HiggsSignals}, are obtained
for~\mbox{$|\phi_{\kappa}|\simeq0.1$--$0.2$}. The~\cp-even singlet and
the heavy doublet states have masses of the order of~$210$\,GeV
and~$1180$\,GeV, respectively. In \reffi{fig:hS_100GeV_CPV}, we show
the branching ratios of the lightest Higgs state as a function of its
mass. Contrarily to the previous cases,~\mbox{$\text{BR}[h_1\to
    b\bar{b}]$} is nearly constant over the whole range; the
tree-level and one-loop results agree reasonably well with each other.
The branching ratios into gauge bosons show an abrupt decrease
near~\mbox{$M_{h_1}\sim102$}\,GeV: this corresponds to
the~\cp-conserving limit of our scenario---in which case~$h_1$ is a
pure~\cp-odd state. The estimated signals in the~\mbox{$e^+e^-\to
  Z(h_1\to b\bar{b})$} and~\mbox{$pp\to h_1\to\gamma\gamma$}~channels
reach~$\simord20$--$25\%$ of their~SM~counterparts
at~\mbox{$M_{h_1}\sim95$}\,GeV. As can be seen in
\refta{tab:examples}, the decays of~$h_1$ for this
point~(\mbox{$|\phi_{\kappa}|\sim0.14$}) approximately stay in~SM-like
proportions. The~$b\bar{b}$ and~$\gamma\gamma$~signals would thus
remain comparable in magnitude. In particular, a diphoton~signal as
large as~$60\%$ of the one for an~SM-Higgs boson could not be
accommodated in this configuration, as the~LEP~limits on~\mbox{$h_1\to
  b\bar{b}$} indirectly constrain the diphoton~rate. Yet, it is
remarkable that a mostly~\cp-odd Higgs state would trigger sizable
signals at both~LEP and the~LHC.

\medskip
To summarize this discussion, a light~\cp-even---or
dominantly~\cp-even and even dominantly~\cp-odd in the~\cp-violating
case---and mostly singlet-like Higgs state in the vicinity
of~$\lsimord100$\,GeV could have interesting consequences for the
phenomenology of the~SM-like state. Sizable signals in
the~\mbox{$e^+e^-\to Z(h_1\to b\bar{b})$} and/or~\mbox{$pp\to
  h_1\to\gamma\gamma$}~channels are possible, independently or
simultaneously. They could thus explain the corresponding~`excesses'
reported by~LEP and~CMS, respectively. Further experimental effort is
required to investigate this scenario.\footnote{While this work was in
  its finalizing stages, preliminary Run\,II~results from~ATLAS
  with~$80$\,fb$^{-1}$ in the~$\ga\ga$~searches below~$125$\,GeV were
  released\,\cite{ATLAS:2018xad}. No significant excess above
  the~SM~expectation was observed in the mass range~$[65,110]$\,GeV.}

\subsection[Discussion concerning the remaining theoretical 
uncertainties]{\tocref{Discussion concerning the remaining 
theoretical uncertainties}}

Below, we provide a summary of the main sources of theoretical
uncertainties from unknown higher-order corrections applying to our
calculation of the~NMSSM Higgs decays. We do not discuss here the
parametric theoretical uncertainties arising from the experimental
errors of the input parameters. For the experimentally
known~SM-type~parameters the induced uncertainties can be determined
in the same way as for the~SM~case (see
\EG~Ref.\,\cite{Denner:2011mq}). The dependence on the
unknown~SUSY~parameters, on the other hand, is usually not treated as
a theoretical uncertainty but rather exploited for setting indirect
constraints on those parameters.

\paragraph{Higgs decays into quarks (\boldmath$h_i\to q\bar{q}$, $q=c,b,t$)}
In our evaluation, these decays have been implemented at full one-loop
order, \IE~at~QCD, electroweak and~SUSY next-to-leading
order~(NLO). In addition, leading~QCD~logarithmic effects have been
resummed within the parametrization of the Yukawa couplings in terms
of a running quark mass at the scale of the Higgs mass. The Higgs
propagator-type corrections determining the mass of the considered
Higgs particle as well as the wave function normalization at the
external Higgs leg of the process contain full one-loop and dominant
two-loop contributions.

For an estimate of the remaining theoretical uncertainties, several
higher-order effects should be taken into account:
\begin{itemize}
\item First, we should assess the magnitude of the
  missing~QCD~NNLO~(two-loop)~effects. We stress that there should be
  no large logarithms associated to these corrections, since these are
  already resummed through the choice of running parameters and the
  renormalization scale. For the remaining~QCD~pieces, we can directly
  consider the situation in the~SM. In the case of the light quarks,
  the~QCD~contributions of higher order have been evaluated and amount
  to~$\simord 4\%$ at~\mbox{$m_{H}=120$}\,GeV (see
  \EG~Ref.\,\cite{Baikov:2005rw}). For the top quark, the uncertainty
  due to missing~QCD~NNLO~effects was estimated
  to~$5\%$\,\cite{Denner:2011mq}.
\item Concerning the electroweak corrections, \reffi{fig:MSSM_hff}
  suggests that the one-loop contribution is small---at the percent
  level---for an~SM-like Higgs, which is consistent with earlier
  estimates in the~SM\,\cite{Denner:2011mq}. For the heavy Higgs
  states, \reffi{fig:MSSM_hff} indicates a larger impact of such
  effects---at the level of~$\simord10\%$ in the considered
  scenario. Assuming that the electroweak~NNLO~corrections are
  comparable to the squared one-loop effects, our estimate for pure
  electroweak higher orders in decays of heavy Higgs states reaches
  the percent level. In fact, for \mbox{multi-TeV}~Higgs bosons, the
  electroweak Sudakov~logarithms may require a
  resummation. Furthermore, mixed electroweak--QCD~contributions are
  expected to be larger than the pure electroweak~NNLO~corrections,
  adding a few more percent to the uncertainty budget. For light Higgs
  states, the electroweak effects are much smaller since the
  Sudakov~logarithms remain of comparatively modest size.
\item Finally, the variations with the squark masses in
  \reffi{fig:MSSM_hff} for the heavy doublet states show that the
  one-loop~SUSY~effects could amount to~$5$--$10\%$ for a
  sub-TeV~stop/sbottom spectrum. In such a case, the two-loop~SUSY and
  the mixed~QCD/electroweak--SUSY~corrections may reach the percent
  level. On the other hand, for very heavy squark spectra, we expect
  to recover an effective singlet-extended Two-Higgs-Doublet~model (an
  effective~SM if the heavy doublet and singlet states also decouple)
  at low energy. However, all the parameters of this low-energy
  effective field theory implicitly depend on the~SUSY~radiative
  effects, since unsuppressed logarithms of~SUSY~origin generate terms
  of dimension~$\leqord4$---\EG~in the Higgs potential or the Higgs
  couplings to~SM~fermions. On the other hand, the explicit dependence
  of the Higgs decay widths on~SUSY~higher-order corrections is
  suppressed for a large~SUSY~scale. In this case, the uncertainty
  from~SUSY~corrections reduces to a parametric effect, that of the
  matching between the~NMSSM and the low-energy lagrangian---\EG~in
  the~SM-limit, the uncertainty on the mass prediction for the~SM-like
  Higgs continues to depend on~SUSY~logarithms and would indirectly
  impact the uncertainty on the decay widths.
\end{itemize}
Considering all these higher-order effects together, we conclude that
the decay widths of the~SM-like Higgs should be relatively well
controlled (up to~$\simord5\%$), while those of a heavy Higgs state
could receive sizable higher-order contributions, possibly adding up
to the level of~$\simord10\%$.

\paragraph{Higgs decays into leptons}
Here,~QCD~corrections appear only at two-loop order in the Higgs
propagator-type corrections as well as in the counterterms of the
electroweak parameters and only from three-loop order onwards in the
genuine vertex corrections. Thus, the theory uncertainty is expected
to be substantially smaller than in the case of quark~final
states. For an~SM-like Higgs, associated uncertainties were estimated
to be below the percent~level\,\cite{deFlorian:2016spz}. For heavy
Higgs states, however, electroweak one-loop corrections are enhanced
by Sudakov~logarithms and reach the~$\simord10\%$~level for Higgs
masses of the order of~$1$\,TeV, so that the two-loop effects could
amount to a few~percent. In addition, light~staus may generate a
sizable contribution of~SUSY~origin, where the unknown corrections are
of two-loop electroweak order.

\paragraph{Higgs decays into \boldmath $WW/ZZ$}
The complexity of these channels is illustrated by our presentation of
two separate estimates, expected to perform differently in various
regimes.
\begin{itemize}
\item In the~SM, the uncertainty of \texttt{Prophecy4f} in the
  evaluation of these channels was assessed at the sub-percent~level
  below~$500$\,GeV, but up to~$\simord15\%$
  at~$1$\,TeV\,\cite{Denner:2011mq}. For an~SM-like Higgs, our
  \reffi{fig:MSSM_hVV} shows that the one-loop electroweak corrections
  are somewhat below~$10\%$, making plausible a
  sub-percent~uncertainty on the results employing
  \texttt{Prophecy4f}. On the other hand, the assumption that the
  decay widths for an~NMSSM Higgs boson can be obtained through a
  simple rescaling of the result for the width in the~SM by tree-level
  couplings, is in itself a source of uncertainties. We expect this
  approximation to be accurate only in the limit of a
  decoupling~SM-like composition of the~NMSSM Higgs boson. If
  these~SM-like characteristics are altered through radiative
  corrections of~SUSY~origins or~NMSSM-Higgs mixing effects---both of
  which may still reach the level of several~percent in a
  phenomenologically realistic setup---the uncertainty on the
  rescaling procedure for the decay widths should be of corresponding
  magnitude.
\item In the case of heavier states, \reffi{fig:MSSM_hVV} and
  \reffi{fig:NMSSM_hVV} indicate that the previous procedure is
  unreliable in the mass range~$\gsimord500$\,GeV. In particular, for
  heavy doublets in the decoupling limit, radiative corrections
  dominate over the---then vanishing---tree-level amplitude, shifting
  the widths by orders of magnitude. In such a case, our one-loop
  calculation captures only the leading order and one can expect
  sizable contributions at the two-loop level: as we already mentioned
  in discussing \reffi{fig:MSSM_hVV}, shifting the quark masses
  between pole and~\MSbar~values---two legitimate choices at the
  one-loop order that differ in the treatment of~QCD~two-loop
  contributions---results in modifications of the widths of
  order~$\simord50\%$. On the other hand, one expects the decays of a
  decoupling heavy doublet into electroweak gauge bosons to remain a
  subdominant channel, so that a less accurate prediction may be
  tolerable. It should be noted, however, that the magnitude of the
  corresponding widths is sizably enhanced by the effects of one-loop
  order, which may be of interest regarding their phenomenological
  impact.
\end{itemize}

\paragraph{Radiative decays into gauge bosons}
As these channels appear at the one-loop order, our~(QCD-corrected)
results represent (only) an improved leading-order evaluation. Yet the
situation is contrasted:
\begin{itemize}
\item In the~SM, the uncertainty on a Higgs decay into~$\gamma\gamma$
  was estimated at the level of~$1\%$ in Ref.\,\cite{Denner:2011mq}:
  however, the corresponding calculation includes both~QCD~NLO and
  electroweak~NLO~corrections. In our case, only~QCD~NLO~corrections
  (with full mass dependence) are taken into account. The comparison
  with~\texttt{NMSSMCALC} in \reffi{fig:NMSSM_hgg} provides us with a
  lower bound on the magnitude of electroweak~NLO
  and~QCD~NNLO~effects: both evaluations are at the same order but
  differ by a few~percent. The uncertainty on the~SUSY~contribution
  should be considered separately, as light charginos or sfermions
  could have a sizable impact. In any case, we expect the accuracy of
  our calculation to perform at the level of~$\gsimord4\%$ (the
  typical size of the deviations in \reffi{fig:NMSSM_hgg}).
\item In the case of the Higgs decays into~gluons, for
  the~SM~prediction---including~QCD~corrections with full mass
  dependence and electroweak two-loop effects---an uncertainty
  of~$3\%$ from~QCD~effects and~$1\%$ from electroweak effects was
  estimated in \citere{Denner:2011mq}. In our case,
  the~QCD~corrections are only included in the heavy-loop
  approximation, and~NLO~electroweak contributions have not been
  considered. Consequently, the uncertainty budget should settle above
  the corresponding estimate for the~SM quoted above. In the case of
  heavy Higgs bosons, the squark spectrum could have a significant
  impact on the~QCD~two-loop corrections, as exemplified in
  Fig.\,\href{https://arxiv.org/pdf/hep-ph/0612254.pdf#page.11}{5} of
  \citere{Muhlleitner:2006wx}.
\item For~\mbox{$h_i\to \gamma Z$}, QCD~corrections are not available
  so far, so that the uncertainty should be above the~$\simord5\%$
  estimated in the~SM\,\cite{Denner:2011mq}.
\end{itemize}

\paragraph{Additional sources of uncertainty from higher orders}
For an uncertainty estimate, the following effects apply to
essentially all channels and should be considered as well:
\begin{itemize}
\item The mixing in the Higgs sector plays a central role in the
  determination of the decay widths. Following the treatment in \FH,
  we have considered~$\mathbf{Z}^{\mbox{\tiny mix}}$ in all our
  one-loop evaluations, as prescribed by the~LSZ~reduction. Most
  public codes consider a unitary approximation in the limit of the
  effective scalar potential~($\mathbf{U}^0$, in our notation). The
  analysis of \citere{Domingo:2017rhb} and our discussion on
  \reffi{fig:MSSM_hff}---employing~$\mathbf{U}^m$, a more reliable
  unitary approximation than~$\mathbf{U}^0$---indicate that the
  different choices of mixing matrices may affect the Higgs decays by
  a few percent (and far more in contrived cases). However, even the
  use of~$\mathbf{Z}^{\mbox{\tiny mix}}$ is of course subject to
  uncertainties from unknown higher-order corrections. While the Higgs
  propagator-type corrections determining the mass of the considered
  Higgs boson and the wave function normalization contain corrections
  up to the two-loop order, the corresponding prediction for the mass
  of the~SM-like Higgs still has an uncertainty at the level of
  about~$2\%$, depending on the~SUSY~spectrum.
\item In this paper, we confined ourselves to the evaluation of the
  Higgs decay widths into~SM~particles and did not consider the
  branching ratios. For the latter an implementation at the full
  one-loop order of many other two-body decays, relevant in particular
  for the heavy Higgs states, would be desirable, which goes beyond
  the scope of the present analysis. Furthermore, in order to consider
  the Higgs branching ratios at the one-loop order, we would have to
  consider three-body~widths at the tree level, for
  instance~\mbox{$h_i\to b\bar{b}Z$}, since these are formally of the
  same magnitude as the one-loop effects for two-body decays. In
  addition, these three-body~decays---typically real~radiation of
  electroweak and Higgs bosons---exhibit Sudakov~logarithms that would
  require resummation in the limit of heavy Higgs states.
\item At decay thresholds, the approximation of free particles in the
  final state is not sufficient, and a more accurate treatment would
  require the evaluation of final-state interactions. Several cases
  have been discussed in
  \EG~\citeres{Domingo:2011rn,Domingo:2016yih,Haisch:2018kqx}.
\end{itemize}

In this discussion we did not attempt to provide a quantitative
estimate of the remaining theoretical uncertainties from unknown
higher-order corrections, as such an estimate would in any case
sensitively depend on the considered region in parameter space.
Instead, we have pointed out the various sources of higher-order
uncertainties remaining at the level of our
state-of-the-art~evaluation of the Higgs decays into~SM~particles in
the~NMSSM. For a decoupling~SM-like Higgs boson one would ideally
expect that the level of accuracy of the predictions approaches the
one achieved in the~SM. However, even in this limit,
missing~NNLO~pieces---that are known for the~SM, but not for
the~NMSSM---give rise to a somewhat larger theoretical uncertainty in
the~NMSSM. Furthermore, uncertainties of parametric nature (for
instance from the theoretical prediction of the Higgs-boson mass) need
to be taken into account as well. For heavy Higgs states, the impact
of electroweak Sudakov~logarithms and~SUSY~corrections add to the
theoretical uncertainty to an extent that is strongly dependent on the
details of the spectrum and the characteristics of the Higgs
state. For a decoupling doublet at~$\simord1$\,TeV, an uncertainty
of~\mbox{$\simord5$--$15\%$} may be used as a guideline for the
fermionic and radiative decays, while the uncertainty may be as large
as~$\simord50\%$ in~\mbox{$h_i\to WW/ZZ$}.

%% file: 04_Conclusions.tex
\section[Conclusions]{\tocref{Conclusions\label{sec:conclusion}}}

In this paper, we have presented our evaluation of neutral Higgs decay
widths into~SM~final states in the~(\cp-conserving
or~\cp-violating)~NMSSM. Full one-loop corrections have been included
for all the considered channels, as well as
higher-order~QCD~corrections to the decays that are generated at the
radiative level. The inclusion of one-loop contributions to the decays
into~SM~fermions or electroweak gauge bosons goes beyond the usual
approximation amounting to a~QCD/QED-corrected~tree level. In
addition,~QCD~corrections to the~digluon and~diphoton~decay widths
have been carefully processed, including the mass~dependence in
the~$\gamma\gamma$~case and corrections beyond~NLO in
the~$gg$~case. In its current form, this
state-of-the-art~implementation of the neutral Higgs decays
into~SM~particles is available as a \texttt{Mathematica} package, but
should also be integrated into the \texttt{FeynHiggs} code in the near
future.

In order to illustrate this calculation of the Higgs decay widths, we
have presented our results in several regimes of the parameter space
of the~NMSSM. In the~MSSM~limit, we were able to recover the
predictions of \texttt{FeynHiggs} and trace the origins of deviations
from our new results. In particular, we emphasized the relevance of
one-loop contributions in the decays of heavy doublet states into
electroweak gauge bosons, for which the usual estimates based on the
tree-level Higgs--gauge~couplings are not appropriate. Minor effects
in the treatment of~QCD and~QED~corrections have also been
noted. Beyond the~MSSM~limit, we have compared our decay widths to the
output of \texttt{NMSSMCALC}. We observed a qualitative agreement
wherever this could be expected. We also gave an account of the
various sources of theoretical uncertainties from higher-order
corrections and discussed the achieved accuracy of our predictions.

As a phenomenological application, we investigated in particular the
case of a mostly singlet-like state with mass in the vicinity
of~$\lsimord100$\,GeV. The decays of such a state can be notably
affected by suppressed couplings to down- or up-type~quarks which can
occur in certain parameter regions as a consequence of the mixing
between the different Higgs states. In particular, an additional Higgs
boson~$h_i$ of this kind could manifest itself via signatures in the
channels~\mbox{$e^+e^- \to Z(h_i\to b\bar{b})$} and/or~\mbox{$pp \to
  h_i \to \gamma\gamma$}.  The presence of such a light Higgs boson
could thus explain the slight deviations from the~SM~predictions
reported by~LEP and~CMS in those channels.